 \documentclass[12pt,draftcls, onecolumn]{IEEEtran}

\usepackage{amssymb}
\usepackage{amsmath}
\usepackage{amsfonts}
\usepackage{graphicx}
\usepackage{algorithm}
\usepackage{algorithmic}
\usepackage{epstopdf}
\usepackage{cite}
\usepackage{stfloats}
\usepackage{url}
\usepackage{CJK}
\usepackage{hyperref}
 
  \usepackage{subfigure}

\usepackage{multirow}

\usepackage{algorithm,algorithmic}
\usepackage{multicol}

\usepackage{caption}
\usepackage{color,xcolor}

\newtheorem{theorem}{\text{Theorem}}

\newtheorem{lemma}{\text{Lemma}}

\newtheorem{remark}{\textbf{Remark}}

\renewcommand\thesubsection{\Alph{subsection}} 
\newtheorem{Prob}{Problem}

\allowdisplaybreaks[4]

\usepackage{verbatimbox}

\newcommand{\mcone}{\textcolor{black}}
\newcommand{\mctwo}{\textcolor{black}}

\newcommand{\mcfour}{\textcolor{black}}
\newcommand{\mcfive}{\textcolor{black}}

\newcommand{\mcele}{\textcolor{black}}
\newcommand{\mctwe}{\textcolor{black}}

\newcommand{\mctht}{\textcolor{black}}
\newcommand{\mcmtwo}{\textcolor{black}}
\newcommand{\mcmthr}{\textcolor{black}}

\newcommand{\mcaone}{\textcolor{black}}

\newcommand{\DeltaQ}{{\Delta_i(Q)}}
\newcommand{\DeltaQf}{{\Delta_{\text{f}}(Q)}}
\newcommand{\DeltaQtf}{{\Delta_{\text{(t,f)}}(Q)}} 

\newcommand{\Tseti}{{\mathcal T_i}}

\newcommand{\GBCD}{{Prop-MLE-Small}}
\newcommand{\FFTBCD}{{Prop-MLE-Large}}
\newcommand{\fsize}{5.8 }
\newcommand{\fsizefourfig}{6 }
\newcommand{\subcap}{-4 }
\newcommand{ \tablecapsize}{0}
\newcommand{  \belowsize}{-4}  
\newcommand{  \figcapsize}{-4}  
 
\newcommand{  \barp}{\bar p}

\begin{document}
\title{MLE-based Device Activity Detection under Rician Fading for Massive Grant-free Access with Perfect and Imperfect Synchronization}


\author{\IEEEauthorblockN{Wang Liu, Ying Cui, Feng Yang, Lianghui Ding, and Jun Sun}\thanks{Wang Liu, Feng Yang, Lianghui Ding, and Jun Sun are with Shanghai Jiao Tong University, China. Ying Cui is with the Hong Kong University of Science and Technology (Guangzhou), China.
{This paper was presented in part at IEEE SPAWC 2022~\cite{SPAWC-22-Liu}.}}}



\maketitle

\begin{abstract}

 Most existing studies on massive grant-free access, proposed to support massive machine-type communications (mMTC) for the Internet of things (IoT), assume Rayleigh fading and perfect synchronization for simplicity. However, in practice, line-of-sight (LoS) components generally exist, and time and frequency synchronization are usually imperfect. 
 This paper systematically investigates maximum likelihood estimation (MLE)-based device activity detection under Rician fading for massive grant-free access with perfect and imperfect synchronization.
 \mcmtwo{We assume that the large-scale fading powers, Rician factors, and normalized LoS components can be estimated offline.}  
\mcmtwo{We} formulate device activity detection in the synchronous case and joint device activity and offset detection in three asynchronous cases (i.e., time, frequency, and time and frequency asynchronous cases) as MLE problems. In the synchronous case, we propose an iterative algorithm to obtain a stationary point of the MLE problem. In each asynchronous case, we propose two iterative algorithms with identical detection performance but different computational complexities. In particular, one is computationally efficient for small ranges of offsets, whereas the other one, relying on fast Fourier transform (FFT) and inverse FFT, is computationally efficient for large ranges of offsets. The proposed algorithms generalize the existing MLE-based methods for Rayleigh fading and perfect synchronization. \mcele{Numerical results show that	
 	the proposed algorithm for the synchronous case can reduce the detection error probability by up to $ 50.4\% $
 	at a $78.6\%$ computation time increase, compared to the MLE-based \mcone{state-of-the-art}, 
 	and 
 	the proposed algorithms for the three asynchronous cases can reduce the detection error probabilities and computation times by up to $ 65.8\% $ and $92.0\%$, respectively, compared to the MLE-based \mcone{state-of-the-arts.}}


\end{abstract}

\begin{IEEEkeywords}
Massive grant-free access, device activity detection, Rician fading, synchronization, time offset, frequency offset, maximum likelihood estimation (MLE), fast Fourier transform (FFT).
\end{IEEEkeywords}

\setlength{\abovedisplayskip}{1.6pt}
\setlength{\belowdisplayskip}{1.6pt}

\section{Introduction}
   Massive machine-type communications (mMTC), one of the three generic services in the fifth
   generation (5G) wireless networks, are expected to provide massive connectivity to a vast number of Internet-of-Things (IoT) devices~\cite{Yaacoub17IWC,ISPM_Liu_2018}.
The fundamental challenge of mMTC is to
enable efficient and timely data transmission from enormous IoT devices that
are cheap, energy-limited, and sporadically active with small packets to send.
Conventional grant-based access
is no longer suitable for mMTC, as the handshaking procedure for obtaining a grant results in excessive control signaling cost and high transmission latency.
Recently, grant-free access,
whose goal is to let devices transmit data in an arrive-and-go manner without waiting for \mcele{the base station (BS)} to schedule a grant,
 has been identified by the Third Generation Partnership Project (3GPP) as a promising solution to support mMTC~\cite{TR-38812}.

\mcele{This paper focuses on a widely investigated grant-free access scheme \mcone{consisting} of a pilot transmission phase and a data transmission phase~\cite{ISPM_Liu_2018}.
	Specifically, devices within a cell are preassigned non-orthogonal pilot sequences that are known to the BS.
	In the pilot transmission phase, 
	active devices send their pilots to the BS \mcfour{simultaneously}, and the BS detects the activity states of all devices and estimates the channel states of the active devices from the received pilot signals.
	In the data transmission phase, all active devices transmit their data to the BS, and the BS detects their transmitted data from the received data signals based on the channels obtained in the pilot transmission phase.}
  The device activity detection and channel estimation at the BS are essential for realizing grant-free access.

  \vspace{-5mm}
 \subsection{\mcmtwo{Related works}}	
 	Existing works \mcmtwo{on device activity detection and channel estimation}
 	mainly rely on compressed sensing (CS), statistical estimation, and machine learning techniques.
 Most works~\cite{TSP_LL1_2018,TWC-21-Li,   TIT_GC_2021,  TSP_Chen_21 ,TWC_Jiang_2021,Arxiv-Jia-22, JSAC_Cui_2020,SPL-Li-19, TWC-22-Shi,JSAC-21-Shao} study only the pilot phase and focus on activity detection and channel estimation.
For instance,~\cite{ TSP_LL1_2018,TWC-21-Li, JSAC_Cui_2020} solve device activity detection
and channel estimation problems using CS techniques, such as approximate message passing (AMP)~\cite{ TSP_LL1_2018} and \mctwe{group least absolute shrinkage and selection operator (LASSO)}~\cite{TWC-21-Li, JSAC_Cui_2020}.
The AMP-based algorithms are computationally efficient but the activity detection and channel estimation accuracies degrade significantly if the number of active devices is larger than the pilot length~\cite{TIT_GC_2021}.
Besides, leveraging statistical estimation techniques, 
\cite{TIT_GC_2021,  TSP_Chen_21 ,TWC_Jiang_2021,Arxiv-Jia-22}  
formulate device activity detection as maximum likelihood estimation (MLE) problems~\cite{TIT_GC_2021,TSP_Chen_21 ,  TWC_Jiang_2021 } and maximum a posterior probability estimation (MAPE) problems~\cite{TWC_Jiang_2021,Arxiv-Jia-22} under flat~\cite{TIT_GC_2021,  TSP_Chen_21 ,TWC_Jiang_2021} and frequency selective~\cite{Arxiv-Jia-22} Rayleigh fading and tackle the non-convex problems using the coordinate descent (CD) method.
The statistical estimation approaches generally outperform the CS-based approaches in activity detection accuracy at the cost of computation complexity increase~\cite{TIT_GC_2021} and have fewer restrictions on the system parameters.
Recently,~\cite{JSAC_Cui_2020,SPL-Li-19, TWC-22-Shi,JSAC-21-Shao} tackle device activity detection and channel estimation using MAPE-based~\cite{JSAC_Cui_2020}, \mctwe{group} LASSO-based~\cite{JSAC_Cui_2020, TWC-22-Shi}, and AMP-based~\cite{JSAC_Cui_2020, JSAC-21-Shao} model-driven neural networks and data-driven neural networks~\cite{SPL-Li-19}.
The machine learning techniques are computational efficient but have no performance guarantee.
 Some works~\cite{BAMP01,BAMP02} consider pilot and data phases and jointly design activity detection, channel estimation, and data detection.
For instance,~\cite{BAMP01} and~\cite{BAMP02} solve the joint activity, channel,
and data estimation problems using \mctwe{bilinear generalized approximate message passing} (BiGAMP).
Note that the BiGAMP-based algorithms~\cite{BAMP01,BAMP02} outperform the AMP-based algorithms in activity detection and channel estimation accuracies at the cost of computation complexity increase.
On the other hand, some works~\cite{ICC_Chen_2019, Wang_ICCASP_21} study grant-free access scheme with data-embedding pilots.
Specifically, there is one phase where each device sends one of its pre-assigned data-embedding pilots, and the BS jointly detects device activity and sent data. For instance,~\cite{ICC_Chen_2019, Wang_ICCASP_21} formulate joint device activity and data detection as MLE problems and tackle the non-convex problems using the CD method~\cite{ICC_Chen_2019 } and projected gradient method~\cite{  Wang_ICCASP_21}. These approaches~\cite{ICC_Chen_2019, Wang_ICCASP_21} are suitable only for very small data payloads as their computational complexities significantly increase with the number of information bits embedded in pilots (or pre-assigned data-embedding pilots per device).

The works mentioned above~\cite{TSP_LL1_2018,TWC-21-Li, JSAC_Cui_2020,TIT_GC_2021,  TSP_Chen_21 ,ICC_Chen_2019,TWC_Jiang_2021,TWC-22-Shi,JSAC-21-Shao,Arxiv-Jia-22  , BAMP01 ,BAMP02, Wang_ICCASP_21} assume that all active devices send their pilots synchronously in time and frequency, referred to as the synchronous case in this paper.
In practice, time synchronization and frequency synchronization are usually imperfect for low-cost IoT devices, due to the lack of coordination between the BS and IoT devices and frequency drifts of cheap crystal oscillators equipped by IoT devices, respectively~\cite{ICM-Yuan-20,IoTJ-CFO2017}. 
If not appropriately handled, the presence of symbol time offsets (STOs) and/or carrier frequency offsets (CFOs) may severely deteriorate the performance
of device activity detection and channel estimation.
\mcele{In the asynchronous cases, the BS has to estimate the unknown activities and channel states together with the unknown offsets of all devices from the same number of observations as in the synchronous case. Besides, the offset of each device lies in a much larger set than the activity state of each device.
	Thus, the device activity detection and channel estimation in the asynchronous cases are more challenging than those in the synchronous case.
Some} recent works primarily investigate
  the time and/or frequency asynchronous cases \mcele{and attempt to address this challenge.}
For instance, 
 \cite{TWC_Tao_2021,ICASSP_LL_2021,SPL-22-Wang} investigate joint device activity detection and channel estimation in the time asynchronous case using an AMP-based model-driven neural network \cite{TWC_Tao_2021}, \mctwe{group} LASSO~\cite{ICASSP_LL_2021}, and an MLE-based method~\cite{SPL-22-Wang}, respectively. For the frequency asynchronous case, \cite{TWC_CFO_2019} deals with joint device activity detection and channel estimation using norm approximation, and \cite{ICCws_CFO_2020} and~\cite{ICC-22-Liu} handle device activity detection using MLE-based methods.
 For the time and frequency asynchronous case,~\cite{TWC-Sun-22} investigates joint device activity detection and channel estimation for an orthogonal frequency division multiplexing (OFDM) system using an AMP-based method.
 The optimization problems in~\cite{SPL-22-Wang, ICASSP_LL_2021,TWC_CFO_2019,ICCws_CFO_2020,ICC-22-Liu} are solved using the majorization-minimization method~\cite{TWC_CFO_2019}, CD method~\cite{ICCws_CFO_2020}, and block coordinate descent (BCD) method~\cite{ICASSP_LL_2021,SPL-22-Wang, ICC-22-Liu}.
Notice that most existing works~\cite{ICASSP_LL_2021,ICCws_CFO_2020,TWC_CFO_2019,TWC-Sun-22} incur much higher computational complexities than those for the synchronous case, 
\mcele{as the dominant terms of their computation costs significantly increase with the offset range.
	When the offset range is large, their computational complexities are exceedingly high and may not be practical.}
To address this issue, our early work~\cite{ICC-22-Liu}  
proposes a low-complexity algorithm, \mcone{whose computation cost (dominant term) does not change with the CFO range}, to solve the MLE problem for the frequency asynchronous case under Rayleigh fading using the BCD method and fast Fourier transform (FFT).

Most existing studies on massive grant-free access that utilize channel statistics to improve estimation performance
assume Rayleigh fading for simplicity~\cite{TSP_LL1_2018,  ICC-22-Liu,JSAC_Cui_2020,TIT_GC_2021,  TSP_Chen_21 ,ICC_Chen_2019,TWC_Jiang_2021,TWC-22-Shi,JSAC-21-Shao, ICCws_CFO_2020, TWC-Sun-22,TWC_Tao_2021,Arxiv-Jia-22 ,Wang_ICCASP_21}.
However, channel measurement
	results have shown that line-of-sight (LoS) components always exist in sub-6 GHz and millimeter wave bands~\cite{APSURSI-Sayeed-11,ICST-Wang-18}.
Hence, Rician fading (including both LoS components and non-line-of-sight (NLoS) components) includes Rayleigh fading (with only NLoS components) as a special case and can be applied to a wider variety of wireless communications systems than Rayleigh fading.
Although device activity detection and channel estimation methods proposed for Rayleigh fading in~\cite{TIT_GC_2021,TWC_Jiang_2021,TSP_Chen_21, SPL-22-Wang,ICC-22-Liu} can be applied to Rician fading in a brute force manner,
	the resulting performance may significantly degrade.
	\mcele{Under Rician fading, the means of the channel states of devices are non-zero and can be significantly different, unlike under Rayleigh fading, where the means of the channel states of devices are zero.
		Thus, \mcone{due to more complex channel statistics}, device activity detection and channel estimation under Rician fading are more challenging than \mcone{those} under Rayleigh fading.}
In~\cite{IoTJ-22-Tian}, the authors attempt to solve the MLE-based device activity detection problem for the synchronous case under Rician fading \mcone{with known Rician factors and LoS components}.
Nevertheless, 
the two-loop iterative algorithm proposed in~\cite{IoTJ-22-Tian} is not computationally efficient, solves only an approximation of the block coordinate optimization problem, and cannot guarantee to obtain a stationary point.
Note that massive grant-free access for the asynchronous cases under Rician fading has not been investigated yet.
 In summary, detecting device activities under Rician fading for massive grant-free access with perfect and imperfect synchronization remains an open problem.\footnote{It is noteworthy that device activity detection is a more fundamental problem as channel conditions of the detected active devices can be subsequently estimated using conventional channel estimation methods~\cite{TIT_GC_2021,TWC_Jiang_2021,TSP_Chen_21}.}
 
 \vspace{-5mm}
 \subsection{\mcmtwo{Contributions}}
In this paper, we would like to shed some light on this problem. 
In particular, we investigate MLE-based device activity detection under Rician fading \mcmtwo{with known large-scale fading powers, Rician factors, and normalized LoS components} for the synchronous case and three asynchronous cases, i.e., time asynchronous case, frequency asynchronous case, and time and frequency asynchronous case.
 The main contributions of this paper are listed as follows.
\begin{itemize}
\item In the synchronous case, we formulate device activity detection under Rician fading as an MLE problem, which is non-convex with a complicated objective function.
Unlike the MLE problem for Rayleigh fading~\cite{TIT_GC_2021,TSP_Chen_21},
the challenge of dealing with the MLE problem for Rician fading lies in how to effectively and efficiently handle the term in the objective function that is related to the LoS components.
 Based on the BCD method, we develop an iterative algorithm (Prop-MLE-Syn), where all block coordinate optimization problems are solved analytically to obtain a stationary point of the MLE problem.
Note that the proposed method successfully generalizes the MLE-based method for Rayleigh fading in the synchronous case~\cite{TSP_Chen_21}.
\item In each of the three asynchronous cases, we formulate joint device activity and offset detection under Rician fading as an MLE problem, which is more challenging than the one in the synchronous case.
Based on the BCD method,  
we develop two iterative algorithms
where all block coordinate optimization problems are solved analytically to tackle the MLE problem.
Note that the two iterative algorithms have identical detection performance but different computational complexities and are suitable for different ranges of offset values.
In particular, one (referred to as \GBCD) is computationally efficient for a small range of offset values, whereas the other one (referred to as \FFTBCD), relying on FFT and inverse fast Fourier transform (IFFT), is computationally efficient for a large range of offset values.
We also analytically compare the computational complexities of the two iterative algorithms.
Note that the proposed algorithms 
successfully generalize the MLE-based methods for Rayleigh fading in the time asynchronous case~\cite{ SPL-22-Wang} and frequency asynchronous case~\cite{ICC-22-Liu}.
\end{itemize}

\mcele{Numerical results show that	
		the proposed algorithm for the synchronous case can reduce the detection error probability by up to $ 50.4\% $
		at a $78.6\%$ computation time increase, compared to the MLE-based \mcone{state-of-the-art}, 
		and 
		the proposed algorithms for the three asynchronous cases can reduce the detection error probabilities and computation times by up to $ 65.8\% $ and $92.0\%$, respectively, compared to the MLE-based \mcone{state-of--the-arts.}}
To our knowledge, this is the first work that utilizes FFT and IFFT techniques to accelerate device activity detection algorithms for massive grant-free access with general imperfect synchronization.
Besides, this is the first work that provides systematic MLE-based device activity detection methods under Rician fading for massive grant-free access with general imperfect synchronization.
The abbreviations in this paper are listed in Table~\ref{tab:abbreviation}.
 
  \begin{table*}[t]
 	  \caption{ \small{Abbreviation}}
 	\centering
    \scriptsize\begin{tabular}{ c|c|c|c }\hline
 	mMTC &   massive machine-type communications     &                   	(B)CD  &    (block) coordinate descent \\ \hline
 	5G    &   the fifth generation                  &                MAPE  &maximum a posterior probability estimation   \\ \hline
 	IoT    &   Internet-of-Things                    &                STO   &   symbol time offset  \\ \hline
 	BS      &  base station                           &                  CFO   &  carrier frequency offset  \\ \hline
 	3GPP   & the Third Generation Partnership Project  &                        OFDM  & orthogonal frequency division multiplexing   \\ \hline
 	CS     &   compressed sensing                       &                 (N)LoS  &    (non)-line-of-sight \\ \hline 	
 	AMP    & approximate message passing                 &                        (I)FFT  &     (inverse) fast Fourier transform \\ \hline
 	   \mctwe{LASSO       }&        \mctwe{least absolute shrinkage and selection operator}                                     &   (I)DFT                         &     (inverse) discrete Fourier  \\ \hline 
 	  \mctwe{BiGAMP}         &        \mctwe{bilinear generalized approximate message passing}                   &           i.i.d.                &    independent and identically distributed  \\ \hline 
 	
 	MLE    &maximum likelihood estimation                 &   \mctwe{p.d.f.}            &   \mctwe{probability density function}     \\ \hline
 	 AWGN    &the additive white Gaussian noise   &       &       \\ \hline
 	\end{tabular}
 	\label{tab:abbreviation} 
 	\vspace{ -6 mm}
 \end{table*}


  \vspace{-2mm}
 \subsection*{Notation}
 	\vspace{-2mm}
 	
 In this paper, unless otherwise stated, we represent scalar constants by non-boldface letters
 	(e.g., $x$ or $X$), vectors
 	by boldface small letters (e.g., $\mathbf x$), matrices by boldface capital letters (e.g., $\mathbf X$),
 	and sets by calligraphic letters (e.g., $\mathcal X$).
 $ \mathbf X^H$, $\mathbf X ^T$, $ \mathbf X^*$, and $\mathrm{tr}(\mathbf X)$  denote the transpose conjugate, transpose, conjugation, and trace of matrix $\mathbf X$, respectively.
 	For $\mathcal A = \{a_1,a_2,...,a_N\}$ where $a_1 <\cdots<a_{N}$ and $\mathcal B = \{b_1,b_2,...,b_{M}\}$ where $b_1 <\cdots<b_{M}$, the notations $( x_a)_{a \in \mathcal A} $, $(\mathbf x_b)_{b \in \mathcal B} $, and $(x_{a,b})_{a\in\mathcal A,b\in\mathcal B} $ represent
 			vector $[   x_{a_1},  x_{a_2},...,  x_{a_N}  ]^T $,
 			matrix $ [\mathbf x_{b_1}, \mathbf x_{b_2},..., \mathbf x_{b_{M}}  ] $, and 
 			a matrix whose $(n,m)$-th element is $  x_{a_n,b_{m}}$,
 			respectively.
 			The notations 
 		$ (\mathbf x)_n$ and $(\mathbf x)_{\mathcal A}$ denote the $n$-th element of vector $\mathbf x$ and a     
 		column vector consisting of elements of vector $\mathbf x$ indexed by $\mathcal A$, respectively,
 		$(\mathbf X)_{n,m}$ denotes the $(n,m)$-th element of matrix $\mathbf X$,
 		$\mathbf X_{n,:}$ and $\mathbf X_{:,m}$ denote the $n$-th row and the $m$-th column of matrix $\mathbf X$, respectively,
 		and $\mathbf X_{ \mathcal A,:} $
 		and $\mathbf X_{:,\mathcal B}$
 		denote matrices consisting of rows of matrix $\mathbf X$ indexed by $\mathcal A$ and columns of matrix $\mathbf X$ indexed by $\mathcal B$, respectively.
 			The complex field, real field, positive real field, and positive integer field are denoted by $\mathbb C$, $\mathbb R$, $\mathbb R^+$, and $\mathbb N^+$, respectively.
  The notations $  \mathbf F_{K}  \triangleq (e^{-j \frac{2\pi}{K} (n-1)(m-1)})_{n,m \in \{1,2,...,K\}}$ and \mcone{$\frac{1}{K}\mathbf F_{K}^H $ denote the $K$-dimensional discrete Fourier transform (DFT) and inverse discrete Fourier transform (IDFT) matrices, respectively,} and $\mathrm{FFT}(\mathbf x) \triangleq \mathbf F_{K}  \mathbf x$ and $\mathrm{IFFT}(\mathbf x) \triangleq  \frac{1}{K}\mathbf F_{K} ^H \mathbf x$ represent the FFT and IFFT of vector $\mathbf x \in \mathbb C^K$, respectively.	
 The notation $\mathrm{diag}(\mathbf x)$ represents the diagonal matrix	with diagonal elements $ \mathbf x  $,
 	$\mathrm{Re}(\cdot)$ denotes the real part of a scalar, a vector, or a matrix,
$\log(\cdot) $ and $\log_2 (\cdot)$ denote the logarithms for base $e$ and base $2$, respectively, 
$ \odot$ represents the element-wise product of matrices of the same size, $\mathbb I(\cdot)$ denotes the indicator function, and $\mathbf I_K$, $\mathbf 1_K$, and $\mathbf 0_K$ denote the $K$-dimensional identity matrix, all-one vector, and zero vector, respectively.


	\vspace{-3mm}
\section{System Model }
	\vspace{-2mm}

We consider the uplink of a single-cell wireless network consisting of one $M$-antenna BS
and $N$ single-antenna IoT devices.
Let $\mathcal M\triangleq\{1,2,...,M\}$ and $\mathcal N\triangleq\{1,2,...,N\}$ denote the sets of antennas and device indices, respectively.
The \mcele{locations of the} BS and devices are assumed to be fixed.\footnote{\mcmtwo{In some IoT scenarios (\mcmthr{e.g., environment sensing}), IoT devices (\mcmthr{e.g., sensors}) are static.}}
 \mcmtwo{As many other papers~\cite{TSP_LL1_2018, TWC-21-Li, JSAC_Cui_2020,TIT_GC_2021, TWC_CFO_2019, TSP_Chen_21, ICCws_CFO_2020,ICC_Chen_2019,TWC_Jiang_2021,TWC-22-Shi,JSAC-21-Shao,Arxiv-Jia-22  , BAMP01 ,BAMP02, Wang_ICCASP_21, TWC_Tao_2021,ICASSP_LL_2021,SPL-22-Wang,ICC-22-Liu,IoTJ-22-Tian},}
 we consider a narrow-band system, assume slow fading, and adopt the block fading model for slow fading.
 \mcmthr{We study one resource block within the coherence time.}\footnote{\mcmtwo{\mcmthr{In practical systems with multiple resource blocks}, IoT devices with significantly different distances from the BS \mcmthr{can be assigned to different resource blocks} to mitigate the near-far problem. Resource allocation for grant-free access is not the focus of this paper.}}
\mcmthr{Let} $\sqrt{g_n} {\bf h}_n \in {\mathbb C}^{ M  } $ denote the channel vector between device $n$ and the BS in the coherence block, where $ g_n \in \mathbb R^+$ represents the large-scale fading power, and $\mathbf h_n \triangleq (h_{n,m})_{m\in\mathcal M }\in {\mathbb C}^{ M  }$ denotes the  small-scale fading coefficients, where $h_{n,m}$ denotes the small-scale fading coefficient between the BS's $m$-th antenna and device $n$.
Denote \mcele{the large scale-fading power vector by $\mathbf g \triangleq (g_n)_{n\in\mathcal N} \in \mathbb R^N$}.
In contrast to most existing works which consider Rayleigh fading~\cite{TSP_LL1_2018,  JSAC_Cui_2020,TIT_GC_2021,  TSP_Chen_21 ,ICC_Chen_2019,TWC_Jiang_2021,TWC-22-Shi,JSAC-21-Shao, ICCws_CFO_2020, TWC-Sun-22,TWC_Tao_2021,Arxiv-Jia-22,ICC-22-Liu ,Wang_ICCASP_21},
we consider the Rician small-scale fading model for $\mathbf h_n$, $n \in \mathcal N$.
 Note that the Rician fading model generalizes the Rayleigh fading model and can capture the LoS components when devices are located in open outdoor or far from the ground. 
Specifically,
\begin{align}
	\mathbf h_{n} =      \sqrt{ \frac{\kappa_n}{1+\kappa_n}  }\overline {\mathbf h}_{n} + \sqrt{ \frac{1}{1+\kappa_n}  }\widetilde {\mathbf h}_{n }, \   n \in \mathcal N ,  \label{small-scale}
\end{align}
where $\kappa_n > 0$ denotes the Rician factor, $\overline {\mathbf h}_{n }\triangleq (\overline h_{n,m})_{m\in\mathcal M }  \in {\mathbb C}^{ M  }$ represents the normalized LoS components with unit-modulus elements (i.e., $|\overline h_{n,m}|=1$, $m   \in \mathcal M$), and $\widetilde {\mathbf h}_{n}  \triangleq  (\widetilde h_{n,m})_{m\in\mathcal M }\in {\mathbb C}^{ M  }$ represents the normalized \mctwe{NLoS} components with elements independent and identically distributed (i.i.d.) according to $\mathcal {CN} (0, 1)$.
Thus, $h_{n,m} \sim \mathcal {CN} (  \sqrt{\frac{\kappa_n}{1+\kappa_n}} \overline h_{n,m} , \frac{1}{1+\kappa_n})$ for all $n \in \mathcal N$ and $m \in \mathcal M$.
\mcele{Let $\boldsymbol\kappa \triangleq (\kappa_n)_{n \in \mathcal N} \in \mathbb R^{N}$ denote the Rician factor vector,} \mcele{and let
	 $ \overline{ \mathbf H}  \triangleq (  \overline {\mathbf h}_n)_{n\in \mathcal N} \in \mathbb C^{M \times N}$,  
	$ \widetilde{ \mathbf H}  \triangleq (\widetilde {\mathbf h}_n)_{n\in\mathcal N} \in \mathbb C^{M \times N}$, and $ { \mathbf H}  \triangleq (    {\mathbf h}_n)_{n\in \mathcal N} \in \mathbb C^{M \times N}$ denote the normalized LoS channel matrix, normalized NLoS channel matrix, and normalized channel matrix, respectively}.
We assume that $\mathbf g$, $\boldsymbol\kappa$, and $\overline {\mathbf H}$ are known constants, \mcele{since they are dominantly determined by the distances and communication environment between the devices and the BS and can be estimated offline (e.g., in the deployment phase or once the communication environment changes significantly) if the locations of devices are fixed.}\footnote{\mcele{By setting device $n$ active and sending its pilot over several fading blocks,
 		we can estimate $g_n$ and $\kappa_n$ by standard channel measurement~\cite{ACCESS-Sunshu-2013} and obtain $\overline {\mathbf h}_{n}$ based on the structure of antenna array at the BS and the angle of arrival of device $n$ estimated by angle estimation techniques~\cite{MUSIC-1986}.}}
 


We study the massive access scenario arising from mMTC,
where very few devices among a large number of potential devices are active and access the BS in each coherence block.
For all $n \in \mathcal N$,
let $a_n \in \{0,1\}$ denote the activity state of device $n$, where $a_n = 1$ indicates that device $n$ is active, and $a_n = 0$ otherwise.
\mcele{Denote the device activity vector by $\mathbf a \triangleq (a_n)_{n\in\mathcal N}$.}
We adopt a grant-free access scheme~\cite{TSP_LL1_2018,TWC-21-Li,   TIT_GC_2021,  TSP_Chen_21 ,TWC_Jiang_2021,Arxiv-Jia-22, JSAC_Cui_2020, TWC-22-Shi,JSAC-21-Shao}.
Specifically, each device $n$ is pre-assigned a specific pilot sequence $ {\bf p}_n    \in {\mathbb C}^{L  }$ consisting of $L $ ($ L\ll N $) pilot symbols.
In the pilot transmission phase, active devices send their length-$L$ pilots to the BS, and the BS detects the activity states of all devices and estimates the channel states of the active devices from the received pilot symbols over the $M$ antennas.
\mcele{In the data transmission phase, all active devices transmit their data to the BS, and the BS detects their transmitted data from the received data symbols based on the channels obtained in the pilot transmission phase.
 In this paper, we study the pilot transmission phase and investigate device activity detection, which is a fundamental problem for massive grant-free access~\cite{ TSP_Chen_21, TIT_GC_2021, TWC_Jiang_2021}.}

Ideally, all active devices
send their pilots and data synchronously in time and frequency, i.e., with the same reception time and carrier frequency.
Nevertheless, time synchronization and frequency synchronization are usually imperfect.
As such, the accuracy of a device activity detection scheme designed for the ideal time and frequency synchronous case may drop significantly.
In this paper, we assume that the device activities are deterministic and unknown and investigate device activity detection under Rician fading in the pilot phase of synchronous and asynchronous cases.
For ease of illustration, in the rest of this paper,  we define $\mathbf a_{ -n}\triangleq (a_{\ell})_{\ell\in \mathcal N, \ell \neq n}$
	and rewrite $\mathbf a$ as $(\mathbf a_{-n}, a_{n})$, for all $n \in \mathcal N$.
	
	\vspace{-5mm}
\subsection{Synchronous Case}
	\vspace{-2mm}

In the synchronous case, all active devices send their length-$L$ pilot sequences synchronously in time and frequency.
Thus, this case is also referred to as the time and frequency synchronous case.
Let $\mathbf Y \in \mathbb C^{L\times M}$ denote the received signal over $L$ signal dimensions and $M$ antennas at the BS. Then, we have:
\begin{align}
	{\mathbf Y} &  =   \sum\limits_{n \in \mathcal N}  a_n  \sqrt{g_n} 	 \mathbf p_{ n}   \mathbf h_n^T     +   \mathbf Z    \nonumber \\
	 &=   {\mathbf P}   \mathbf A{\mathbf \Gamma}^{\frac{1}{2}}  ( \mathbf {K}^{\frac{1}{2}} \overline{\bf H}^T   +  \widetilde{\bf H}^T  ) 	+ {\bf Z}   , \label{Y_syn}
\end{align}	
where the last equality is due to~(\ref{small-scale}).
Here, $\mathbf P   \triangleq  ( \mathbf p _{ n})_{n \in \mathcal N}\in \mathbb C^{{L  \times N}}   $
\mcele{represents the pre-assigned pilot matrix,}
${\bf A} \triangleq  \mathrm{diag}({\bf a}) \in \mathbb R^{N \times N} $ \mcele{is the
diagonal matrix with diagonal elements $\mathbf a$},
$\mathbf \Gamma \triangleq  \mathrm{diag} \Big(  \big(\frac{g_n}{1+\kappa_n}  \big)_{n\in \mathcal N} \Big) \in \mathbb R^{N\times N}$ \mcele{is the 
	diagonal matrix with diagonal elements $ \frac{g_n}{1+\kappa_n} $, $n\in\mathcal N$},
$\mathbf K \triangleq \mathrm{diag}\left( \boldsymbol \kappa\right) \in \mathbb R^{N\times N}$ \mcele{is the diagonal matrix with diagonal elements $\boldsymbol \kappa$}, and
$  {\bf Z}  \in \mathbb C^{ L  \times M} $ is the additive white Gaussian noise (AWGN) at the BS with all elements i.i.d. according to $\mathcal {CN}(0, \sigma^2)$.
In the synchronous case, the BS detects $\bf a $ from $	{\bf Y} $ given in~(\ref{Y_syn}), based on the values of $\mathbf P$, $\mathbf g$, $\mathbf K$, $\overline {\mathbf H}$, and the distributions of $\mathbf Z $ and $\widetilde { \mathbf H }$.

	\vspace{ -5mm}
\subsection{Asynchronous Cases}
	\vspace{ -2mm}

Following the convention~\cite{TWC_Tao_2021,ICASSP_LL_2021,SPL-22-Wang}, we assume that pilot signals sent by the devices are symbol synchronous for tractability, i.e., the time delays of pilot signals are multiples of the symbol duration.
Consequently, let $t_n \in  \mathcal D \triangleq \{0 ,1,...,D\}$ represent the STO for the pilot signal of device $n \in \mathcal N$, where $D \in \mathbb N^+$ denotes the maximum STO.
Let $\omega_n \in [-\Omega, \Omega] $ denote the accumulated phase shift in one symbol duration, also called CFO~\cite{TWC_CFO_2019, ICCws_CFO_2020}, where $\Omega \in [0,\pi]$ denotes the maximum absolute value of CFO.
 Denote \mcele{the STO vector and the CFO vector by} $\mathbf t \triangleq (t_n)_{n \in \mathcal N}$ and $ {\boldsymbol \omega} \triangleq ( \omega_n)_{n \in \mathcal N} $, \mcele{respectively}.
In the following, we introduce three asynchronous cases.\footnote{The time and frequency asynchronous case ($D\geq 1  ,\Omega \in (0,\pi] $) cannot reduce to the time asynchronous case ($D\geq 1 , \Omega =0$) and the frequency asynchronous case ($D=0 , \Omega \in (0,\pi] $).}

\begin{itemize}
	\item 	\textbf{Time asynchronous case ($D\geq 1 , \Omega =0$):}
In this case, the time synchronization is imperfect, i.e., $ D\geq 1  $, whereas the frequency synchronization is perfect, i.e., $\Omega =0   $ (implying $\boldsymbol\omega= \mathbf 0$).
	We assume that $\bf t$ is a deterministic but unknown constant.
 \item  \textbf{Frequency asynchronous case ($D=0 , \Omega \in (0,\pi] $):}
	In this case, the time synchronization is perfect, i.e., $D=0 $ (implying $\mathbf t = \mathbf 0$), whereas the frequency synchronization is imperfect, i.e., $\Omega \in (0,\pi]$.
	We assume that $\boldsymbol\omega$ is a deterministic but unknown constant.
 \item  \textbf{Time and frequency asynchronous case ($D\geq 1  ,\Omega \in (0,\pi] $):}
In this case, both the time synchronization and frequency synchronization are imperfect, i.e., $\Omega \in (0,\pi]  $ and $\Omega \geq 1  $.
	We assume that $\mathbf t$ and $\boldsymbol\omega$ are deterministic but unknown constants.
\end{itemize}

The time asynchronous case, frequency asynchronous case, and time and frequency asynchronous case are also called asynchronous case-t, asynchronous case-f, and asynchronous case-(t,f), respectively, for ease of exposition.
Define:
 \begin{align}
 	x_{i,n } \triangleq \left\{
 	\begin{array}{ll}
 		t_n   \!\!\!\!&,  \ i = \text{t} \\
 		\omega_n  \!\!\!\!&,  \ i = \text{f} \\
 		(t_n	, \omega_n)  \!\!\!\!&,  \ i = \text{(t,f)} 
 	\end{array}
 	\right.,
 	\quad 	
 	L_{i } \triangleq \left\{
 	\begin{array}{ll}
 		L+D \!\!\!\!&,  \ i = \text{t} \\
 		L \!\!\!\!&,  \ i = \text{f} \\
 		L+D \!\!\!\!&,  \ i = \text{(t,f)} 
 	\end{array}
 	\right.  ,  \label{L_i}
 	\end{align}
 and
 \begin{align}
 	\boldsymbol\tau_{i}(\omega ) \triangleq 
 	\left(e^{j (\ell-1) \omega }\right)_{\ell \in \mathcal L_i}  ,  \ i \in \{ \text{f, (t,f)}\},
 	 \label{tau}
 \end{align}	
where $\mathcal L_{i} \triangleq \{1,2,...,L_{i}\}$ for $i\in$ \{t, f, (t,f)\}.
For all $i \in \{\text{t, f, (t,f)}\}$, let $\mathbf Y_i \in \mathbb C^{L_i\times M}$ denote the received signal over $L_i$ signal dimensions and $M$ antennas at the BS in asynchronous case-$i$. Then, we have~\cite{TWC_CFO_2019}:
	\begin{align}
		{\bf Y}_{i}& =  \sum\limits_{n \in \mathcal N}  a_n  \sqrt{g_n} 	 {\bf p}_{i,n}(x_{i,n})   \mathbf h_n^T   +  \mathbf Z_{i}     ,  \ i \in \{\text{t, f, (t,f)}\} ,\label{Y0}
	\end{align}
 where
\begin{align}
	 &{\bf p}_{i,n}(x_{i,n})\triangleq     \left\{
	\begin{array}{ll}
 \!\! \!\!	\left[ {\bf 0}_{t_n}^T,{\bf p}_n^T,  {\bf 0}_{D -t_n}^T    \right]^T \in \mathbb C^{ L+D }   \!\!\!\!\!\!&,  \ i = \text{t} \\
 \!\! \!\!	\mathrm{diag}( \boldsymbol\tau_{\text{f}}(  \omega_n) )   {\bf p}_{n} \in \mathbb C^{L } \!\! \!\!\!\! &,  \ i = \text{f} \\
 \!\! \!\!	\mathrm{diag}(\boldsymbol\tau_{ \text{(t,f)} }(\omega_n)) 	\left[ {\bf 0}_{t_n}^T,{\bf p}_n^T,  {\bf 0}_{D -t_n}^T    \right]^T  \in \mathbb C^{L+D}   \!\!\!\!\!\!&,  \ i = \text{(t,f)} 
	\end{array}
	\right.    \label{EF_pilot}
\end{align}
can be interpreted as the equivalent transmitted pilot of device $n$ and $  {\bf Z}_{i} \in \mathbb C^{ L_{i} \times M} $ is the additive AWGN at the BS with all elements i.i.d. according to $\mathcal {CN}(0, \sigma^2)$.
It is clear that $ {\bf p}_{\text{f},n}(\omega_n)$ and $	 {\bf p}_{\text{(t,f)},n}(t_n,\omega_n)$ are periodic functions with respect to $\omega_n$ with period $2\pi$, i.e., $  {\bf p}_{\text{f},n}(\omega_n +2\pi) = {\bf p}_{\text{f},n}(\omega_n)$,  $  {\bf p}_{\text{(t,f)},n}(t_n,\omega_n +2\pi) =  {\bf p}_{\text{(t,f)},n}(t_n,\omega_n)$, for all $\omega_n \in [-\Omega,\Omega]$.
		Thus, the range of $  \omega_n$ can be transformed to $  [0, \Omega] \cup[2\pi-\Omega, 2\pi  ]$ for tractability.
In addition, from~(\ref{Y0}), we have:
\begin{align}
	{\bf p}_{i,n}(x_{i,n}) \in  \mathcal S_{i,n} \triangleq \{	 	{\bf p}_{i, n}(x_{i,n }): x_{i,n } \in \mathcal X_{i} \}  \label{PilotSet},  
\end{align}
where \mcele{$ \mathcal S_{i,n}$ represents the set of all possible equivalent transmitted pilots of device $n$ with}
\begin{align}
	\mathcal X_i  \triangleq \left\{
	\begin{array}{ll}
		\mathcal D   &\!\!\!\!,  \ i = \text{t} \\
		\text{$[0, \Omega] \cup[2\pi-\Omega, 2\pi  ]$}&\!\!\!\!, \ i = \text{f}  \\
		\mathcal D \times \text{$[0, \Omega] \cup[2\pi-\Omega, 2\pi  ]$} &\!\!\!\!,  \ i = \text{(t,f)} 
	\end{array}
	\right.   \label{Set_X}.
\end{align}
More compactly, by~(\ref{small-scale}), we can rewrite
$\mathbf Y_{i} $ in~(\ref{Y0}) as:
\begin{align}
  \!\!\! \mathbf Y_{i} = {\mathbf P}_{i}( \mathbf x_i) \mathbf A{\mathbf \Gamma}^{\frac{1}{2}}  ( \mathbf  K^{\frac{1}{2}} \overline{\bf H}^T +  \widetilde{\bf H}^T  )	+ {\bf Z}_{i }   ,   \ i \in \{ \text{t, f, (t,f)}\}, \label{Y_asyn}
	\end{align}
where $\mathbf x_i \triangleq (x_{i,n})_{n \in \mathcal N} $ and $ {\bf P}_{i} (\mathbf x_{i }) \triangleq  ({\mathbf p}_{i,n}(x_{i,n}))_{n\in\mathcal N} \in \mathbb C^{{L_{i} \times N}}   $ \mcele{denote the offset vector and equivalent transmitted pilot matrix in asynchronous case-$i$, respectively}.
In asynchronous case-$i$, the BS has to estimate $\mathbf a$ together with $\mathbf x_i$ from $\mathbf Y_i$, based on the values of $\mathbf P$, $D$, $\Omega$, $\mathbf g$, $\mathbf K$, $ \overline{\mathbf H}$, and the distributions of $\mathbf Z_i$ and $\widetilde {\mathbf H}$, due to the presence of STOs and/or CFOs.
  For ease of illustration, we define $\mathbf x_{i,-n}\triangleq (x_{i,\ell})_{\ell\in \mathcal N, \ell \neq n}$
  	and rewrite $\mathbf x_i$ as $(\mathbf x_{i,-n}, x_{i,n})$, for all $n \in \mathcal N$ and all $i \in \{\text{t, f, (t,f)}\}$.

 	\vspace{ -3.9mm}
 \section{MLE-based Device Activity Detection in Synchronous Case}
 	\vspace{ -2mm}
 \subsection{MLE Problem Formulation}
 	\vspace{ -2mm}
  
  In this part~\cite{SPAWC-22-Liu},\footnote{\mcele{The conference version of the paper~\cite{SPAWC-22-Liu} presents some preliminary results for the synchronous case. This paper provides more details for the synchronous case and \mctht{additionally} investigates three asynchronous cases.}} we formulate the MLE problem for device activity detection in the synchronous case. 
   First, we derive the log-likelihood of $\mathbf Y $.
   For all $m\in \mathcal M$, 
  by the distributions of $  {\bf H}$ and $\mathbf Z $, the distribution of the $m$-th column of $\mathbf Y$ given in~(\ref{Y_syn}) is given by~\cite{TIT_GC_2021}:
   \begin{align}
   	\!\!\! 	{ \mathbf Y}_{ :,m} \sim \ 	&  \mathcal {CN}\big( \overline{\mathbf Y}_{:,m}(\mathbf a),     {\bf\Sigma}({\mathbf a}) \big),   \label{Dist_y1m_Ri_syn}
   \end{align}
   where $  \overline {\mathbf Y}(\mathbf a) \triangleq \mathbf P  {\bf A} {\bf \Gamma}^{\frac{1}{2}} \mathbf K^{\frac{1}{2}}\overline {\mathbf H}^T$ 
   represents the mean of $\mathbf Y$ and 
$
   {\bf\Sigma}(\mathbf a)  \triangleq  \mathbf P \mathbf A^2  \mathbf \Gamma      \mathbf P ^H   + \sigma^2{\bf I}_{L }   =  \mathbf P \mathbf A \mathbf \Gamma      \mathbf P ^H   + \sigma^2{\bf I}_{L } 
   $ (as $\mathbf A^2 = \mathbf A$) represents the covariance matrix of $\mathbf Y_{:,m}$, $m \in \mathcal M$.
   Based on~(\ref{Dist_y1m_Ri_syn}) and the fact that $ { \mathbf Y}_{:, m}, m \in \mathcal M$ are i.i.d., 
   the \mctwe{probability density function (p.d.f.)} of $\mathbf Y $ is given by:
   \begin{align}
   	&   	p (  {\bf Y} ;{\bf a }  )   =\frac{1}{\pi^{ ML }	{ \mid {\bf\Sigma}({\mathbf a})  \mid^M   } } 	 \exp \!\Big( \!\! -  \!\!\sum\limits_{m=1}^{M}  \!\!\left({\mathbf Y}_{ :,m}  \!\!-  \!\!\overline {\mathbf Y}_{ :, m}(\mathbf a) \right) ^H    {\bf\Sigma}  ^{-1} ({\mathbf a})  \left({\mathbf Y}_{:, m} -\overline {\mathbf Y}_{:, m}(\mathbf a) \right)    \!\!    \Big)  .
   	\nonumber
   \end{align}
   Thus, 
   the log-likelihood of $\mathbf Y $ is given by:
   \begin{align}
   	\log  	p  (  {\bf Y}  ;{\bf a }  )  =  & -M\log |  {\bf\Sigma}(\mathbf a)      |   -  ML \log\pi   
   	  -     \mathrm{tr} \left(     {\bf\Sigma}  ^{-1}(\mathbf a)       \widetilde {\mathbf Y}(\mathbf a) \widetilde {\mathbf Y}^H(\mathbf a)   \right)   ,\nonumber
   \end{align}
where $\widetilde {\mathbf Y}(\mathbf a) \triangleq \mathbf Y - \overline{\mathbf Y}(\mathbf a)$. 
   Define:
   \begin{align}
   	f  (\mathbf a )   \! \triangleq &  -\frac{1}{M} \log   	p  (  {\bf Y}  ;{\bf a }   )   \! -   \!  L\log\pi   \nonumber \\
   	   =  & \log |  {\bf\Sigma}(\mathbf a)   |   \! +  \!   \frac{1}{M}   \mathrm{tr} \left(     {\bf\Sigma}  ^{-1} (\mathbf a)      \widetilde {\mathbf Y}(\mathbf a) \widetilde {\mathbf Y}^H(\mathbf a)     \right)
   	. \label{LikelySy_Ri_syn}
   \end{align}
Note that $ f  (\mathbf a )$ can be rewritten as: 
\begin{align}
	f(\mathbf a) = \frac{1}{M}	f_{\text{Ray}}(\mathbf a) + \Delta(\mathbf a),  \nonumber 
\end{align}
where $ 	f_{\text{Ray}}(\mathbf a) \triangleq  M\log |  {\bf\Sigma}(\mathbf a)   | +    \mathrm{tr} \left(     {\bf\Sigma}(\mathbf a)     ^{-1}   \mathbf Y   \mathbf Y^H    \right)$ is the negative log-likelihood function omitting the constant $ML \log \pi$ under Rayleigh fading~\cite{TIT_GC_2021} with large-scale fading powers $\frac{g_n}{1+\kappa_n}$, $n \in \mathcal N$, and $ \Delta(\mathbf a) \triangleq  \frac{1}{M}    \mathrm{tr} \left(    {\bf\Sigma}^{-1}(\mathbf a)         	\overline {\mathbf Y} ({\mathbf a})	\overline {\mathbf Y}^H (\mathbf a)     \right) + \frac{2}{M} \mathrm{Re}  \left(\mathrm{tr} \left(    {\bf\Sigma}  ^{-1}   ({\mathbf a} )   	  {\mathbf Y} 	\overline {\mathbf Y}^H (\mathbf a)     \right)  \right)$ is an additional
non-convex term which solely reflects the influence of normalized LoS components $  	\overline {\mathbf H}  $. 

 Then,  
 the MLE of $ \bf a   $ for the synchronous case can be formulated as follows.\footnote{For tractability, binary condition $a_n \in \{0,1\}$ is relaxed to continuous condition $a_n \in [0,1]$, and the
 	activity of device $n$ can be constructed by performing thresholding on $a_n$ obtained by solving Problem~\ref{Prob:_rlx_Ri_syn}~\cite{TSP_Chen_21 ,TWC_Jiang_2021,ICC-22-Liu}.} 
 \begin{Prob}[MLE in Synchronous Case]\label{Prob:_rlx_Ri_syn}
 	\begin{align}
 		\min\limits_{ {\mathbf a} } & \quad  f  (\mathbf a )    \nonumber  \\
 		\ 	\mathrm{s.t.} &\quad
 		0 \leq a_n \leq 1  ,\  n \in \mathcal N \label{Ca}.  
 	\end{align}
 \end{Prob}

 Since $f  (\mathbf a )$ is non-convex with respect to $ \mathbf a $ and the constraints in~(\ref{Ca}) are convex, Problem~\ref{Prob:_rlx_Ri_syn} is a non-convex problem over a convex set.\footnote{Obtaining a stationary point is the classic goal for dealing with a non-convex problem over a convex set.}
Note that Problem~\ref{Prob:_rlx_Ri_syn} generalizes the MLE problems for device activity detection in massive grant-free access under Rayleigh fading in~\cite{TSP_Chen_21,TIT_GC_2021}.

	\vspace{ -4mm}
\subsection{CD Algorithm}
In the following, we obtain a stationary point of Problem~\ref{Prob:_rlx_Ri_syn} using the CD method and careful matrix manipulation.\footnote{The CD (BCD) method usually leads to iterative algorithms that have closed-form updates without adjustable parameters and can converge to a stationary point under some mild conditions. The resulting algorithms are computationally efficient for solving the MLE-based activity detection problems~\cite{TIT_GC_2021,  TSP_Chen_21 ,ICC_Chen_2019,TWC_Jiang_2021,Arxiv-Jia-22, SPL-22-Wang,ICCws_CFO_2020,ICC-22-Liu }.}
Specifically, in each iteration, all coordinates are updated once. 
At the $n$-th step of each iteration, we optimize $a_n$ to minimize $f(\mathbf a_{-n}, a_n)$ for current $\mathbf a_{-n}$.
The coordinate optimization $	\min_{ a_n  \in[0,1 ] }	 f   ( \mathbf a_{-n} , a_n )$ is equivalent to the optimization of
the increment $d_n$ in $a_n$ for current $(\mathbf a_{-n},a_n)$~\cite{TIT_GC_2021, TSP_Chen_21,TWC_Jiang_2021}:
\begin{align}
d_n^* (\mathbf a)\triangleq  \mathop{\arg\min}\limits_{d_n \in[-a_n,1-a_n]}	 	\quad & f   ( \mathbf a_{-n} , a_n +d_n ). \label{Prob:cd_Ri_syn}
\end{align}
Note that it is more challenging to obtain a closed-form solution of the coordinate optimization problem in~(\ref{Prob:cd_Ri_syn}) than in~\cite{TIT_GC_2021, TSP_Chen_21} due to the existence of $\Delta(\mathbf a)$.
For notation convenience, define:     
\begin{align}
	&\!\! \!\!  \!\! \alpha_{n }(\mathbf a)    \triangleq
	\frac{g_n}{1+ \kappa_n} { \bf p}_n^H
	{\bf\Sigma} ^{-1}(\mathbf a)       { \bf p}_n,    \label{alpha_syn}
	\\
	& \!\!  \!\! \!\! 	 \beta_{n } (\mathbf a)   \triangleq
	\frac{g_n}{ M(1+ \kappa_n)} { \bf p}_n^H {\bf\Sigma} ^{-1}(\mathbf a)     \widetilde {\mathbf Y}(\mathbf a) \widetilde {\mathbf Y}^H(\mathbf a)  {\bf\Sigma}^{-1}({\mathbf a} )  { \bf p}_n	,    \label{beta_syn} 
	\\
	&	\!\! \!\!    	\eta_{n }(\mathbf a)   \triangleq 	  \frac{ 2}{M}		\sqrt{\frac{g_n\kappa_n}{1+ \kappa_n}}
	\mathrm{Re} \left(
	\overline {\bf h}_n^T    \widetilde {\mathbf Y}^H(\mathbf a)        {\bf\Sigma} ^{-1} ({\mathbf a} ) { \bf p}_n   \right) .     \label{eta_syn}
\end{align}
For notation simplicity, we omit the argument $\mathbf a$ in $\widetilde {\mathbf Y}(\mathbf a)$, ${\mathbf\Sigma}(\mathbf a)$, $d_n^* (\mathbf a)$, $ \alpha_{n }(\mathbf a)$, $ \beta_{n }(\mathbf a)$, and $ \eta_{n }(\mathbf a)$ in what follows. 
Based on the structural properties of the problem in~(\ref{Prob:cd_Ri_syn}),
we have the following result.

\begin{theorem}[Optimal Solution of Problem in~(\ref{Prob:cd_Ri_syn})]\label{Tem:OptpointBCD_Ri_syn}
	For all $n \in \mathcal N$,
	the optimal solution of the problem in~(\ref{Prob:cd_Ri_syn}) is given by:
		\begin{align}
	d_n^{*} =	\left\{
		\begin{array}{ll}
			\min\{ \max\{ \hat d_n,-a_n  \} ,    1-a_n\}       \!\!\!\!&,  \    \mctwo{ 4 \kappa_n  (\kappa_n + \beta_n + \eta_n)      + \alpha_{n }^2> 0} \\    		    
			-a_n         &,  \   \mctwo{ 4 \kappa_n  (\kappa_n + \beta_n + \eta_n)      + \alpha_{n }^2  \leq 0}
		\end{array} \right.       ,  \label{Opt_bc_a_Ri_syn}
	\end{align}
	where  
	\begin{align}
	& \hat	d_{n  }  \triangleq  \frac{-\alpha_n - 2\kappa_n + \sqrt{     4 \kappa_n  (\kappa_n + \beta_n + \eta_n)  +  \alpha_{n }^2 }}{2 \kappa_n \alpha_n }   \label{d_n0}
	\end{align}
			with $\alpha_{n } $, $\beta_{n } $, and, $\eta_{n }  $ given by~(\ref{alpha_syn}),~(\ref{beta_syn}), and~(\ref{eta_syn}), respectively.
\end{theorem}
	\begin{IEEEproof} 
	Please refer to Appendix A.
\end{IEEEproof}


\begin{algorithm}[t]
	\caption{{Prop-MLE-Syn}}
	\label{Alg:Ri_syn}
\scriptsize{	\begin{algorithmic}[1]
		\STATE Initialize: set $\mathbf a =  \mathbf 0 $, ${\bf \Sigma}^{-1}  =\frac{1}{\sigma^2} {\bf I}_{ L }$, $  \widetilde{ \mathbf Y}= \mathbf Y$, $\mathbf \barp =  \sqrt{\frac{g_n}{1+\kappa_n}} \mathbf p_n $, $n\in\mathcal N$, and choose $\epsilon>0$.
		\STATE \textbf{repeat}
		\STATE Set $\mathbf a_{\mathrm{last}} =\mathbf a $.
		\STATE \textbf{for $n \in \mathcal N$ do}
		\STATE \quad  Compute $\mathbf c  =  {\bf \Sigma}^{-1}  {\bf \barp}_{n}   $ and $\mathbf d =  \widetilde{ \mathbf Y}^H \mathbf c $.
		\STATE \quad  Compute $\alpha_n =   \mathbf \barp_n^H \mathbf c$, $\beta_n=	\frac{1}{ M} \mathbf d^H\mathbf d $, and $\eta_{n} =\frac{ 2	\sqrt{ \kappa_n} }{M}	 
		\mathrm{Re} \left(
		\overline {\bf h}_n^T \mathbf d \right)  $. 
		\STATE  \quad  Compute $d_n^*$ according to~(\ref{Opt_bc_a_Ri_syn}). 
		\STATE \quad Compute $
		{\bf \Sigma} ^{-1} = {\bf \Sigma} ^{-1}   -  
		\frac{ 	d_n^*   }
		{1 +  d_n^*\alpha_n  } \mathbf c\mathbf c^H
		$, 
		$
	 \widetilde{ \mathbf Y}= \widetilde { \mathbf Y}  -  d_{n}^* \sqrt{ \kappa_{n}  }   {\bf \barp}_{ n}  	\overline {\mathbf h}_{n}^T
		$, 
		and \\
		\quad $a_n =a_n +d_n^*$.
		\STATE \textbf{end for}
		\STATE \textbf{until} $ \vert f(\mathbf a)-f(\mathbf a_{\mathrm{last}}) \vert < \epsilon  {\vert f(\mathbf a_{\mathrm{last}  } )\vert }$.
	\end{algorithmic}}%
 
\end{algorithm}

\begin{remark}[Connection to Rayleigh Fading]
	 If $\kappa_n \rightarrow 0$, then
	\begin{align}
		& \mctwo{ 4 \kappa_n  (\kappa_n + \beta_n + \eta_n)      + \alpha_{n }^2}   \overset{(a)}\rightarrow  g_n ^2 \left({ \bf p}_n^H
		{\bf\Sigma} ^{-1}       { \bf p}_n \right)^2 \overset{(b)}>0, \nonumber \\
		&\hat d_n  \overset{\mcfour{(c)}}\rightarrow \frac{  \frac{1}{ M } { \bf p}_n^H {\bf\Sigma} ^{-1}      \mathbf Y     \mathbf Y ^H  	 {\bf\Sigma} ^{-1}   { \bf p}_n   -  { \bf p}_n^H
			{\bf\Sigma} ^{-1}     { \bf p}_n}{ g_n (  { \bf p}_n^H
			{\bf\Sigma} ^{-1}     { \bf p}_n)^2}  \triangleq    \tilde d_{n}   
		\nonumber,
	\end{align}
	where (a) is due to  $\alpha_n \rightarrow g_n { \bf p}_n^H
	{\bf\Sigma} ^{-1}   { \bf p}_n $, $\beta_n \rightarrow \frac{g_n}{M} { \bf p}_n^H {\bf\Sigma} ^{-1}      \mathbf Y     \mathbf Y ^H  	 {\bf\Sigma} ^{-1}    { \bf p}_n $, and $\eta_n \rightarrow 0$, (b) is due to ${\bf\Sigma}   \succ \mathbf 0$, and (c) is obtained by L'Hospital's rule.
Thus, by~(\ref{Opt_bc_a_Ri_syn}), we have:
	\begin{align} 
		d_n^* \rightarrow  	\min\{ \max\{  \tilde  d_{n}  ,-a_n  \} ,    1-a_n\}    , \ \text{as } \kappa_n \rightarrow 0, \nonumber
	\end{align} 
Note that $ 	\min\{ \max\{  \tilde  d_{n}  ,-a_n  \} ,    1-a_n\}  $ is the optimal coordinate increment under Rayleigh fading in~\cite[Eq. (11)]{TSP_Chen_21}.
That is, the optimal coordinate increment in this paper turns to that under Rayleigh fading
as $\kappa_n \rightarrow 0$.
\end{remark}

The details of the proposed algorithm are summarized in Algorithm~\ref{Alg:Ri_syn} (Prop-MLE-Syn).  
In the $n$-th step of each iteration,
we compute $\alpha_n$, $\beta_n$, and $\eta_n$ according to~(\ref{alpha_syn}),~(\ref{beta_syn}), and~(\ref{eta_syn}), respectively, in two steps, i.e., Steps~5 and~6, to avoid redundant calculations and computationally expensive matrix-matrix multiplications as in~\cite{Asilomar_Henriksson_2020};
in Step~8,
we compute $\mathbf \Sigma^{-1} $ instead
of $\mathbf \Sigma$ according to the Sherman-Morrison rank-1 update identity as in~\cite{TSP_Chen_21,TIT_GC_2021}, and
 update $ \widetilde {\mathbf Y}$ 
 instead of $\overline {\mathbf  Y} $
  to simplify computation.
For all $n \in \mathcal N$, the computational complexities of Steps~5,~6,~7, and~8 are $\mathcal O(L(L+M)  )$, $\mathcal O(L + M )$, $\mathcal O(1)$, and $\mathcal O(L(L+M)  )$, respectively, as $L,M \rightarrow \infty$.
Thus, the overall computational complexity of each iteration of Algorithm~\ref{Alg:Ri_syn} is $\mathcal O (   NL(L+M)  )$, as $N,L,M \rightarrow \infty$, same as the one in~\cite{IoTJ-22-Tian} and higher than the one in~\cite{TIT_GC_2021}, i.e., $\mathcal O (NL^2)$, as $N,L \rightarrow \infty$, if $M$ is large enough, due to the extra operation for dealing with the LoS components $ 	\overline {\mathbf H}  \mathbf K^{\frac{1}{2}}\in \mathbb C^{M\times N}$.\footnote{It is still unknown how to analyze the number of iterations required to reach certain stopping criteria. Therefore, we provide only per-iteration computational complexity analysis to reflect the computation cost to some extent as in~\cite{TSP_LL1_2018, TIT_GC_2021,  TSP_Chen_21 ,ICC_Chen_2019,TWC_Jiang_2021,Arxiv-Jia-22,    TWC_CFO_2019, ICC-22-Liu,BAMP02}. Later in Section V, the overall computation time will be evaluated numerically.}
Note that Algorithm~\ref{Alg:Ri_syn} successfully generalizes MLE-based activity detection under Rayleigh fading~\cite{TIT_GC_2021,TSP_Chen_21 } to MLE-based activity detection under Rician fading.
We can show the following result based on~\cite[Proposition 3.7.1]{Bertsekas1998NP}.
\begin{theorem}[Convergence of Algorithm~\ref{Alg:Ri_syn}] \label{Thm_SP_Ri_syn}
	Algorithm~\ref{Alg:Ri_syn} converges to a stationary point of Problem~\ref{Prob:_rlx_Ri_syn}.
\end{theorem}

	\begin{IEEEproof} 
	Please refer to Appendix B.
\end{IEEEproof}

  \section{MLE-based Device Activity Detection in Asynchronous Cases}
  
  \subsection{MLE Problem Formulation}

  In this part, we formulate the MLE problem for device activity detection in the three asynchronous cases.
 First, we derive the log-likelihood of $\mathbf Y_i$, for all $i \in$ \{t, f, (t,f)\}.
 For all $m \in \mathcal M$, by the distributions of $  {\bf H}$ and $\mathbf Z_i $, the distribution of the $m$-th column of $\mathbf Y_i$ given in~(\ref{Y_asyn}) is given by~\cite{TIT_GC_2021}:
\begin{align}
 	    	\setcounter{equation}{18}
 		\!\!\! 	{ \bf Y}_{ i,:,m} \sim \ 	&  \mathcal {CN}\big( \overline{\mathbf Y}_{i,:,m} (\mathbf a, \mathbf x_i),     {\bf\Sigma}_i (\mathbf a, \mathbf x_i) \big),   \ m\in  \mathcal M, \label{Dist_y1m_Ri_asyn}
 	\end{align}
 	where 
 	\begin{align}
 	 \overline {\mathbf Y}_{ i} (\mathbf a, \mathbf x_i) \triangleq   {\bf P}_i(\mathbf x_i){\bf A} {\bf \Gamma}^{\frac{1}{2}} \mathbf K^{\frac{1}{2}}  \overline {\bf H} ^T   \label{overlineY_Ri_asyn}
 	 \end{align}
 	represents the mean of $\mathbf Y_{i}$ and 
 	\begin{align}
 	{\bf\Sigma}_i  (\mathbf a, \mathbf x_i)  &\triangleq  \mathbf P_{i}( \mathbf x_i)  \mathbf A^2  \mathbf \Gamma      \mathbf P_{i}^H( \mathbf x_i)    + \sigma^2{\bf I}_{L_{i}} 
  \nonumber \\
  	&= \mathbf P_{i}( \mathbf x_i)  \mathbf A  \mathbf \Gamma      \mathbf P_{i}^H ( \mathbf x_i)   + \sigma^2{\bf I}_{L_{i}} 
 	   \label{Sigma_Ri_asyn}
 	\end{align}
 	represents the covariance matrix of $\mathbf Y_{i,:,m}$, $m\in\mathcal M$.
 Analogously, based on (\ref{Dist_y1m_Ri_asyn}) and the fact that $\mathbf Y_{i,:,m}$, $m \in \mathcal M$ are i.i.d., 
 the p.d.f. of $\mathbf Y _{i}$ is given given by:
 \begin{align}
 	&  \!\!\! 	p_i (  {\bf Y}_i ;{\bf a }  ,\mathbf x_i)   = \!\frac{\exp\Big( - \sum\limits_{m=1}^{M}  \left({\bf Y}_{i,:, m} -\overline {\mathbf Y}_{ i, :,m} (\mathbf a, \mathbf x_i)\right) ^H    {\bf\Sigma}_i  ^{-1}  (\mathbf a, \mathbf x_i)  \left({\bf Y}_{i, :,m} -\overline {\mathbf Y}_{i,:, m} (\mathbf a, \mathbf x_i)\right)       \Big)}{\pi^{ ML } 	{ \mid {\bf\Sigma}_i  (\mathbf a, \mathbf x_i)  \mid^M   } } 	.
 \label{pdfasyn}
 \end{align}
 	Thus, 
 	the log-likelihood of $\mathbf Y_i $ is given by:
 	\begin{align}
 	 	\log  	p_i  (  {\mathbf Y_i}  ;{\bf a } ,\mathbf x_i )  = &  -M\log |  {\bf\Sigma}_ i (\mathbf a, \mathbf x_i)      |   -  ML_i \log\pi  -  \mathrm{tr} \left(     {\bf\Sigma}_i^{-1}  (\mathbf a, \mathbf x_i)        \widetilde {\mathbf Y}_{i}(\mathbf a, \mathbf x_i)   \widetilde {\mathbf Y}_{i} ^H(\mathbf a, \mathbf x_i)     \right)  , \nonumber
 	\end{align}
 where 
 	\begin{align}
 			\setcounter{equation}{22}
 		 \widetilde {\mathbf Y}_{i}(\mathbf a, \mathbf x_i) \triangleq \mathbf Y_i-	\overline{\mathbf Y }_i (\mathbf a, \mathbf x_i).   \label{widetilde Y_i}
 		\end{align}
Define:
 	\begin{align}
 		f_i (\mathbf a, \mathbf x_i)  \! \triangleq  &  \log |  {\bf\Sigma}_ i (\mathbf a, \mathbf x_i)      |  + \frac{1}{M}    \mathrm{tr} \left(     {\bf\Sigma}_i ^{-1}  (\mathbf a, \mathbf x_i)          \widetilde {\mathbf Y}_{i}(\mathbf a, \mathbf x_i)   \widetilde {\mathbf Y}_{i} ^H(\mathbf a, \mathbf x_i)     \right)   
 		.  \label{LikelySy_Ri_asyn}
 	\end{align}

To reduce the computational complexities for solving the minimization of $f_{ \text{f}} (\mathbf a,\boldsymbol\omega)$ or $f_{ \text{(t,f)}} (\mathbf a, \mathbf t,\boldsymbol\omega)$ with respect to $\omega_n \in    {\mathcal X}_{\text{f}}$ (where $   {\mathcal X}_{\text{f}}$ is given in~(\ref{Set_X})) for some $n\in\mathcal N$ (with the other variables being fixed), which does not have an analytical solution, we consider a finite set $	\hat {\mathcal X}_{\text{f}}  \triangleq  
\left\{ \omega^{(1)} ,\omega^{(2)} ,...,\omega^{(Q)}   \right\} 	\cap \mathcal X_{\text{f}}$
in place of the interval $   {\mathcal X}_{\text{f}}$, where $   \omega^{(q)} \triangleq \frac{q-1  }{Q}2\pi $, $q \in \{1,2,...,Q\}$, and focus on discrete $\omega_n \in  \hat{\mathcal X}_{\text{f}}	 $ rather than continuous $\omega_n \in   {\mathcal X}_{\text{f}}	 $ in asynchronous case-f and asynchronous case-(t,f).
 Let $\mathcal Q \triangleq \{q: \omega^{(q)} \in  \hat {\mathcal X}_{\text{f}}\}$.
 For ease of illustration, let
 $\hat{\mathcal X}_{\text{(t,f)}} \triangleq   \mathcal X_{\text{t}} \times \hat{\mathcal X_{\text{f}}}	$ and $\hat{\mathcal X}_{\text{t}} \triangleq \mathcal X_{\text{t}}	$. 
 For all $\mathbf a$, we approximately represent $f_{ \text{f}} (\mathbf a,  \boldsymbol\omega )  $, $f_{ \text{(t,f)}} (\mathbf a, \mathbf t, \boldsymbol\omega )  $, $\boldsymbol\omega \in \mathcal X_{\text{f}}$ with $f_{ \text{f}} (\mathbf a,  \boldsymbol\omega )  $, $f_{ \text{(t,f)}} (\mathbf a, \mathbf t, \boldsymbol\omega )  $, $\boldsymbol\omega \in \hat {\mathcal X}_{\text{f}}$.

Then, for all $i =$ t, f, (t,f),
the MLE of $ \bf a   $ and $\mathbf x_i$ for asynchronous case-$i$ under Rician fading can be formulated as follows.
\begin{Prob}[Joint MLE in Asynchronous Case-$i$]\label{Prob:Ri_asyn}
	\begin{align}
		\min\limits_{ {\mathbf a, \mathbf x_i } } & \quad f_{ i} (\mathbf a, \mathbf x_i )    \nonumber  \\
		\ 	\mathrm{s.t.} &\quad
		(\ref{Ca}) , \nonumber \\
		&  \quad   x_{i,n}\in \hat{\mathcal X_{i}},  
		\ n \in \mathcal N .   \label{Cftf_app}
	\end{align}
\end{Prob}

Note that the value of $ x_{i,n}$ is meaningful only if $a_n>0$, as it affects the received pilot signal $\mathbf Y_i$ in~(\ref{Y_asyn}) only if $a_n>0$. 
That is, the value of $x_{i,n}$ is vacuous if $a_n =0$.\footnote{If $a_n=0$, no matter which value $x_{i,n} $ takes in $\hat {\mathcal X}_i$, the objective function $f_{i} (\mathbf  a_{-n} , a_n, \mathbf x_{-n} , x_{i,n})$ of Problem~2 does not change.}
As a result, the constraints on $ a_n $ and $\mathbf x_{i,n}$ can be decoupled, as shown in Problem~\ref{Prob:Ri_asyn}. 
Different from Problem~\ref{Prob:_rlx_Ri_syn} for the MLE of $\mathbf a$ in the synchronous case,
Problem~\ref{Prob:Ri_asyn} deals with the MLE of $(\mathbf a ,\mathbf x_{i})$ in asynchronous case-$i$ and hence is more challenging.
	Since $f_i  (\mathbf a, \mathbf x_i)$ is non-convex with respect to $(\mathbf a ,\mathbf x_{i})$, the constraints in~(\ref{Ca}) are convex, and the constraints in~(\ref{Cftf_app}) are non-convex, Problem~\ref{Prob:Ri_asyn} is a non-convex problem.

Note that the finite sets $	\hat {\mathcal S}_{\text{f},n} \triangleq  \{ 	{\bf p}_{\text{f},  n}(  \omega_n ):  \omega_n\in \hat{\mathcal X_{\text{f}}} \}  $ and $	\hat {\mathcal S}_{\text{(t,f)},n} \triangleq  \{ 	{\bf p}_{\text{(t,f)},  n}( t_n,\omega_n): t_n \in \mathcal X_{\text{t}}, \omega_n\in \hat{\mathcal X_{\text{f}}} \} $
are discrete approximations of $\mathcal S_{\text{f},n}$ and $\mathcal S_{\text{(t,f)},n}$ in~(\ref{PilotSet}), respectively. 
For ease of illustration, we also write $	  {\mathcal S}_{\text{t},n} $ as $	\hat {\mathcal S}_{\text{t},n} $.
Thus, in asynchronous case-$i$, we can view $\hat {\mathcal S}_{i,n}$ as the set of data-embedding pilots maintained at device $n$ in massive grant-free access with data embedding~\cite{ICC_Chen_2019}, view the device activity detection {for asynchronous cases} as the joint device activity and data detection in the synchronous case in~\cite{ICC_Chen_2019}, 
and treat Problem~\ref{Prob:Ri_asyn} as the MLE problem for joint device activity and data detection in~\cite{ICC_Chen_2019}.
However, different from the MLE problem in~\cite{ICC_Chen_2019}, Problem~2 for asynchronous case-$i$ has discrete variables $(\mathbf a, \mathbf x_{i})$, whose number does not increase with offset ranges $D$ and $\Omega$ or $\vert \hat {\mathcal X}_i\vert$, where
		\begin{align}
		|\hat {\mathcal X}_{i}| = \left\{\begin{array}{ll}
			D+1\!\!\!&,   \   i = \text{t} \\
			Q+\mathbb{I}(\Omega\neq\pi )	(2  \lfloor \frac{Q\Omega}{2\pi}  \rfloor+1-Q)   \!\!\!&,   \ i = \text{f} \\
			(D+1) (Q+\mathbb{I}(\Omega\neq\pi )	(2  \lfloor \frac{Q\Omega}{2\pi}  \rfloor+1-Q)  ) \!\!\!&,  \ i = \text{(t,f)} 
		\end{array}
		\right. .
		\label{hatX}
	\end{align}
	In other words, Problem~2 does not face the same scalability issue as the MLE problem in~\cite{ICC_Chen_2019}. Instead, the size of the feasible set of $(\mathbf a, \mathbf x_{i})$ increases with the range of offset values, which is also challenging to deal with.

\subsection{BCD Algorithms}

In this part, by carefully considering the impact of the range of offset values, we propose two computationally efficient iterative algorithms to tackle Problem~\ref{Prob:Ri_asyn}.
The idea is to divide the variables $(\mathbf a , \mathbf x_i    )$ into $N$ blocks, i.e., $( a_n, x_{i,n})$, $n \in \mathcal N$, and update them sequentially by solving the corresponding block coordinate optimization problems analytically for asynchronous case-$i$.
 
Specifically, for all $n \in \mathcal N$, given $(\mathbf a_{-n}, \mathbf x_{i,-n} )$ obtained in the previous step, the block
coordinate optimization with respect to $(a_n, x_{i,n} ) $ is formulated as follows:
\begin{align}
&(a_n^*(\mathbf a_{-n}, \mathbf x_{i,-n}), x_{i,n}^*(\mathbf a_{-n}, \mathbf x_{i,-n}) ) \triangleq  \mathop{\arg\min}_{ a
		  \in[0,1], x  \in \hat{\mathcal X}_{i} }		 	\quad f_i  ( \mathbf a_{-n} , a , \mathbf x_{i,-n} , x ). \label{Prob:bcd_ray}
\end{align}
For notation convenience, for all $x \in \mathcal X_{i}$, define:
\begin{align}
   \alpha_{i,n }(\mathbf a_{-n}, \mathbf x_{i,-n},x ) & \triangleq  
	\frac{g_n   { \bf p}_{i,n}^H \!(x)	
		{\bf\Sigma}_{i ,n}^{-1}   (\mathbf a_{-n},  \mathbf x_{i,-n})  { \bf p}_{i,n} \!(x)  }{1+ \kappa_n}  , 
	   \label{C1_asyn}  \\
 	  	 \beta_{i,n } (\mathbf a_{-n}, \mathbf x_{i,-n},x ) &\triangleq
	\frac{g_n}{ M(1+ \kappa_n)}  { \bf p}_{ i,n}^H(x)		   {\bf\Sigma}_{i,n }^{-1}  (\mathbf a_{-n}, \mathbf x_{i,-n})  \widetilde {\mathbf Y}_{i ,n}  (\mathbf a_{-n}, \mathbf x_{i,-n})  \widetilde {\mathbf Y}_{i,n }^H  (\mathbf a_{-n}, \mathbf x_{i,-n})   \nonumber \\
	  &\quad\times    
	   {\bf\Sigma}_{i ,n}^{-1} (\mathbf a_{-n}, \mathbf x_{i,-n})    { \bf p}_{ i,n} (x)  
	  \label{C2_asyn} , \\
	   	\eta_{i,n }(\mathbf a_{-n}, \mathbf x_{i,-n},x)  & \triangleq 	  \frac{ 2}{M}		\sqrt{\frac{g_n\kappa_n}{1+ \kappa_n}}
	  \mathrm{Re} \left(
	  \overline {\bf h}_n^T \widetilde {\mathbf Y}_{i,n}^H      (\mathbf a_{-n}, \mathbf x_{i,-n})\   {\bf\Sigma}_{i ,n}^{-1}   (\mathbf a_{-n}, \mathbf x_{i,-n})\ {  { \bf p}}_{i,n}(x)    \right)  , 
	  \label{C3_asyn}
\end{align}
where
\begin{align}
	&	{\bf\Sigma}_{  i,n} (\mathbf a_{-n}, \mathbf x_{i,-n})\triangleq  	{\bf\Sigma}_{ i}(\mathbf a_{-n}, 0, \mathbf x_{i,-n},0),
	\label{Sigma_in} \\
	& \widetilde	{\mathbf Y}_{  i,n} (\mathbf a_{-n}, \mathbf x_{i,-n}) \triangleq  \widetilde	{\mathbf Y}_i(\mathbf a_{-n}, 0, \mathbf x_{i,-n},0),
	\label{Y_in}
\end{align}
with ${\bf\Sigma}_{i} (\cdot) $ and $\widetilde {\mathbf Y}_{i}(\cdot)$ given by~(\ref{Sigma_Ri_asyn}) and (\ref{widetilde Y_i}), respectively.
For $i \in$ \{t, f, (t,f)\}, we omit the argument $( \mathbf a_{-n},\mathbf x_{i,-n} )$ in  $a_n^*(\mathbf a_{-n}$,  $\mathbf x_{i,-n}), x_{i,n}^*(\mathbf a_{-n}, \mathbf x_{i,-n}) $, $ \alpha_{i,n }(\mathbf a_{-n}, \mathbf x_{i,-n},x)$, $ \beta_{i,n }(\mathbf a_{-n}, \mathbf x_{i,-n},x)$, $ \eta_{i,n }(\mathbf a_{-n}, \mathbf x_{i,-n},x)$,  ${\mathbf\Sigma}_{ i, n}(\mathbf a_{-n}, \mathbf x_{i,-n})$, and $\widetilde {\mathbf Y}_{i,n}(\mathbf a_{-n}, \mathbf x_{i,-n})$ in what follows for notation simplicity.
Then, based on the structural properties of the problem in (\ref{Prob:bcd_ray}),
we have the following result.

\begin{theorem}[Optimal Solution of Problem in~(\ref{Prob:bcd_ray}) for Asynchronous Case-$i$]\label{Tem:OptpointBCD_app}
	For all $n \in \mathcal N$,
	the optimal solution of the block coordinate optimization problem in~(\ref{Prob:bcd_ray}) for asynchronous case-$i$ is given by:
	\begin{align}
	&a_{n}^* =   d_{i,n}(x_{i,n}^* ),    \quad   x_{i,n}^*  \in  \mathop{\arg\min}_{x\in \hat{ \mathcal X}_{i} }  \  h_{i,n}(x),  \label{Opt_time_t_Ri_asyn}
\end{align}
where 
	\begin{align}
& d_{i,n}(x)  \triangleq	\left\{
	\begin{array}{ll}
		\min\{ \max\{ {\hat d}_{i,n}(x),0  \} ,    1 \}   \!\!\!\! &,  \      4 \kappa_n  (\kappa_n + \beta_{i,n }(x) + \eta_{i,n }(x)) + \left( \alpha_{i,n }(x)\right)^2 >0   \\    		    
	0  &,  \  	    4 \kappa_n  (\kappa_n + \beta_{i,n }(x) + \eta_{i,n }(x)) + \left( \alpha_{i,n }(x)\right)^2 \leq 0  
	\end{array} \right. ,\label{d_n0_asyn} \\
 	&h_{i,n}(x) \triangleq    \log( 1+      d_{i,n}(x )      	\alpha_{i,n }(x    ))   
  + \frac{ \kappa_n
	\alpha_{i,n }(x)   \left(d _{i,n} (x) \right)^2     \!     -  \!
	(  \beta_{i,n }(x)  \! +\!   	\eta_{i,n }(x )  )    d_{i,n}(x ) }{1+      d_{i,n}(x)   	\alpha_{i,n } (x)   }  \label{h_i_asyn},
\end{align}
with
\begin{align}
&{\hat d}_{i,n} (x) \triangleq  \frac{- \alpha_{i,n }(x ) - 2\kappa_n + \sqrt{ 4 \kappa_n  (\kappa_n + \beta_{i,n }(x) + \eta_{i,n }(x)) + \left( \alpha_{i,n }(x)\right)^2}  }{2 \kappa_n \alpha_{i,n }(x ) }     \nonumber 
  \end{align}
and  $\alpha_{i,n }(x)$, $\beta_{i,n }(x)$, and $\eta_{i,n }(x)$ given by~(\ref{C1_asyn}),~(\ref{C2_asyn}) and~(\ref{C3_asyn}), respectively.
\end{theorem}

\begin{IEEEproof} 
	Please refer to Appendix C.
\end{IEEEproof}

 \begin{remark}[Connection to Rayleigh Fading]
 	If $\kappa_n \rightarrow 0$, then
 	\begin{align}
 		& 4 \kappa_n  (\kappa_n + \beta_{i,n }(x) + \eta_{i,n }(x)) + \left( \alpha_{i,n }(x)\right)^2 \overset{(a)} \rightarrow  g_n ^2 \left({ \bf p}_{i,n}^H(x)	
 		{\bf\Sigma}_{i ,n}^{-1}    { \bf p}_{i,n}(x)  \right)^2  \overset{(b)} >0 ,   \nonumber \\
 		&{\hat d}_{i,n} (x)   \overset{(c)}\rightarrow \frac{  \frac{1}{ M } { \mathbf p}_{i,n}^H( x ){\bf\Sigma}_{i,n} ^{-1}       \mathbf Y_i    \mathbf Y_i ^H  	 {\bf\Sigma} ^{-1} _{i,n}  { \mathbf p}_{i,n}( x )   - 
 			{ \bf p}_{i,n}^H(x)		{\bf\Sigma}_{i ,n}^{-1}    { \bf p}_{i,n}(x)}{ g_n   \left({ \bf p}_{i,n}^H(x)	
 			{\bf\Sigma}_{i ,n}^{-1}    { \bf p}_{i,n}(x)  \right)^2} \triangleq    {\tilde d}_{i,n} (x)  \label{hatdinx},
 	\end{align}
 	where (a) is due to  $\alpha_{i,n }(x ) \rightarrow 
 	g_n { \bf p}_{i,n}^H(x)		{\bf\Sigma}_{i ,n}^{-1}    { \bf p}_{i,n}(x)$, $\beta_{i,n }(x ) \rightarrow
 	\frac{g_n}{M} { \mathbf p}_{i,n}^H( x ){\bf\Sigma}_{i,n} ^{-1}       \mathbf Y_i    \mathbf Y_i ^H  	 {\bf\Sigma} ^{-1} _{i,n}  { \mathbf p}_{i,n}( x ) $, and $\eta_{i,n }(x ) \rightarrow 0$, 
 	(b) is due to ${\bf\Sigma}_{i ,n}  \succ \mathbf 0$, and (c) is 
  obtained by L'Hospital's rule.
  Thus, by~(\ref{d_n0_asyn}) and~(\ref{h_i_asyn}),
		we have	
	$
		d_{i,n} (x) \rightarrow  	\min\{ \max\{  {\tilde d}_{i,n} (x)  ,0  \} ,    1 \}   $ and
	$	
	 	 h_{i,n}(x) \rightarrow  \tilde h_{i,n}(x),
	 $
	 	as $ \kappa_n  \rightarrow 0  $, where ${\tilde d}_{i,n} (x)$ is defined in  (\ref{hatdinx}) and
	 	\begin{align}
	 		&\tilde h_{i,n}(x) \triangleq  \log( 1+          	g_n { \bf p}_{i,n}^H(x)		{\bf\Sigma}_{i ,n}^{-1}    { \bf p}_{i,n}(x)	\min\{ \max\{  {\tilde d}_{i,n} (x)  ,0  \} ,    1 \}  )
	 		\nonumber \\
	 		& 
	 		+ \frac{ \frac{g_n}{M} { \mathbf p}_{i,n}^H( x ){\bf\Sigma}_{i,n} ^{-1}       \mathbf Y_i    \mathbf Y_i ^H  	 {\bf\Sigma} ^{-1} _{i,n}  { \mathbf p}_{i,n}( x )
	 			\min\{ \max\{  {\tilde d}_{i,n} (x)  ,0  \} ,    1 \}
	 		}{1+          	g_n { \bf p}_{i,n}^H(x)		{\bf\Sigma}_{i ,n}^{-1}    { \bf p}_{i,n}(x)	\min\{ \max\{  {\tilde d}_{i,n} (x)  ,0  \} ,    1 \} }. \nonumber 
	 		\end{align}
By comparing with~\cite[Eq. (20)]{SPL-22-Wang} and~\cite[Eq. (21)]{ICC-22-Liu}, we know that
the optimal coordinates in this paper reduce to those under Rayleigh fading, as $\kappa_n \rightarrow 0$.	
\end{remark}%

As shown in Theorem~\ref{Tem:OptpointBCD_app}, the optimal solution $(a_n^*, x_{i,n}^*)$ of the problem in~(\ref{Prob:bcd_ray}) in asynchronous case-$i$ is expressed with $\alpha_{i,n  }(x)$, $\beta_{i,n  }(x)$, $\eta_{i,n  }(x)$, $x \in \hat {\mathcal X}_{i}$.
In the rest of this section, for asynchronous case-$i$, where $i\in $ \{t, f, (t,f)\}, we propose two BCD algorithms
 \mcone{which rely} on two different methods for computing
	$\alpha_{i,n  }(x)$, $\beta_{i,n  }(x)$, $\eta_{i,n  }(x)$, $x \in \hat {\mathcal X}_{i}$ and have different computational complexities.

 \subsubsection{BCD Algorithm for Small \mcone{Offset Ranges}}
 \begin{algorithm}[t]
 	\caption{\GBCD-$i$}
 	\label{Alg:Ri_asyn_ts}
 	\scriptsize{\begin{algorithmic}[1]
 			\STATE initialize: Set $\mathbf a =  \mathbf 0 $, $\mathbf x_i = \mathbf 0$, ${\bf \Sigma}^{-1}_i  =\frac{1}{\sigma^2} {\bf I}_{ L_i }$, $ \widetilde {\mathbf Y}_i = \mathbf Y_i$, $  {\mathbf {\barp}}_{i,n}(x) = \sqrt{\frac{g_n}{1+\kappa_n} }\mathbf p_{i,n}(x)$, $x \in \hat{\mathcal X}_i$, $n\in\mathcal N$, and choose $\epsilon>0$.
 			\STATE \textbf{repeat}
 			\STATE Set $\mathbf a_{\mathrm{last}} =\mathbf a $ and $\mathbf x_i = \mathbf x_{i,\mathrm{last}}$.
 			\STATE \textbf{for $n \in \mathcal N$ do}
 			\STATE \quad  Compute $\mathbf c_{i,n} = {\bf \Sigma}^{-1}_i   {\mathbf \barp}_{i,n}(x_{i,n})  $ and $q_{i,n} =    {\mathbf \barp}_{i,n}^H(x_{i,n})  \mathbf c_{i,n}  $.
 			\STATE \quad Compute
 			${\bf \Sigma} ^{-1}_{i,n} = {\bf \Sigma} ^{-1}_i+ 
 			\frac{ 	a_n    }
 			{1 -  a_nq_{i,n}    } \mathbf c_{i,n} \mathbf c_{i,n} ^H
 			$ and
 			$
 			\widetilde {\mathbf Y}_{i,n}  =	 \widetilde {\mathbf Y}_i  + a_{n} \sqrt{  \kappa_{n}  }  {\mathbf \barp}_{i,n}(x_{i,n})  \overline {\mathbf h}_{n}^T
 			$.
 			\STATE \quad  \textbf{for $x  \in \hat{\mathcal X_i}$ do}
 			\STATE \quad \quad  Compute $\mathbf c_{i,n,x}  =  {\bf \Sigma}^{-1}_{i,n}   {\bf \barp}_{i,n}(x )   $ and $\mathbf d_{i,n,x} =  \widetilde{\mathbf Y}_{i,n}^H \mathbf c_{i,n,x} $. 
 			\STATE \quad \quad  Compute $\alpha_{i,n}(x ) =   {\mathbf \barp}_{i,n}^H(x )  \mathbf c_{i,n,x}  $, $\beta_{i,n}(x ) = \frac{1}{M}	 \mathbf d^H_{i,n,x}\mathbf d_{i,n,x}  $, and
 			 $\eta_{i,n}(x ) =    	2\frac{\sqrt{ \kappa_n }}{M}   	\mathrm{Re} \left(
 			 	\overline {\bf h}_n^T \mathbf d_{i,n,x} \right)   
 		 $. 
 			\STATE \quad \textbf{end for}
 			\STATE  \quad  Compute $(a_{i,n}^*,x_{i,n}^*) $ according to~(\ref{Opt_time_t_Ri_asyn}), and 
 			update $(a_n, x_{i,n}) =  (a_{i,n}^*,x_{i,n}^*)$.
 			\STATE \quad Compute $\mathbf c_{i,n}  = {\bf \Sigma}^{-1}_{i,n}   {\mathbf \barp}_{i,n}(x_{i,n})  $ and $q_{i,n} =    {\mathbf \barp}_{i,n}^H(x_{i,n})  \mathbf c_{i,n}  $.
 				\STATE \quad Compute
 			$
 			{\bf \Sigma} ^{-1}_{i} = {\bf \Sigma} ^{-1}_{i,n} -  
 			\frac{ 	a_n    }
 			{1 +  a_nq_{i,n}   } \mathbf c_{i,n} \mathbf c_{i,n} ^H
 			$ and
 			$
 			\widetilde {\mathbf Y}_i  =	 \widetilde {\mathbf Y}_{i,n}    -  a_{n} \sqrt{  \kappa_{n}  }     {\mathbf \barp}_{i, n}(x_{i,n})  \overline {\mathbf h}_{n}^T
 			$.
 			\STATE \textbf{end for}
 			\STATE \textbf{until} $ {\vert f_i(\mathbf a, \mathbf x_i)-f_i(\mathbf a_{\mathrm{last}},\mathbf x_{i,\mathrm{last}}) \vert } < \epsilon {\vert f_i(\mathbf a_{\mathrm{last}},\mathbf x_{i,\mathrm{last}}) \vert }$.
 	\end{algorithmic}}
 
 \end{algorithm}
 
 \mcele{In this part,
 we obtain a BCD algorithm with $\alpha_{i,n}(x)$, $\beta_{i,n}(x)$, and $\eta_{i,n}(x)$, $x \in \hat {\mathcal X}_{i}$ being computed according to~(\ref{C1_asyn}),~(\ref{C2_asyn}), and~(\ref{C3_asyn}), respectively, using basic matrix operations.
Later in Section IV. C, we shall see that this algorithm is more computationally efficient for small ranges of offset.
Thus, we also refer it to as the BCD algorithm for small offset ranges.}
  The details of the proposed algorithm are summarized in Algorithm~\ref{Alg:Ri_asyn_ts} (\GBCD-$i$).  
Specifically,
in Steps~5 and~6, 
 we compute $\mathbf \Sigma^{-1}_{i,n}$ and $ \widetilde {\mathbf Y}_{i,n}$ based on $\mathbf \Sigma^{-1}_{i} $ and $ \widetilde {\mathbf Y}_i$, respectively, similar to the updates of $\mathbf \Sigma^{-1}  $ and $\widetilde {\mathbf Y}$ in Step~8 of Algorithm~\ref{Alg:Ri_syn};
in Steps~8 and~9, we compute $\alpha_{i,n}(x)$, $\beta_{i,n}(x)$, and $\eta_{i,n}(x)$ according to~(\ref{C1_asyn}),~(\ref{C2_asyn}), and~(\ref{C3_asyn}), respectively, for all $x \in \hat{\mathcal X}_i$, similar to the computation of $\alpha_n$, $\beta_n$, and $\eta_n$ in
Steps~5 and~6 of 
Algorithm~\ref{Alg:Ri_syn};
in Steps~12 and~13, we compute $\mathbf \Sigma^{-1}_{i} $ and $ \widetilde {\mathbf Y}_i$ based on $\mathbf \Sigma^{-1}_{i,n}$ and $ \widetilde {\mathbf Y}_{i,n}$, respectively, similar to Steps~5 and~6.
For all $n \in \mathcal N$, the computational complexities of 
Step~5, Step~6, Steps~7-10, Step~11, Step~12, and Step~13 are 
$\mathcal O (L_i^2)$, $\mathcal O (L_i(L_i+ M))$, $\mathcal O( \vert \hat { \mathcal X}_{i}\vert   L_i(L_i+ M)) $, $\mathcal O (  \hat{\mathcal X}_{i})$, $\mathcal O (L_i^2)$, and $\mathcal O (L_i(L_i+ M))$, respectively, 
as $L,M,Q \rightarrow \infty$,
where $L_i$ and $ \vert \hat { \mathcal X}_{i}\vert $ are given in~(\ref{L_i}) and~(\ref{hatX}), respectively.
Thus, the overall computational complexity of each iteration of Algorithm~\ref{Alg:Ri_asyn_ts} is $\mathcal O(\vert \hat { \mathcal X}_{i}\vert N    L_i (L_i + M)) $,
as $N,L,M,Q \rightarrow \infty$.
Substituting $L_i$ in~(\ref{L_i}) and $ \vert \hat { \mathcal X}_{i}\vert$ in~(\ref{hatX}) into $\mathcal O(\vert \hat { \mathcal X}_{i}\vert N    L_i (L_i + M)) $, we can obtain
the detailed computational complexities of Algorithm~\ref{Alg:Ri_asyn_ts} for the three asynchronous cases given in Table~\ref{tab1:Complexity}.
Following the proof
for~\cite[Proposition 3.7.1]{Bertsekas1998NP}, we can show that the objective values of the iterates $  (\mathbf a, \mathbf x_i)$
generated by Algorithm~\ref{Alg:Ri_asyn_ts} converge.

 \begin{table}[t] 
 	\caption{\small\upshape{Computational complexities of Algorithm~\ref{Alg:Ri_asyn_ts} and 	Algorithm~\ref{Alg:Ri_asyn_IM} as $N,M,L,Q\rightarrow\infty$}.}
 	\centering
 	\scriptsize{\begin{tabular}{ c|c |c|c}\hline
 		\multirow{2}*{Estimation Methods} &	\multicolumn{3}{|c}{Computational Complexity of Each Iteration }      \\ 
 		\cline{2-4}
 		&  Asynchronous case-t ($i=$ t)  & Asynchronous case-f ($i=$ f) & Asynchronous case-(t,f) ($i=$ (t,f)) \\ \hline
 		\GBCD-$i$ \mcele{(Alg.~2)}     &  $ \mathcal O\big( (D+1)N L(L +M) \big)$       & $\mathcal O\Big(  \Omega Q  N   L (L+M) \Big)$ & $\mathcal O\Big((D+1)\Omega Q N    L(L+M) \Big)$  \\ \hline
 		\multirow{2}* {\FFTBCD-$i$ \mcele{(Alg.~3)}}       & 	\multirow{2}*{$ \mathcal O\big(N  L(L\log_2 L+M) \big) $  }     &   	\multirow{2}*{$\mathcal O \big(N   L(L+M) +N Q\log_2 Q \big)   $}   &  $ \mathcal O\Big(N  L( L\log_2 L +M)   $\\  
 		&  &   &    $  + (D+1)NQ \log_2 Q\Big)$	\\
 		\hline
 	\end{tabular}}
 	\label{tab1:Complexity}
 	\vspace{ \belowsize mm}
 \end{table}


\subsubsection{BCD Algorithm for \mcone{Large Offset Ranges}}

\mcele{In this part, we obtain a BCD algorithm with $\boldsymbol\alpha_{i,n} \triangleq (\alpha_{i,n}(x))_{x\in  \hat {\mathcal X}_i}$, $\boldsymbol\beta_{i,n} \triangleq (\beta_{i,n}(x))_{x\in  \hat {\mathcal X}_i}$, 
and $\boldsymbol\eta_{i,n} \triangleq (\eta_{i,n}(x))_{x\in  \hat {\mathcal X}_i}$ being computed according to their equivalent forms, which
are products of a matrix and a submatrix of a DFT/IDFT matrix and can be efficiently computed via FFT/IFFT.
Here, $\boldsymbol\alpha_{i,n}, \boldsymbol\beta_{i,n} ,\boldsymbol\eta_{i,n}\in \mathbb R^{	\vert\hat {\mathcal X}_{i}\vert} $, $i\in$ \{t, f\} and $ \boldsymbol \alpha_{\text{(t,f)},n} ,  \boldsymbol \beta_{\text{(t,f)},n}, \boldsymbol \eta_{\text{(t,f)},n}   \in \mathbb R ^{\vert\hat {\mathcal X}_{{\text{t}},n}\vert \times \vert\hat {\mathcal X}_{\text{f},n}\vert}$.
 Later in Section~IV.~C, we shall see that this algorithm is more computationally efficient for large offset ranges.
Thus, we also refer to it as the BCD algorithm for large offset ranges.}

 	 
 
 In what follows, we first introduce the equivalent forms of $ \boldsymbol \alpha_{i,n}$, $ \boldsymbol \beta_{i,n}$, and $ \boldsymbol \eta_{i,n}$, $i \in $ \{t, f, (t,f)\}.
For notation convenience, for arbitrary $\mathbf X  \in \mathbb C^{L_i\times L_i}$ and $  \mathbf p \in \mathbb C^{L_i}$ where $i\in$ \{t, f, (t,f)\}, define:
 \begin{align}
	& \boldsymbol  \Xi_i (\mathbf X , \mathbf p )	\triangleq    (   \mathbf U_i^H \boldsymbol\Psi(\mathbf X)) \odot (\mathbf U_i \boldsymbol\Psi(\mathbf p^*  \mathbf p ^T ))  )  \in \mathbb C^{L_i\times L_i},    i \in \text{\{t, f, (t,f)\}}, \label{Mfun}
\end{align}
 where $\mathbf U_i \triangleq \mathbf F_{L_i} \mathbb I (i \neq\text{f}) + \mathbf I_{L_i} \mathbb I (i =\text{f})  $ and  
 $\boldsymbol\Psi (\mathbf B) \triangleq (\psi_k(\mathbf B))_{k\in\mathcal K}$ with $\mathbf B \in \mathbb C^{K\times K}$, $\mathcal K \triangleq \{1,2,...,K\}$, and 
	\mcele{\begin{align}
	&\boldsymbol	\psi_k(\mathbf B) \triangleq  \left(1   + \left(\frac{\sqrt{2}}{2} -1 \right)\mathbb I(k =1) \right)  \left[ ((  (\mathbf B)_{\ell, \ell+k-1} )_{\ell\in\{1, ...,K-k+1\}})^T , \mathbf 0_{k-1}^T \right]^T .   \label{k_column}
\end{align}}%
	\mcele{Note that $ \sqrt{2} \boldsymbol	\psi_1(\mathbf B)$ represents the main diagonal of $\mathbf B$, and $ \boldsymbol	\psi_{k}(\mathbf B)$ represents the $(k-1)$-th super-diagonal of $\mathbf B$, for all $k  \in {\mathcal K} \setminus\{1\} $.}
In asynchronous case-t,
the equivalent forms of $\boldsymbol \alpha_{\text{t},n} $, $\boldsymbol \beta_{\text{t},n} $, and $\boldsymbol \eta_{\text{t},n} $ are given as follows.
\begin{lemma}[Equivalent Forms of $\boldsymbol\alpha_{\textup{t},n}$, $\boldsymbol\beta_{\textup{t},n} $, and $\boldsymbol\eta_{\textup{t},n} $]\label{Lem:Compute_t}
	\begin{align}
		& \boldsymbol\alpha_{\text{t},n}=  	\frac{ 2g_n/L_{\text{t}}}{ (1+ \kappa_n)} \left( \mathrm {Re} \Big(  \mathbf
		F_{L_{\text{t}}}
			  \boldsymbol\Xi_{\text{t}}\left(  {\bf\Sigma}_{\text{t},n }^{-1},    \mathbf p_{\text{t},n}(0 )   \right) 
			   \mathbf 1_{L_{\text{t}}}    \Big) \right)_{ \widetilde{\mathcal D}  },  
		\label{C1t_Ri_n} \\
		&	 \boldsymbol\beta_{\text{t},n}= \frac{ 2g_n/L_{\text{t}}}{  M(1+ \kappa_n)}  
		  \Big( \mathrm {Re}  \Big( 
		\mathbf  	F_{L_{\text{t}}}
		 	    \boldsymbol\Xi_{\text{t}}\left( {\bf\Sigma}_{\text{t},n }^{-1}   \widetilde {\mathbf Y}_{\text{t} ,n}    \widetilde {\mathbf Y}_{\text{t},n }^H      {\bf\Sigma}_{\text{t} ,n}^{-1}  ,     \mathbf p_{\text{t},n}(0 )   \right)  
              \mathbf 1_{L_{\text{t}}}
		\Big)  \Big)_{ \widetilde{\mathcal D}} ,
		\label{C2t_Ri_n}  \\
		& 	 \boldsymbol\eta_{\text{t},n}=  \frac{ 2/L_{\text{t}}}{ M }		\sqrt{\frac{g_n\kappa_n}{1+ \kappa_n}}  \Big(\mathrm {Re} \Big(
		\mathbf 	F_{L_{\text{t}}}
	     \left(
		\mathbf
		F_{L_{\text{t}}}^{H}
		(	\overline {\bf h}_n^T  \widetilde {\mathbf Y}_{\text{t},n}^H        {\bf\Sigma}_{\text{t} ,n}^{-1} )^T \right)  
		\odot \Big (\mathbf F_{L_{\text{t}}}  \mathbf p_{\text{t},n}(0 ) \Big)
		\Big)\Big)_{\widetilde{\mathcal D}} ,
		\label{C3t_Ri_n} 	
	\end{align}
	where $ \widetilde{\mathcal D} \triangleq \{1,2,...,D+1\}$. 
\end{lemma}
\begin{IEEEproof} 
	Please refer to Appendix D.
\end{IEEEproof}


Since $\mathbf F_{L_{\text{t}}} $ is a $L_{\text{t}}$-dimensional DFT matrix,
 the matrix multiplications with $\mathbf F_{L_{\text{t}}} $ and $\frac{1}{L_{\text{t}}} \mathbf F^H_{L_{\text{t}}} $
in~(\ref{C1t_Ri_n})-(\ref{C3t_Ri_n}) can be efficiently computed using $L_{\text{t}}$-dimensional FFT and IFFT, respectively. 



 In asynchronous case-f, 
the equivalent forms of $\boldsymbol \alpha_{\text{f},n} $, $\boldsymbol \beta_{\text{f},n} $, and $\boldsymbol \eta_{\text{f},n} $ are given as follows.

\begin{lemma}[Equivalent Forms of $\boldsymbol\alpha_{\textup{f},n}$, $\boldsymbol\beta_{\textup{f},n} $, and $\boldsymbol\eta_{\textup{f},n} $]\label{Lem:Compute_f}
	\begin{align}
		& \boldsymbol\alpha_{\text{f},n} \!= \!  	\frac{2g_n}{1+ \kappa_n} \! \left( \!\mathrm {Re} \left( \! \mathbf
		T_{\text{f}} 
   \left( \boldsymbol\Xi _{\text{f}}\left( {\mathbf\Sigma}_{ \text{f},n}^{-1},   \mathbf p_{\text{f},n} (0)   \right)  \right)^T   \!\! 	\!
  \mathbf 1_{L_{\text{f}}} \right) \!\right)_{\mathcal Q} \!  ,  
		\label{C1f_Ri_n} 
	\\
		&	 \boldsymbol\beta_{\text{f},n}= \frac{2g_n}{ M(1+ \kappa_n)}   \Big(  \mathrm {Re} \Big(
		\mathbf  T_{\text{f}} 
			   \left( \boldsymbol\Xi _{\text{f}}\left(   {\bf\Sigma}_{\text{f},n }^{-1}    \widetilde {\mathbf Y}_{\text{f} ,n}    \widetilde {\mathbf Y}_{\text{f},n }^H      {\bf\Sigma}_{\text{f} ,n}^{-1},   \mathbf p_{\text{f},n} (0)   \right)  \right)^T  \!\!\! \mathbf 1_{L_{\text{f}}} 
		\Big)\Big)_{\mathcal Q},
		\label{C2f_Ri_n}  
	\\
	& 	 \boldsymbol\eta_{\text{f},n}=\frac{ 2}{M}		\sqrt{\frac{g_n\kappa_n}{1+ \kappa_n}}  \Big(  \mathrm {Re} \Big(
	\mathbf  T_{\text{f}}  
	 \left((\overline {\bf h}_n^T  \widetilde{\mathbf Y}_{\text{f},n}^H         {\bf\Sigma}_{\text{f} ,n}^{-1})^T    \odot \mathbf p_{\text{f},n} (0)    \right)
	\Big)  \Big)_{\mathcal Q},
	\label{C3f_Ri_n} 	
	\end{align}
	where 
	\begin{align}
		&	\mathbf T_{\text{f}}    \triangleq [ \boldsymbol\tau_{\text{f}} (   \omega^{(1)}), \boldsymbol\tau_{\text{f}} (  \omega^{(2)})  ,...,
		\boldsymbol\tau_{\text{f}}  ( \omega^{(Q)})   ]^T \in \mathbb C^{Q\times L_{\text{f}} } \label{DFT}.
	\end{align} 
	
\end{lemma}
\begin{IEEEproof} 
	Please refer to Appendix E.  
\end{IEEEproof}

Note that 
	the matrix multiplications with $\mathbf	T_{\text{f}}$ in~(\ref{C1f_Ri_n})-(\ref{C3f_Ri_n}) 
	can be efficiently computed using a zero-padded $ \DeltaQf$-dimensional IFFT, where $\DeltaQf\triangleq Q\left\lceil  \frac{L_{\text{f}}}{Q}      \right\rceil   \geq L_{\text{f}} $.
	The details for computing matrix multiplications with $\mathbf	T_{\text{f}}$ by IFFT are provided in Appendix F.


In asynchronous case-(t,f), the 
equivalent forms of $ \boldsymbol \alpha_{\text{(t,f)},n}$, $ \boldsymbol \beta_{\text{(t,f)},n}$, and $\boldsymbol \eta_{\text{(t,f)},n}$ are given as follows.
 
 \begin{lemma}[Equivalent Forms of $\boldsymbol\alpha_{\textup{(t,f)},n}$, $\boldsymbol\beta_{\textup{(t,f)},n}$, and $\boldsymbol\eta_{\textup{(t,f)},n}$]\label{Lem:Compute_tf4}
 	\begin{align}
 		&  \boldsymbol\alpha_{\text{(t,f)},n}= \frac{ 2g_n/L_{\text{(t,f)}}}{ (1+ \kappa_n)}   \bigg( \mathrm {Re}   \Big(  \mathbf F_{L_{\text{(t,f)}}}  
 			\Big(  
 	 \boldsymbol\Xi_{\text{(t,f)}}\left(  {\bf\Sigma}_{\text{(t,f)},n }^{-1},      \mathbf p_{\text{(t,f)},n}(0,0)   \right)  \Big)_{ \widetilde {\mathcal D},: }\mathbf T_{\text{(t,f)}}^{T}       \Big)  \bigg)_{:,\mathcal Q }    ,  
 	\label{C1tf4_Ri_n} ,\\
 	& \boldsymbol\beta_{\text{(t,f)},n}= \frac{ 2g_n/L_{\text{(t,f)}}}{ M(1+ \kappa_n)} \! \bigg( \! \mathrm {Re}     \Big(  \mathbf F_{L_{\text{(t,f)}}}   
 	  	\Big( 
 	 \boldsymbol\Xi_{\text{(t,f)}}\left( {\bf\Sigma}_{\text{(t,f)},n }^{-1}    \widetilde {\mathbf Y}_{\text{(t,f)} ,n}    \widetilde {\mathbf Y}_{\text{(t,f)},n }^H      {\bf\Sigma}_{\text{(t,f)} ,n}^{-1}  ,      \mathbf p_{\text{(t,f)},n}(0,0)   \right) 
 	  \!\Big)_{ \widetilde {\mathcal D},: } \! \! \mathbf T_{\text{(t,f)}}^{T}       \Big) \! \bigg) _{ \!\!:,\mathcal Q }   ,
 	\label{C2tf4_Ri_n}
 \\& 
 		 \boldsymbol\eta_{\text{(t,f)},n}=\frac{ 2}{M}		\sqrt{\frac{g_n\kappa_n}{1+ \kappa_n}}  \bigg( \mathrm {Re} \bigg( \Big( \left( (\overline {\bf h}_n^T  \widetilde{\mathbf Y}_{\text{(t,f)},n}^H         {\bf\Sigma}_{\text{(t,f)} ,n}^{-1})^T \odot \mathbf p_{\text{(t,f)},n}(t,0) \right)_{t \in \mathcal D}\Big)^T  	\mathbf  T_{\text{(t,f)}}^T     \bigg)\bigg)_{:,\mathcal Q}, \label{C3tf4_Ri_n}
 	\end{align}%
 	Here,
 	\begin{align}
 	&	\mathbf T_{\text{(t,f)}}   	 \triangleq [ \boldsymbol\tau_{\text{(t,f)}}  (   \omega^{(1)}), \boldsymbol\tau_{\text{(t,f)}}  (  \omega^{(2)})  ,...,
 		\boldsymbol\tau _{\text{(t,f)}}  ( \omega^{(Q)})   ]^T \in \mathbb C^{Q\times  L_\text{(t,f)}  }.  \label{DFTtf}
 		\end{align}
 	\end{lemma}
 \begin{IEEEproof} 
 	Please refer to Appendix G.
 \end{IEEEproof}
 
Analogously,
the matrix multiplications 
with $\mathbf F_{L_{\text{(t,f)}}} $, $\frac{1}{L_{\text{(t,f)}}} \mathbf F^H_{L_{\text{(t,f)}}} $, and $\mathbf	T_{\text{(t,f)}}$
in~(\ref{C1tf4_Ri_n})-(\ref{C3tf4_Ri_n}) can be efficiently computed using $L_{\text{(t,f)}}$-dimensional FFT, $L_{\text{(t,f)}}$-dimensional IFFT, and zero-padded $\DeltaQtf$-dimensional IFFT, respectively, where $\DeltaQtf\triangleq Q\left\lceil  \frac{L_{\text{(t,f)}}}{Q}      \right\rceil \geq L_{\text{(t,f)}}$.
 The details for computing matrix multiplications with $\mathbf	T_{\text{(t,f)}}$ using IFFT are given in Appendix F.

\begin{algorithm}[t]
	\caption{\FFTBCD-$i$}
	\label{Alg:Ri_asyn_IM}
	\scriptsize{	\begin{algorithmic}[1]
			\STATE initialize: Set $\mathbf a =  \mathbf 0  $, $\mathbf t = \mathbf 0 $, $\boldsymbol\omega =\mathbf 0 $, 
			${\bf \Sigma}^{-1}_{i} =\frac{1}{\sigma^2} {\bf I}_{ L_i}$, $	 \widetilde {\mathbf Y}_{i } = \mathbf Y_{i}$,
			$   \boldsymbol\Phi_i = \frac{1}{ \sigma^4}  {\mathbf Y}_{i }    {\mathbf Y}_{i }^H   $, $  {\mathbf {\barp}}_{i,n}(x) = \sqrt{\frac{g_n}{1+\kappa_n} }\mathbf p_{i,n}(x)$, $x \in \hat{\mathcal X}_i$, $n\in\mathcal N$,
			and choose $\epsilon >0$.
			\STATE \textbf{repeat}
			\STATE Set $\mathbf a_{\mathrm{last}} =\mathbf a $ and $\mathbf x_i = \mathbf x_{i,\mathrm{last}}$.
			\STATE \textbf{for $n \in \mathcal N$ do}
			\STATE \quad  Compute $\mathbf c_{i,n}  = {\bf \Sigma}^{-1}_i   {\mathbf \barp}_{i,n}(x_{i,n})  $ and $q_{i,n}  =   {\mathbf p}_{i,n}^H(x_{i,n})  \mathbf c_{i,n}  $.
			\STATE \quad Compute
			${\bf \Sigma} ^{-1}_{i,n} = {\bf \Sigma} ^{-1}_i+ 
			\frac{ 	a_n    }
			{1 -  a_n q_{i,n}   } \mathbf c_{i,n} \mathbf c_{i,n} ^H
			$,
			$
			\widetilde {\mathbf Y}_{i,n}  =	 \widetilde {\mathbf Y}_i  + a_{n} \sqrt{ \kappa_{n}  }  {\mathbf \barp}_{i,n}(x_{i,n})  \overline {\mathbf h}_{n}^T
			$.
			\STATE  \quad   
			Compute $\boldsymbol \varphi_{i,n} = 	{\bf \Sigma}_{i  }^{-1}   (\widetilde {\mathbf Y}_{i } \overline {\mathbf h}_{n}^{*} ) $, $\boldsymbol \phi_{i,n} ={\bf \Sigma}_{i,n  }^{-1}   (\widetilde {\mathbf Y}_{i,n } \overline {\mathbf h}_{n}^{*} ) $,
			$\mathbf D _{i,n}= 	        
			\boldsymbol \Phi_{i } +  a_n \sqrt{  \kappa_n }  \left( {\mathbf c_{i,n}}  { \boldsymbol \varphi_{i,n}^H}  +  {\boldsymbol \varphi_{i,n}}  \mathbf c_{i,n}^H \right) +  a_n^2   M   \kappa_n   { \mathbf c}_{i,n}  { \mathbf c}_{i,n}^H   $, \\ \quad $\mathbf s_{i,n} =     \mathbf D _{i,n}{\bf \barp}_{i,n}(x_{i,n }) $, and $  \boldsymbol \Phi_{i,n }=  \mathbf D_{i,n} + \frac{ a_n    }{1 -   a_n q_{i,n} } \left( {\mathbf c}_{i,n} {\mathbf s}_{i,n}^H +   {\mathbf s}_{i,n}  {\mathbf c}_{i,n}^H \right) + 
			\frac{    a_n ^2    {\bf \barp}_{i,n}^H(x_{i,n }) {\mathbf s}_{i,n} }
			{(1 -   a_n q_{i,n})^2 }  {\mathbf c}_{i,n}   {\mathbf c}_{i,n}^H $. 
			\STATE \quad Compute $\boldsymbol\alpha_{i,n}$, $\boldsymbol\beta_{i,n}$, and $\boldsymbol\eta_{i,n}$ according to their equivalent forms in Lemmas~\ref{Lem:Compute_t}-\ref{Lem:Compute_tf4} where the matrix multiplications with $\mathbf F_{L_i}$, $\mathbf F_{L_i}^H$,\\ \quad and $\mathbf T_i$ are computed using FFT/IFFT.
			\STATE \quad Compute $(a_{i,n}^*,x_{i,n}^*) $ according to~(\ref{Opt_time_t_Ri_asyn}), and update $(a_n, x_{i,n}) =  (a_{i,n}^*,x_{i,n}^*)$.
			\STATE \quad  Compute $\mathbf c_{i,n}  = {\bf \Sigma}^{-1}_{i,n}   {\mathbf \barp}_{i,n}(x_{i,n})  $ and $q_{i,n}  =    {\mathbf \barp}_{i,n}^H(x_{i,n})  \mathbf c_{i,n}  $. 
			\STATE \quad Compute
			${\bf \Sigma} ^{-1}_{i } = {\bf \Sigma} ^{-1}_{i,n}- 
			\frac{ 	a_n  }
			{1 +  a_n q_{i,n}   } \mathbf c_{i,n} \mathbf c_{i,n} ^H
			$,
			$
			\widetilde {\mathbf Y}_{i }  =	 \widetilde {\mathbf Y}_{i,n}  - a_{n} \sqrt{ \kappa_{n}} {\mathbf \barp}_{i,n}(x_{i,n})  \overline {\mathbf h}_{n}^T
			$.
			\STATE \quad Compute 
			$\mathbf D _{i,n}= 	 \boldsymbol\Phi_{i,n}       
			- a_n \sqrt{  \kappa_n}  \left( {\mathbf c_{i,n}}  \boldsymbol \phi_{i,n}^H +  {\boldsymbol \phi_{i,n}}   \mathbf c_{i,n}^H\right)  +  a_n^2  M   \kappa_n { \mathbf c}_{i,n}  { \mathbf c}_{i,n}^H   $,  
			$\mathbf s_{i,n} =      \mathbf D _{i,n}{\bf \barp}_{i,n}(x_{i,n }) $
			and $ \boldsymbol \Phi_{i} 
			=  \mathbf D_{i,n} $\\ \quad $-\frac{ a_n   }{1 +   a_n q_{i,n} } \left({\mathbf c}_{i,n} {\mathbf s}_{i,n}^H +   {\mathbf s}_{i,n}  {\mathbf c}_{i,n}^H\right)   + 
			\frac{   a_n^2     {\bf \barp}_{i,n}^H(x_{i,n }) {\mathbf s}_{i,n} }
			{ (1+  a_n q_{i,n} )^2}  {\mathbf c}_{i,n}   {\mathbf c}_{i,n}^H $.
			\STATE \textbf{end for} 
			\STATE \textbf{until} 
			$  {\vert f_i(\mathbf a, \mathbf x_i)-f_i(\mathbf a_{\mathrm{last}},\mathbf x_{i,\mathrm{last}}) \vert} < \epsilon{\vert f_i(\mathbf a_{\mathrm{last}},\mathbf x_{i,\mathrm{last}}) \vert } $.
	\end{algorithmic}}%
\end{algorithm}

Next, we propose a BCD algorithm where $ \boldsymbol \alpha_{i,n}$, $ \boldsymbol \beta_{i,n}$, and $ \boldsymbol \eta_{i,n}$, in the equivalent forms given by Lemmas~\ref{Lem:Compute_t}-\ref{Lem:Compute_tf4}, are computed using FFT/IFFT.
The details of the corresponding BCD algorithm are summarized in Algorithm~\ref{Alg:Ri_asyn_IM} (\FFTBCD-$i$).\footnote{Bruteforcely applying Algorithm~3 for asynchronous case-(t,f) in asynchronous case-t (or asynchronous case-f) yields a higher computation cost than applying Algorithm~3 for asynchronous case-t (or asynchronous case-f).}
In Algorithm~\ref{Alg:Ri_asyn_IM}, define 
$\boldsymbol \Phi_{i } \triangleq {\mathbf \Sigma}_{i  }^{-1}   \widetilde {\mathbf Y}_{i }   \widetilde{\mathbf Y}_{i }^H   {\mathbf \Sigma}_{i   }^{-1} $ and
$\boldsymbol \Phi_{i,n} \triangleq  {\mathbf \Sigma}_{i ,n }^{-1}   \widetilde {\mathbf Y}_{i,n }   \widetilde {\mathbf Y}_{i,n }^H  {\mathbf \Sigma}_{i   ,n}^{-1}  $, $n \in \mathcal N$.
Specifically, in Steps~5-7, 
besides computing $\mathbf \Sigma^{-1}_{i,n}$ and $ \widetilde {\mathbf Y}_{i,n}$ based on $\mathbf \Sigma^{-1}_{i} $ and $ \widetilde {\mathbf Y}_i$, respectively, as in Steps~5 and~6 of Algorithm~\ref{Alg:Ri_asyn_ts}, 
 we compute $\boldsymbol \Phi_{i ,n}$ based on $\boldsymbol \Phi_{i }$ by matrix manipulations and the methods for calculating $\mathbf \Sigma^{-1}_{i,n}$ and $ \widetilde {\mathbf Y}_{i,n}$;
in Step~8, we compute $\boldsymbol\alpha_{i,n} $, $\boldsymbol\beta_{i,n} $, and $\boldsymbol\eta_{i,n}$ according to their equivalent forms given by Lemmas~\ref{Lem:Compute_t}-\ref{Lem:Compute_tf4};
in Steps 10-12, we compute $\mathbf \Sigma^{-1}_{i} $, $ \widetilde {\mathbf Y}_i$, and $\boldsymbol \Phi_{i }$ based on $\mathbf \Sigma^{-1}_{i,n}$, $ \widetilde {\mathbf Y}_{i,n}$, and $\boldsymbol \Phi_{i ,n}$, respectively, similar to the updates of $\mathbf \Sigma^{-1}_{i,n}$, $ \widetilde {\mathbf Y}_{i,n}$, and $\boldsymbol \Phi_{i ,n}$ in Steps~5-7.
For all $n \in \mathcal N$, the computational complexities of  
Steps~5,~6,~7,~9,~10,~11, and~12 are $\mathcal O(L_i^2)$, $\mathcal O(L_i(L_i+M))$, $\mathcal O( L_i(L_i+M))$, $\mathcal O (\vert \hat {\mathcal X}_i\vert )$, $\mathcal O(L_i^2)$, $\mathcal O(L_i(L_i+M))$, and $\mathcal O( L_i^2)$, respectively, as $L ,M\rightarrow \infty$;
and
 the computational complexities of Step~$8$ for 
 asynchronous case-t,
  asynchronous case-f,
  and 
asynchronous case-(t,f)
  are 
  $\mathcal O(  L_{\text{t}} (L_{\text{t}} \log_2 L_{\text{t}} +M)   )$,
  $\mathcal O( L_{\text{f}}(L_{\text{f}} +M) + Q \log_2 Q)$, 
  and
  $\mathcal O(L_{\text{(t,f)}}(L_{\text{(t,f)}}\log_2 L_{\text{(t,f)}}+M ) + (D+1)Q \log_2 Q     )$, respectively,
  as $L,M,Q \rightarrow \infty$. 
   \mcone{Note that the complexity analysis for Step~$8$ is summarized in appendix H.}
   Substituting $L_i$ in~(\ref{L_i}) into the overall computational complexity above for each asynchronous case-$i$, we can obtain
the computational complexities of Algorithm~\ref{Alg:Ri_asyn_IM} for the three asynchronous cases given in Table~\ref{tab1:Complexity}.\footnote{
	For Algorithm~3 for asynchronous case-$i$ where $i \in$ \{f, (t,f)\}, $Q\geq L_i$ is more preferable than $Q <L_i$, as $Q <L_i$ always yields lower detection accuracy than some $Q'\geq L_i$ at the same computational complexity.  
	\label{footnoteQ}}   
Following the proof
for~\cite[Proposition 3.7.1]{Bertsekas1998NP}, we can show that the objective values of the iterates $  (\mathbf a, \mathbf x_i)$
generated by Algorithm~\ref{Alg:Ri_asyn_IM} converge.%


  \subsection{\mctht{Comparisons of Algorithm~\ref{Alg:Ri_asyn_ts} and Algorithm~\ref{Alg:Ri_asyn_IM}}}

 \begin{table}[t]
	\caption{\small\upshape{Computational complexities of Algorithm~\ref{Alg:Ri_asyn_ts} and 	Algorithm~\ref{Alg:Ri_asyn_IM} for $M =\Theta(L^s)$ and $Q=\Theta(L^q)$ ($s\geq 0, q\geq 1$), as $N, L \rightarrow\infty$. For ease of exposition, we define $\bar s \triangleq \max(1,s)$.}}
	\vspace{ \tablecapsize mm}
	\centering
	\scriptsize{\begin{tabular}{ c|c |c|c}\hline
			\multirow{2}*{Estimation Methods} &	\multicolumn{3}{|c}{Computational Complexity of each Iteration}      \\ 
			\cline{2-4}
			&  Asynchronous case-t ($i=$ t)    &Asynchronous case-f ($i=$ f) &  Asynchronous case-(t,f) ($i=$ (t,f)) \\ \hline
			\GBCD-$i$ \mcele{(Alg.~2)}    &  $ \mathcal O\left( (D+1)N L^{1+\bar s} \right)$       & $\mathcal O\Big( \Omega NL^{1+q + \bar s} \Big)$ & $\mathcal O\Big((D+1)\Omega N  L^{1+ q+ \bar s} \Big)$  \\ \hline
			\multirow{2}* {\FFTBCD-$i$ \mcele{(Alg.~3)}}         & \multirow{2}*{$ \mathcal O\big(N  L^{1+\bar s} (1 + \mathbb{I}(s\leq 1)\log_2 L )  \big) $  }     &   	{$\mathcal O \Big(N    L^{1+\max(\bar s, q-1)}   $}   &  $ \mathcal O\Big(N  L^{1+ \max(\bar s, q-1)}     $\\ 
			&     &  $\times(1 +    \mathbb{I}(\bar s \leq q -1)  \log_2 L  ) \Big) $   &  	$\times(1 +    \mathbb{I}( \bar s\leq q-1) (D+1) \log_2 L  ) \Big) $	  \\
			\hline
	\end{tabular}}
	\label{tab1:Complexity_sup} 
	\vspace{ -2 mm}
\end{table}

\mctht{In this subsection, we compare Algorithm~\ref{Alg:Ri_asyn_ts} and Algorithm~\ref{Alg:Ri_asyn_IM}.}
Note that Algorithm~\ref{Alg:Ri_asyn_ts} and Algorithm~\ref{Alg:Ri_asyn_IM} differentiate from each other only in the computation methods for $\boldsymbol\alpha_{i,n}, \boldsymbol\beta_{i,n}, \boldsymbol\eta_{i,n}$ (cf. Steps~7-10 of Algorithm~\ref{Alg:Ri_asyn_ts} and Steps~7,~8,~12 of Algorithm~\ref{Alg:Ri_asyn_IM}). 
Thus, Algorithm~\ref{Alg:Ri_asyn_ts} and Algorithm~\ref{Alg:Ri_asyn_IM} can generate identical iterates $(\mathbf a, \mathbf x_i)$ and achieve the same detection accuracy.
 \mcone{Therefore, we select the algorithm with the lower computational complexity.}
In the following, we \mcone{analytically} compare the computational complexities of Algorithm~\ref{Alg:Ri_asyn_ts} and Algorithm~\ref{Alg:Ri_asyn_IM}.

First, we compare the computational complexities at arbitrary $N,L,M,Q \in \mathbb N^+$ with $Q\geq L \geq 6$ by analyzing the flop counts.
Note that we define a flop as one addition, subtraction,
 multiplication, or division of two real numbers, or one logarithm operation of a real number.
 \begin{lemma}[Computational Complexity Comparisons of Algorithm~\ref{Alg:Ri_asyn_ts} and Algorithm~\ref{Alg:Ri_asyn_IM} for Arbitrary Parameters]\label{Lem:FCC}
 	For any $N, M\in \mathbb N^+$ and $Q\geq L \geq 6$,
 	the flop count of Algorithm~\ref{Alg:Ri_asyn_ts} is smaller (larger) than that of Algorithm~\ref{Alg:Ri_asyn_IM} 
 	if the following conditions hold: 
 		(i) Asynchronous case-t: $   D <  \overline D_{\text{t}}$ ($  D > \underline D_{\text{t}}$) for some $ \overline D_{\text{t}}>0$ ($\underline D_{\text{t}}>0$);
 (ii) Asynchronous case-f: $\Omega <   \overline \Omega_{\text{f}}$ ($\Omega >  \underline \Omega_{\text{f}}$) for some $ \overline \Omega_{\text{f}}>0$ ($\underline \Omega_{\text{f}}>0$);
  (iii) Asynchronous case-(t,f): $  D< \overline D_{\text{(t,f)}} $ or $\Omega <  \overline \Omega_{\text{(t,f)}}$ ($ D>  \underline D_{\text{(t,f)}}$ or $\Omega >  \underline \Omega_{\text{(t,f)}}$) for some $ \overline D_{\text{(t,f)}},\overline \Omega_{\text{(t,f)}}>0$ ($\underline D_{\text{(t,f)}},\underline \Omega_{\text{(t,f)}}>0$).\footnote{\mcele{$\overline D_{\text{t}}$, $\underline D_{\text{t}}$, $\overline \Omega_{\text{f}}$, $\underline \Omega_{\text{f}}$,
  		$\overline D_{\text{(t,f)}}$, $ \overline \Omega_{\text{(t,f)}} $,
  		$\underline D_{\text{(t,f)}}$, and $	\underline \Omega_{\text{(t,f)}}$ depend on system parameters $L$, $M$, $Q$, $D$, and $\Omega$ in a complicated manner, and \mctht{their expressions} are given in Appendix~H.}}
 \end{lemma}
 \begin{IEEEproof} 
   	Please refer to Appendix I.
\end{IEEEproof}%

Next, we compare the computational complexities based on the order analysis shown in Table~\ref{tab1:Complexity} at large $N,L,M,Q \in \mathbb N^+$ with $Q\geq L$.
	 Suppose $M = \Theta(L^{s})$ for some $s\geq 0$ in each asynchronous case and $Q = \Theta (L^{q}) $ for some $ q\geq 1$ in asynchronous case-f and case-(t,f).
Then, the computational complexities of Algorithm~\ref{Alg:Ri_asyn_ts} and Algorithm~\ref{Alg:Ri_asyn_IM} for large parameters in Table~\ref{tab1:Complexity} become those in Table~\ref{tab1:Complexity_sup}.
		From Table~\ref{tab1:Complexity_sup}, we have the following result.
	\begin{lemma}[Computational Complexity Comparisons of Algorithm~\ref{Alg:Ri_asyn_ts} and Algorithm~\ref{Alg:Ri_asyn_IM} for Large Parameters]\label{Lem:FCCL}
Suppose $M = \Theta(L^s)$ for some $s\geq 0$ in each asynchronous case and $Q = \Theta (L^q) $ for some $ q\geq 1$ in asynchronous case-f and case-(t,f),
(i) In asynchronous case-t, the computational complexity order of Algorithm~\ref{Alg:Ri_asyn_IM}  	
	 is higher than (equivalent to) that of Algorithm~\ref{Alg:Ri_asyn_ts} if $s\leq1$ ($s>1$);
(ii) In asynchronous case-f and asynchronous case-(t,f), the computational complexity order of Algorithm~\ref{Alg:Ri_asyn_IM}  	
is lower than that of Algorithm~\ref{Alg:Ri_asyn_ts} 
for all $s\geq 0$ and $q \geq 1$.	
	\end{lemma} 
	

 \begin{IEEEproof} 
	Please refer to Appendix J.
\end{IEEEproof}%

\section{Numerical Results}

 In this section, we evaluate the performances of Prop-MLE-Syn, Prop-MLE-Small-$i$, and Prop-MLE-Large-$i$, $i\in$ \{t, f, (t,f)\}.\footnote{\mcaone{Source code for the experiment is avaliable at~\cite{gitcode11}.}}
As Prop-MLE-Small-$i$ and Prop-MLE-Large-$i$ have the same detection performance, they are both referred to as Prop-MLE-$i$ when showing detection performance.
We adopt the existing methods for the synchronous case~\cite{TSP_Chen_21,JSAC_Cui_2020, IoTJ-22-Tian,TSP_LL1_2018}, asynchronous case-t\cite{ICASSP_LL_2021 ,SPL-22-Wang}, and asynchronous case-f~\cite{TWC_CFO_2019,ICC-22-Liu} and the straightforward extensions of the existing methods for the synchronous case~\cite{TSP_Chen_21,JSAC_Cui_2020,  TSP_LL1_2018} to three asynchronous cases as baseline schemes, as illustrated in Table~\ref{tab_BL}. 
In each extension for asynchronous case-$i$, where $i\in$ \{t, f, (t,f)\}, we construct $N \vert \hat {\mathcal X}_{i } \vert $ perfectly synchronized virtual devices based on $N$ asynchronous actual devices. In particular, each actual asynchronous device $n$ corresponds to $\vert \hat {\mathcal X}_{i } \vert$ synchronous virtual devices which are indexed by $(n-1)\vert \hat {\mathcal X}_{i } \vert  +1, ...,n \vert \hat {\mathcal X}_{i } \vert $ and have preassigned pilots ${\mathbf p}_{i,n}(x)$, $x \in \hat {\mathcal X}_{i}  $ and separate activity states and channel vectors.
To obtain the binary activity state of device $n$ based on the estimated results, we perform the following thresholding rules: 
	$ \mathbf 1( {\hat a}_n \geq \theta ) $ for the proposed MLE-based methods and existing MLE-based methods~\cite{TSP_Chen_21,SPL-22-Wang, ICC-22-Liu}, 
    $   \mathbf 1( \max_{\ell\in \{(n-1)\vert \hat {\mathcal X}_{i } \vert  +1, ...,n \vert \hat {\mathcal X}_{i } \vert \}}  \hat {a}_{ \ell}^{\text{vir}} \geq\theta  ) $ for the extension of the existing MLE-based method~\cite{TSP_Chen_21}, 
    $   \mathbf 1(   \frac{\| \hat {\mathbf h}_{n }  \|_2^2 }{M}\geq\theta   ) $ for the existing AMP~\cite{TSP_LL1_2018}, \mctwe{group} LASSO~\cite{JSAC_Cui_2020,ICASSP_LL_2021}, and norm approximation-based method~\cite{TWC_CFO_2019}, 
 and 
 $   \mathbf 1( \frac{\max_{\ell\in \{(n-1)\vert \hat {\mathcal X}_{i } \vert  +1, ...,n \vert \hat {\mathcal X}_{i } \vert \}} \| \hat {\mathbf h}_{\ell}^{\text{vir}}  \|_2^2}{M}\geq\theta  ) $ for the extensions of existing AMP~\cite{TSP_LL1_2018} and \mctwe{group} LASSO~\cite{JSAC_Cui_2020}.
Here, $\theta > 0$ represents a threshold, ${\hat a}_n$ and $\hat {\mathbf h}_{n }$ denote device $n$'s estimated activity state and channel vector, respectively,
and ${\hat a}_{\ell}^{\text{vir}}$ and $ \hat {\mathbf h}_{\ell}^{\text{vir}}$ denote virtual device $\ell$'s estimated activity state and channel vector, respectively, where $n \in \mathcal N$ and $\ell \in \{1,...,N\vert \hat {\mathcal X}_{i } \vert\}$.
The iterative algorithms designed for solving optimization problems for MLE, \mctwe{group} LASSO, and norm approximation adopt the same convergence criterion,
i.e, the relative difference between the objective values at two consecutive iterations is smaller
than $10^{-7}$. The iterative algorithms based on AMP run 100 iterations.\footnote{Due to the differences between the AMP-based algorithms and optimization-based algorithms, we cannot use the same stopping criteria. Here, we choose to show the best detection performances of the AMP-based algorithms, which are mostly achieved within $100$ iterations.}
\mcele{All methods follow the same naming format, i.e., $\mu$-$i$, where $\mu\in$ \{Prop-MLE, AMP, MLE-Ra, MLE-Ri, LASSO, NA\} and $i\in$ \{Syn, t, f, (t,f)\} denote the device activity detection approach and the asynchronous case, respectively.
	For the same asynchronous case-$i$, the methods represented by different $\mu$ have different detection accuracies indicating their different abilities \mcone{to detect} device activities. However, for the same approach $\mu$, the methods represented by different~$i$ have different detection accuracies indicating the different difficulties of detecting device activities in different asynchronous cases.}

 \begin{table}[t]
	\caption{\small\upshape{Computational complexities (as $N,L,M,Q \rightarrow \infty)$. 
Here, GL and NA represent \mctwe{group} LASSO and norm approximation, respectively, and Ra} and Ri represent Rayleigh fading and Rician fading, respectively.
	}
\vspace{\tablecapsize mm}
	\centering
	\scriptsize{\begin{tabular}{ c|c |c|c}  \hline
				\multirow{1}*{Proposed Methods} & 	\multicolumn{3}{|c}{Computational Complexity of Each Iteration }     
			   \\ \hline
	Prop-MLE-Syn \mcele{(Alg.~1)}	&         	\multicolumn{3}{|c}{$\mathcal O(NL(L+M))$ }                  \\ \hline	
  $-$	&  	Asynchronous case-t ($i=$ t)  &Asynchronous case-f ($i=$ f) & Asynchronous case-(t,f) ($i=$ (t,f)) \\ \hline
			\GBCD-$i$ \mcele{(Alg.~2)}     &  $ \mathcal O\big( (D+1)N L(L +M) \big)$       & $\mathcal O\Big(\Omega Q N  L (L+M) \Big)$ & $\mathcal O\Big((D+1)\Omega Q N   L(L+M) \Big)$  \\ \hline
				\multirow{2}* {\FFTBCD-$i$ \mcele{(Alg.~3)}}       & 	\multirow{2}*{$ \mathcal O\big(N  L(L\log_2 L+M) \big) $  }     &   	\multirow{2}*{$\mathcal O \big(N   L(L+M) +N Q\log_2 Q \big)   $}   &  $ \mathcal O\Big(N  L( L\log_2 L +M)   $\\  
				&  &   &    $  + (D+1)NQ \log_2 Q\Big)$	\\       		
	\end{tabular}}
	\scriptsize{\begin{tabular}{ c|c|c|c|c }\hline
		\multirow{2}*{Cases} &	\multicolumn{2}{|c}{Existing Methods}  & 	\multicolumn{2}{|c}{Extensions of Existing Methods}    \\ 
		\cline{2-5}
		&  Name &Complexity    &Name  &Complexity    \\ \hline
		&  MLE-Ra-Syn~\cite{TSP_Chen_21} & $\mathcal O(NL^2)$     & \multirow{4}*{$-$} &  \multirow{4}*{$-$}    \\ 
		\cline{2-3}
			Synchronous 	&  MLE-Ri-Syn~\cite{IoTJ-22-Tian} &$\mathcal O(NL(L+M))$   &  & \\
		\cline{2-3} 
		case	&  AMP-Syn~\cite{TSP_LL1_2018} & $\mathcal O(NLM)$   &  &\\
		\cline{2-3}
		&GL-Syn~\cite{JSAC_Cui_2020} &  $\mathcal O(NLM)$    &  & \\
		\hline
		Asynchronous          &MLE-Ra-t~\cite{SPL-22-Wang} &   $\mathcal O((D+1)NL^2)$  &   \multirow{2}*{AMP-t (extensions of~\cite{TSP_LL1_2018})}   &   \multirow{2}*{$\mathcal O((D+1)NLM)$ }  \\ 
		\cline{2-3}
		case-t	& GL-t~\cite{ICASSP_LL_2021} &  $\mathcal O((D+1)NLM)$ & &  \\ \hline
		Asynchronous        &MLE-Ra-f~\cite{ICC-22-Liu} &   $\mathcal O(N(L^2+ Q\log_2 Q))$      &   \multirow{1}*{AMP-f (extensions of~\cite{TSP_LL1_2018})} & \multirow{1}*{$\mathcal O(\Omega QNLM )$}  \\  
		\cline{2-5}
		case-f	&NA-f~\cite{TWC_CFO_2019} &   $\mathcal O(\Omega QN LM )$   &  \multirow{1}*{GL-f (extensions of~\cite{JSAC_Cui_2020})}  &  \multirow{1}*{$\mathcal O(\Omega Q NLM )$} \\	
		\cline{2-3}
		\hline
		Asynchronous	       &\multirow{3}*{$-$} &   \multirow{3}*{$-$}  &   MLE-Ra-(t,f) (extensions of~\cite{TSP_Chen_21}) &    $\mathcal  O((D+1)\Omega QN L^2 )$  \\	  
		\cline{4-5}
		case-(t,f)	& &    &AMP-(t,f) (extensions of~\cite{TSP_LL1_2018}) & $\mathcal  O ((D+1)\Omega Q N LM )$  \\
		\cline{4-5}
		&   &  & GL-(t,f) (extensions of~\cite{JSAC_Cui_2020}) & $\mathcal  O ((D+1)\Omega Q N LM )$\\
		\hline 	                      
	\end{tabular}}
	\label{tab_BL} 
	\vspace{ \belowsize mm}
\end{table}

In the simulation, we adopt Gaussian pilot sequences (i.e., $\mathbf p_n$, $n\in\mathcal N$ are i.i.d. generated according to $\mathcal {CN}( \mathbf 0, \mathbf I_L)$)~\cite{TSP_LL1_2018, TIT_GC_2021,  TSP_Chen_21 ,ICC_Chen_2019,TWC_Jiang_2021,Arxiv-Jia-22,TWC-22-Shi,JSAC-21-Shao,TWC_Tao_2021,ICASSP_LL_2021,SPL-22-Wang, TWC_CFO_2019,ICCws_CFO_2020,ICC-22-Liu,TWC-Sun-22,Wang_ICCASP_21,BAMP02}.
Set $\overline {\mathbf h}_{n} =  (  e^{j(m-1)\phi_n })_{m\in\mathcal M}$~\cite{TCOM_19_Ozdogan}, 
where $\phi_n $ is uniformly chosen at random from the interval $[0,2\pi]$.
Set $g_n=1,n \in \mathcal N$ and $\kappa_n= \kappa,n \in \mathcal N$.\footnote{The proposed methods apply to any $g_n \in \mathbb R^+$, $n\in \mathcal N$ \mcmtwo{and $\kappa_n \in \mathbb R^+$, $n\in\mathcal N$}. In the simulation, we assume $g_n=1$, $n \in\mathcal N$ \mcmtwo{and $\kappa_n = \kappa$, $n\in\mathcal N$} for simplicity.}
For all $n\in\mathcal N$, $t_n$ and $\omega_n$ are uniformly chosen at random from the set $\mathcal D$~\cite{TWC_Tao_2021} and the interval $[- \Omega, \Omega]$\cite{TWC_CFO_2019}, respectively.
The device activities are i.i.d. with probability $p$ of being active.
We independently generate $3000$ realizations for $\mathbf p_n$, $a_n$, $t_n$, $\omega_n$, $ \overline {\mathbf h}_n$, $ \widetilde {\mathbf h}_n, n\in \mathcal N$, perform device activity detection in each realization, and evaluate the average error probability over all $3000$ realizations.
For all the considered methods, we evaluate the average error probabilities for $\theta \in \{0.01,0.02,...,1\}$ and choose
the optimal threshold which achieves the minimum average error probability. 
In the following, we set $N=1000$, $p=0.08$, $\sigma^2=2$ and choose $M=48$, $L=60$, $\kappa=-10 $~dB, $D=4$, $\Omega = \pi $, $Q = 128$ unless otherwise stated.\footnote{\mcele{The choices of system parameters $N$, $M$, $L$, $D$, $\Omega$, and $\kappa$ in our simulation are representative~\cite{TSP_LL1_2018,TIT_GC_2021, TWC_Jiang_2021, TSP_Chen_21,ICC_Chen_2019, ICASSP_LL_2021,SPL-22-Wang, TWC_CFO_2019,ICCws_CFO_2020, TWC-Sun-22, IoTJ-22-Tian }.}}

 \begin{figure}[t]
 	\begin{center}
 		
 		\subfigcapskip = \subcap pt
 		
 		\subfigure[\scriptsize{$\kappa$ at $M =32$ and $L=48 $.}] 
 	{\resizebox{\fsizefourfig  cm}{!}{\includegraphics{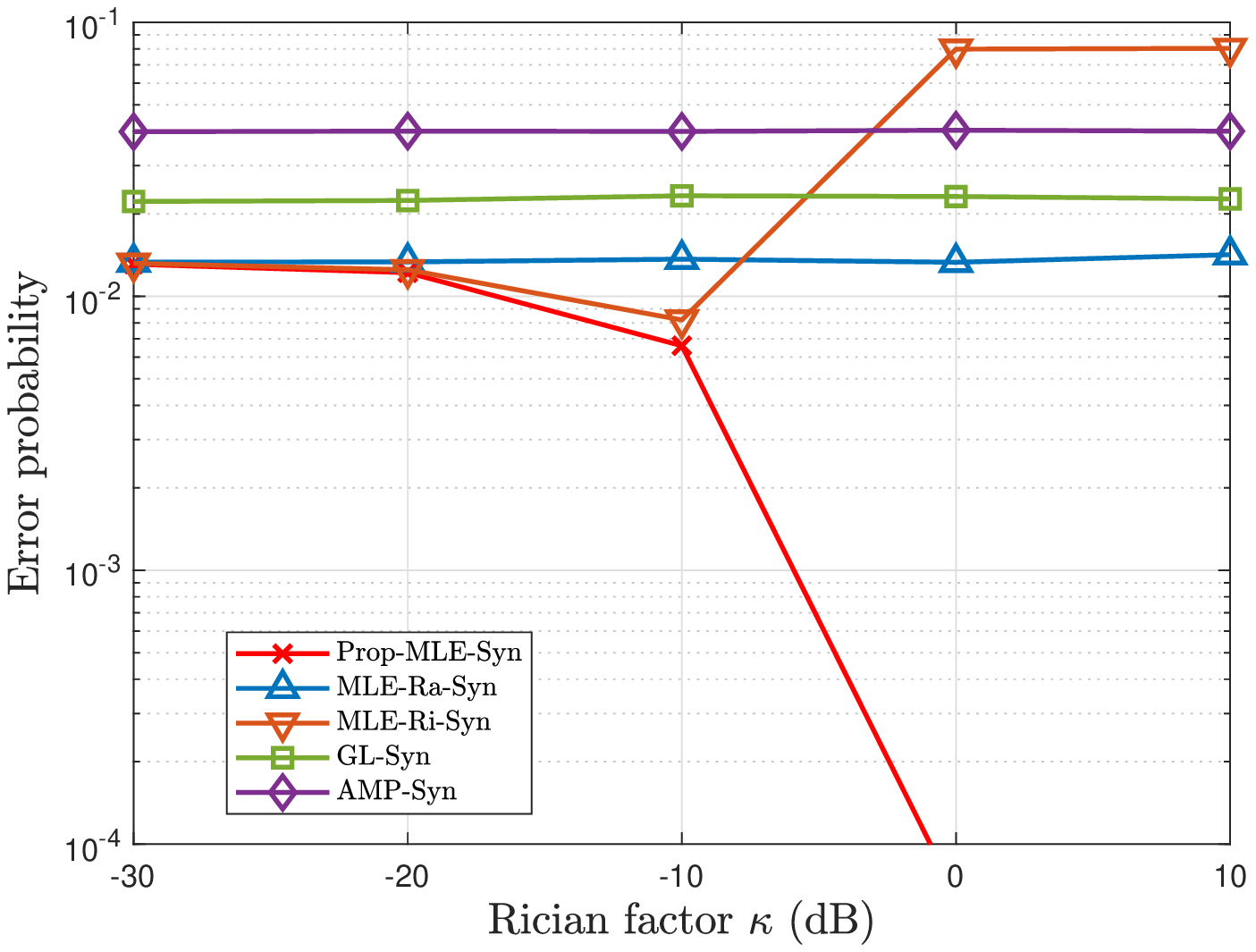}}} 
 	\subfigure[\scriptsize{$L$ at $M =32$.}]
 {\resizebox{\fsizefourfig cm}{!}{\includegraphics{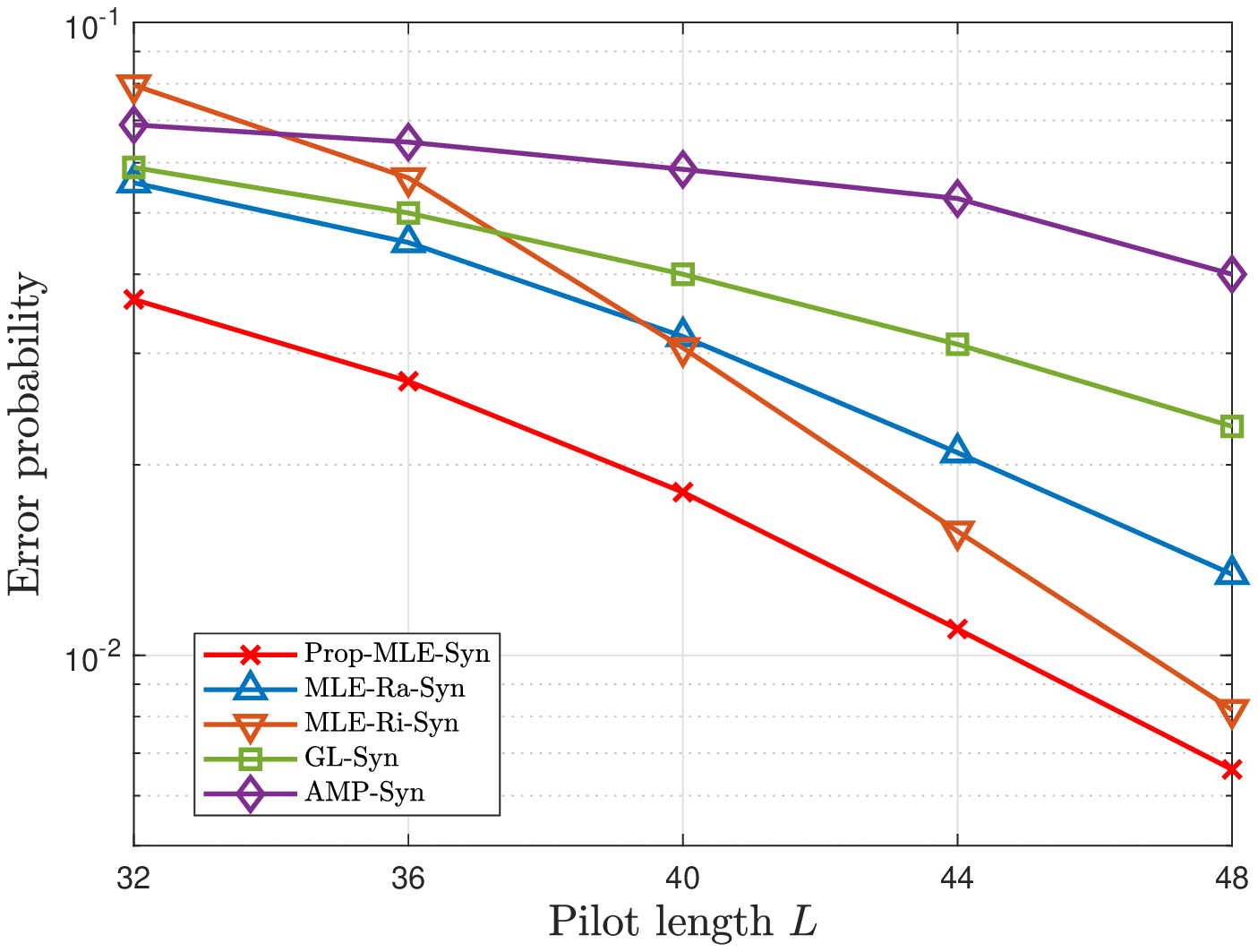}}}   
 \\
 \vspace{-3.5mm}
 \subfigure[\scriptsize{$M$ at $L =48$.}]
{\resizebox{\fsizefourfig  cm}{!}{\includegraphics{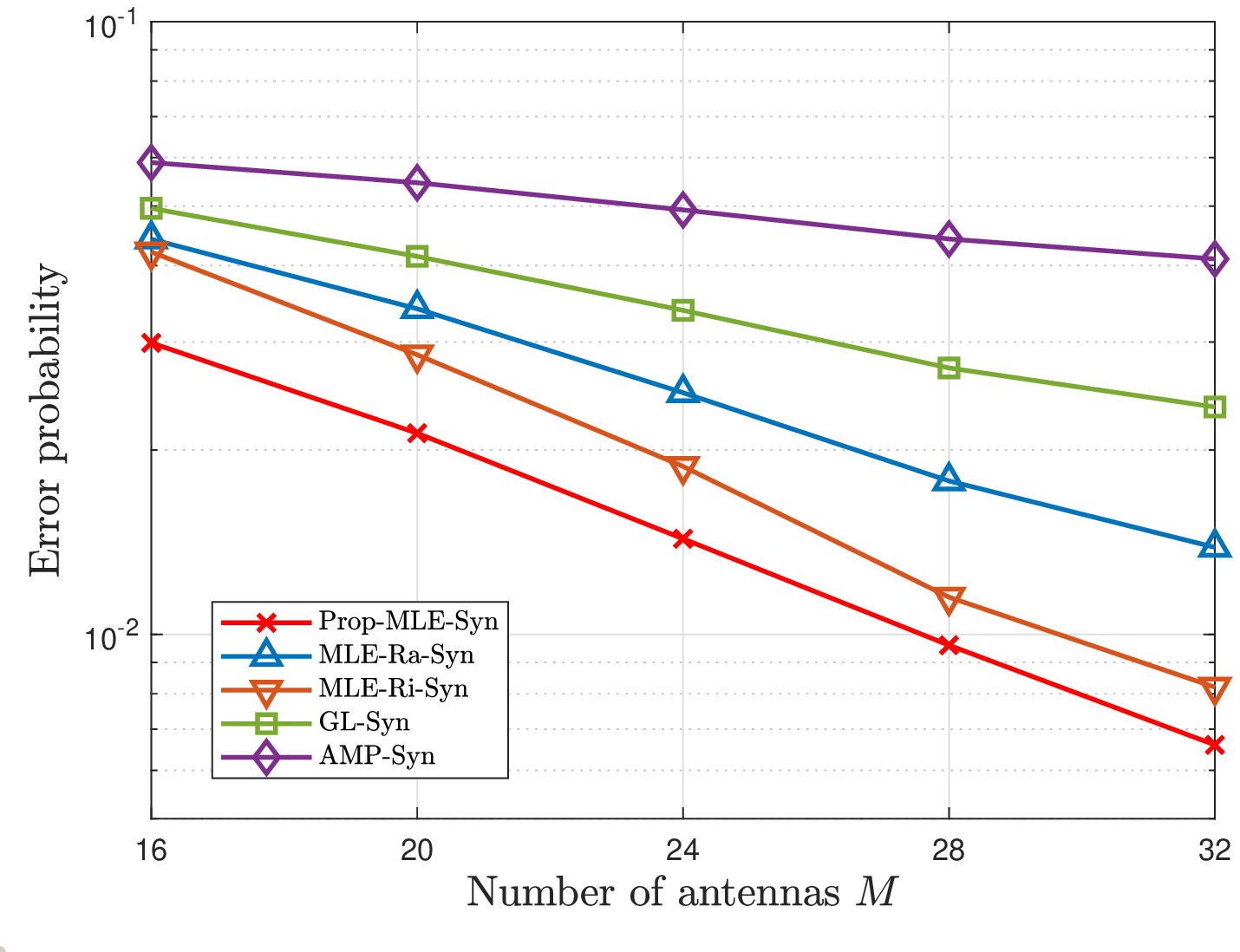}}}
\subfigure[\footnotesize{$p$ at $N=2000$, $M=256$, and $\sigma=20$.}]
{\resizebox{\fsizefourfig  cm}{!}{\includegraphics{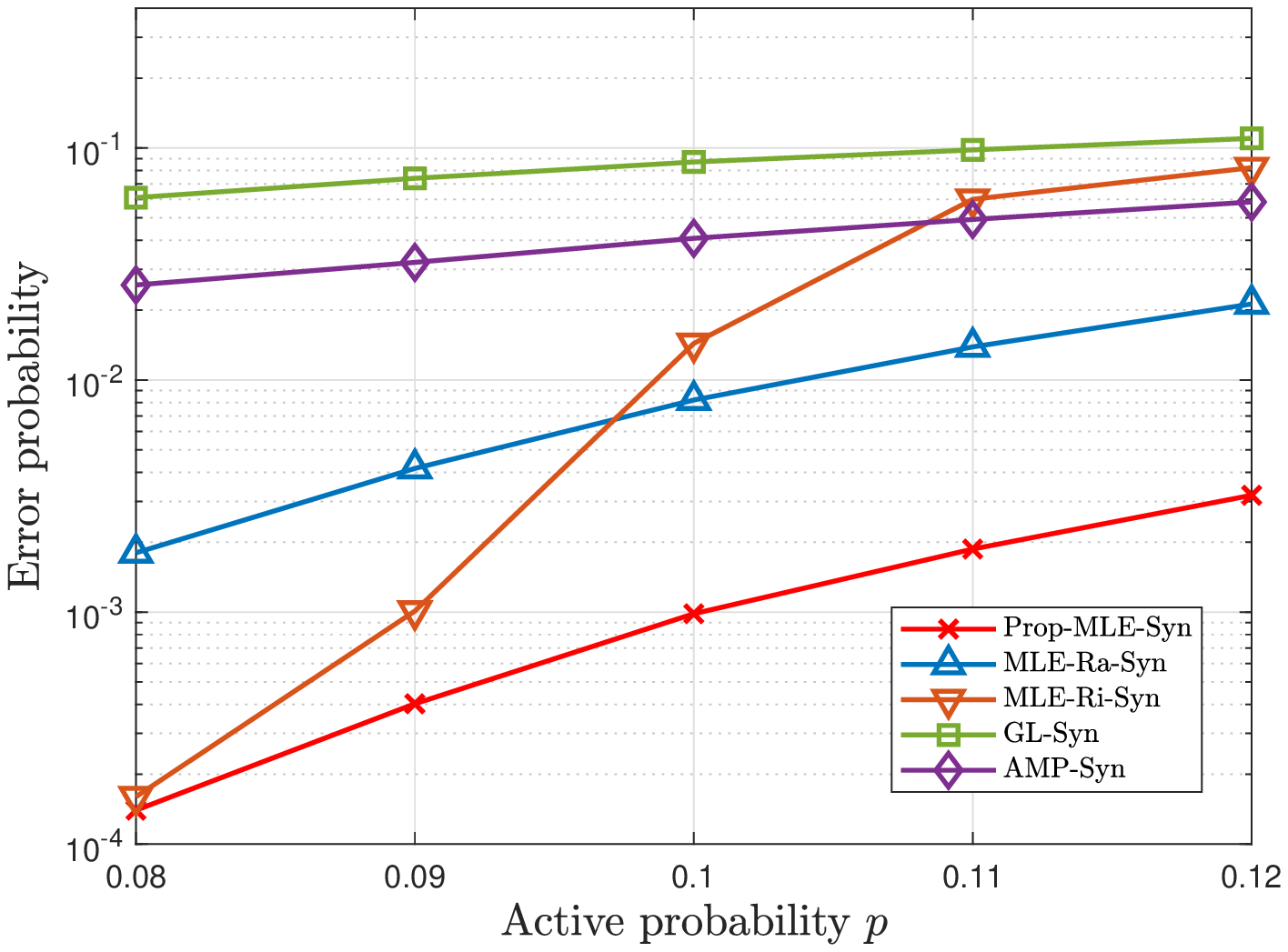}}} 

\end{center}
\vspace{-5mm}
\caption{\small{Error probability versus $\kappa$, $L$, $M$, and $p$ in the synchronous case.}}
\vspace{-5 mm}
\label{Figure:1}
\end{figure}

\begin{figure}[t]
\begin{center}
\subfigcapskip = \subcap pt

\subfigure[\scriptsize{$\kappa$.}] 
{\resizebox{\fsizefourfig  cm}{!}{\includegraphics{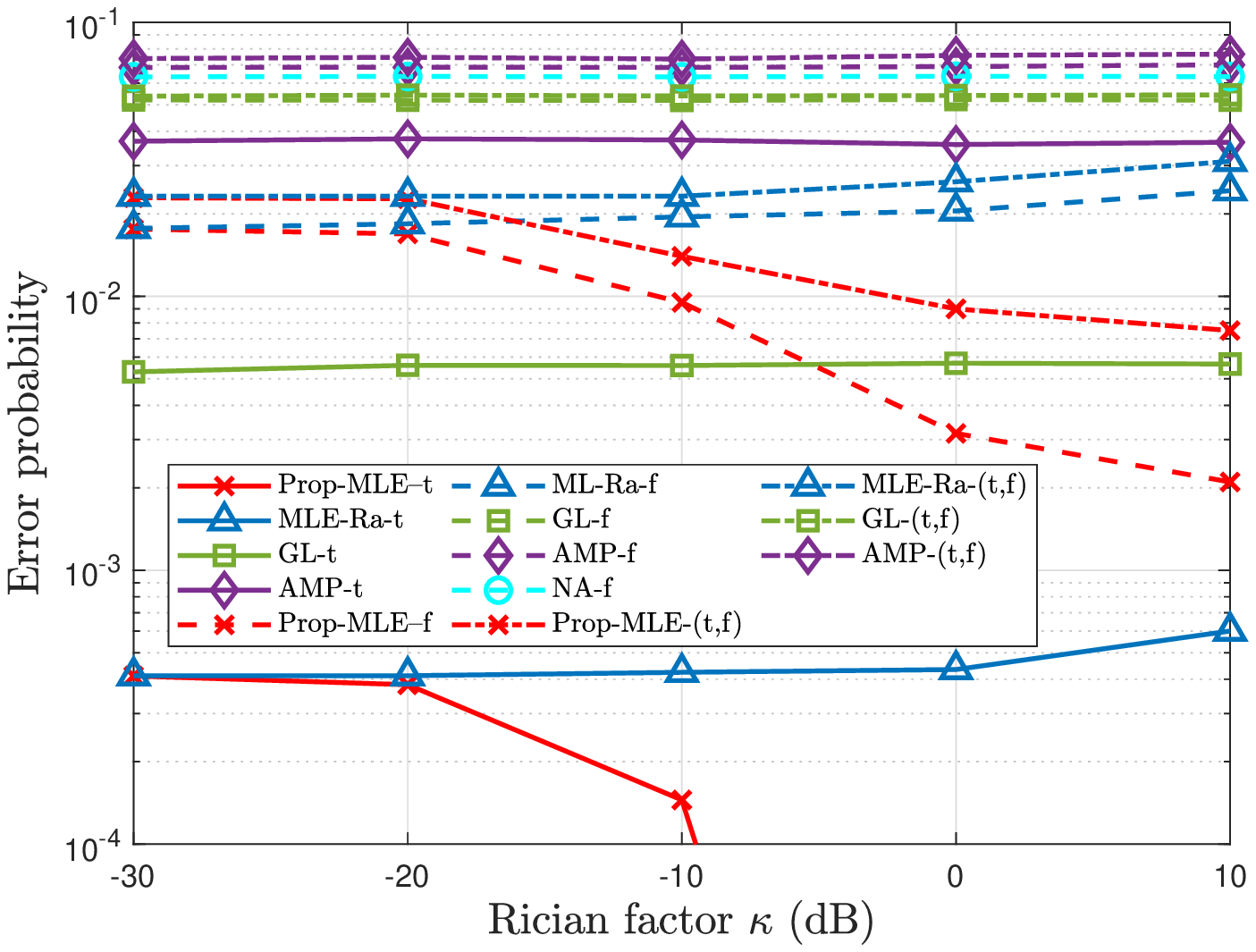}} } 
\subfigure[\scriptsize{$L$.}]
{\resizebox{\fsizefourfig cm}{!}{\includegraphics{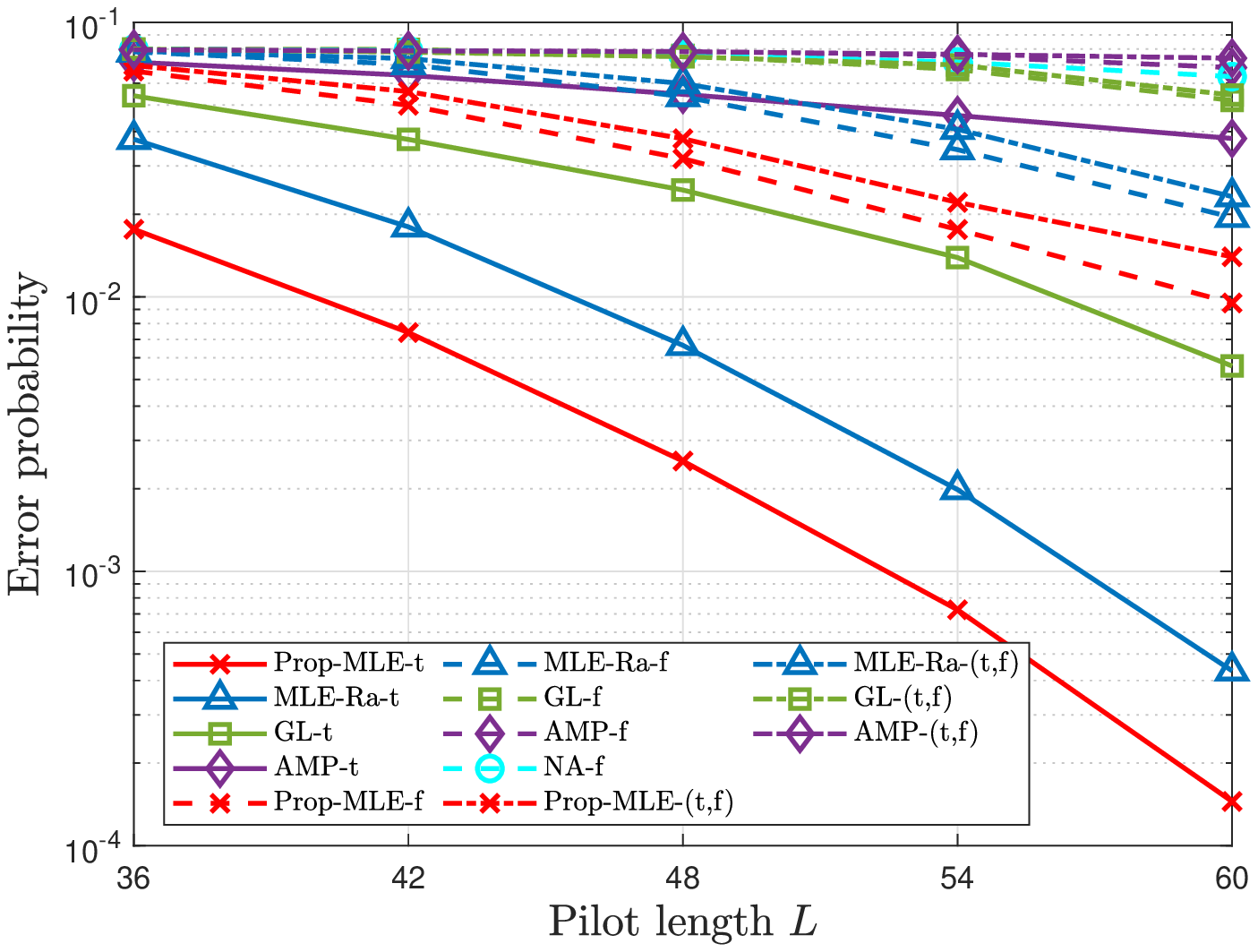}}}  
 \\
\vspace{-3.5mm}
\subfigure[\scriptsize{$M$.}]
{\resizebox{\fsizefourfig cm}{!}{\includegraphics{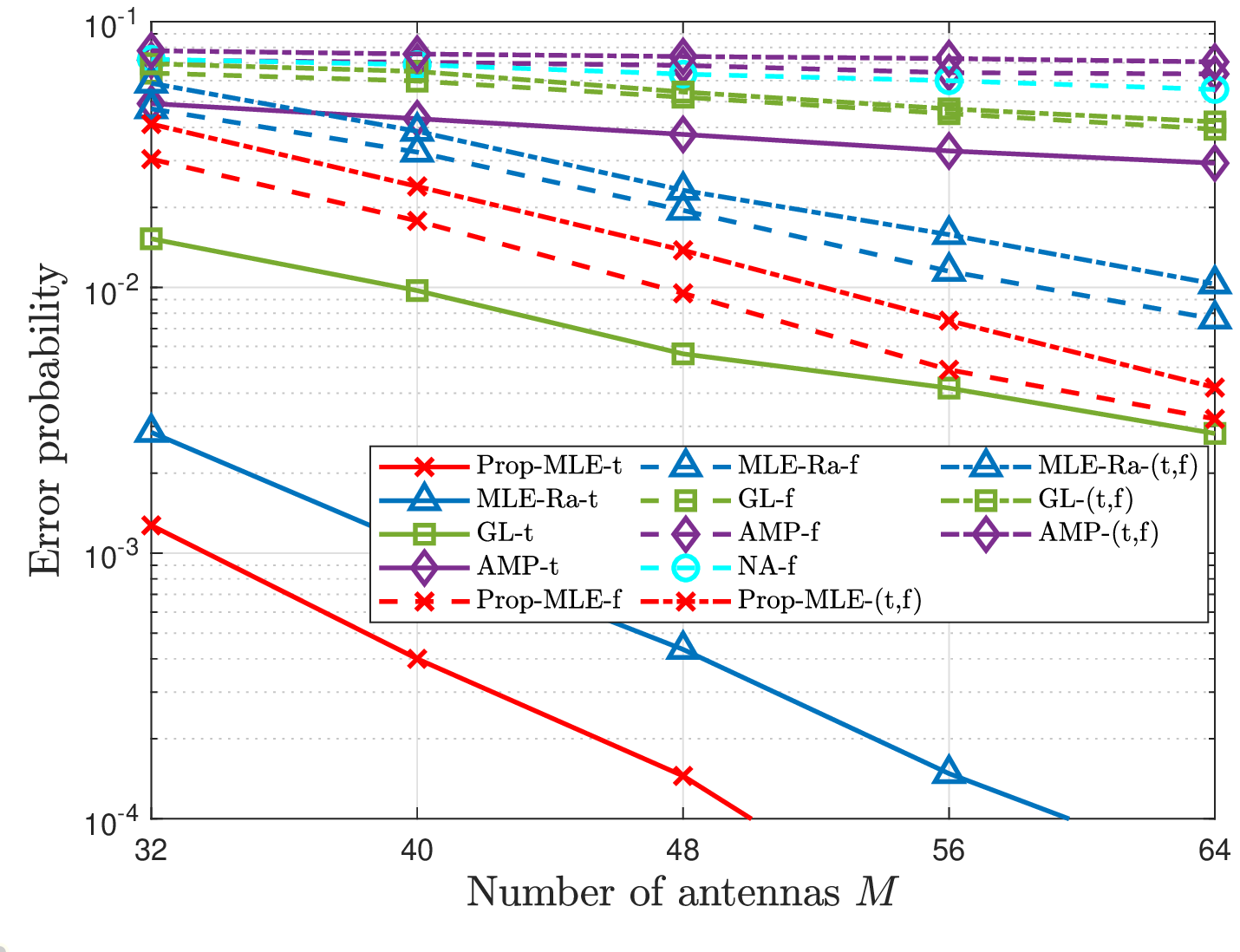}}}
\subfigure[\scriptsize{$p$ at $N=2000$, $M=256$, and $\sigma=10$.}]
{\resizebox{\fsizefourfig cm}{!}{\includegraphics{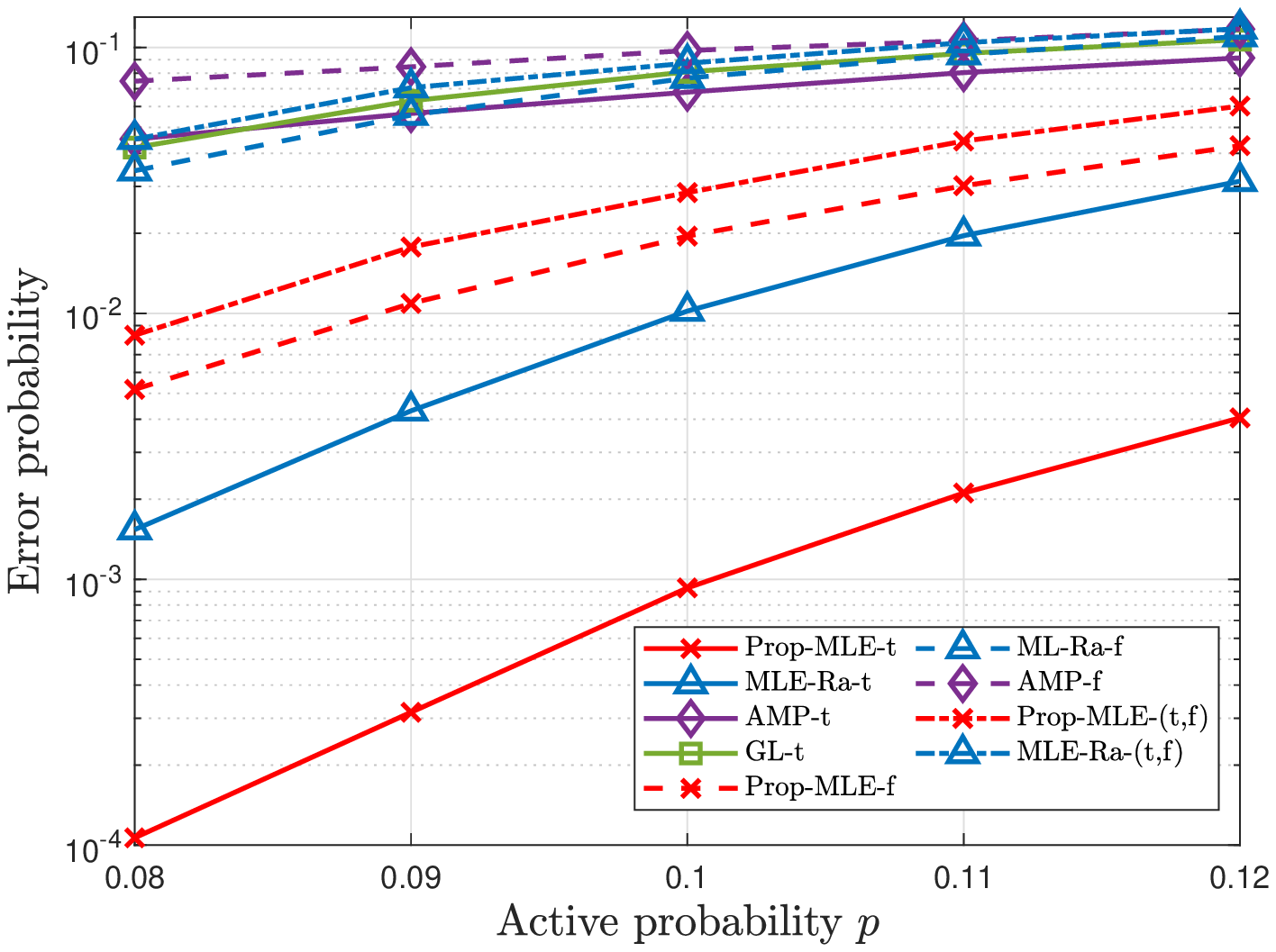}}}
\end{center}
\vspace{-5mm}
\caption{\small{Error probability versus $\kappa$, $L$, $M$, and $p$ in the asynchronous cases.}}
\vspace{\figcapsize mm}
\label{Figure:2}
\end{figure}

Fig.~\ref{Figure:1} and Fig.~\ref{Figure:2} plot the error probability versus the Rician factor $\kappa$,  pilot length $L$, and number of antennas $M$ in the synchronous case and three asynchronous cases, respectively.
We can make the following observations from Fig.~\ref{Figure:1}~(a) and Fig.~\ref{Figure:2}~(a). Firstly, the error probabilities of Prop-MLE-$i$, $i\in $ \{Syn, t, f, (t,f)\} significantly decrease with $\kappa$. This is because the influences of the device activities and offsets on the p.d.f.s of received signals	are larger when $\kappa $ is larger. Secondly, the error probabilities of MLE-Ra-$i$, $i\in $ \{Syn, t, f, (t,f)\} slightly increase with $\kappa$, as the errors caused by approximating Rician fading with Rayleigh fading increase with $\kappa$. Thirdly, the error probabilities of GL-$i$, AMP-$i$, $i\in $ \{Syn, t, f, (t,f)\}, and NA-f hardly change with $\kappa$, as \mctwe{group} LASSO~\cite{JSAC_Cui_2020,ICASSP_LL_2021} and the norm approximation-based method~\cite{TWC_CFO_2019} do not rely on the channel model.
Fourthly, when $\kappa\geq 1 $, the error probabilities of MLE-Ri-Syn reach the active probability $p$, meaning that MLE-Ri-Syn does not work reasonably in this regime.\footnote{The detection accuracy of MLE-Ri-Syn is determined by the approximation error, which is zero at $ \kappa =  0$ and is generally large at large $\kappa$.}
From Fig.~\ref{Figure:1}~(b), Fig.~\ref{Figure:2}~(b), Fig.~\ref{Figure:1}~(c), and Fig.~\ref{Figure:2}~(c), we can see that the error probability of each method decreases with $L$, as the number of measurement vectors increases with $L$; and the error probability of each method decreases with $M$, as the number of observations increases with $M$.
From Fig.~\ref{Figure:1}~(d) and Fig.~\ref{Figure:2}~(d), we can see that the error probability of each method decreases with $p$, as the interference between active devices increases with $p$.

  \begin{figure}[t]
  	\begin{center}
  		\subfigcapskip = \subcap pt
  		
  		\subfigure[\scriptsize{$D$.}]
  		{\resizebox{\fsize cm}{!}{\includegraphics{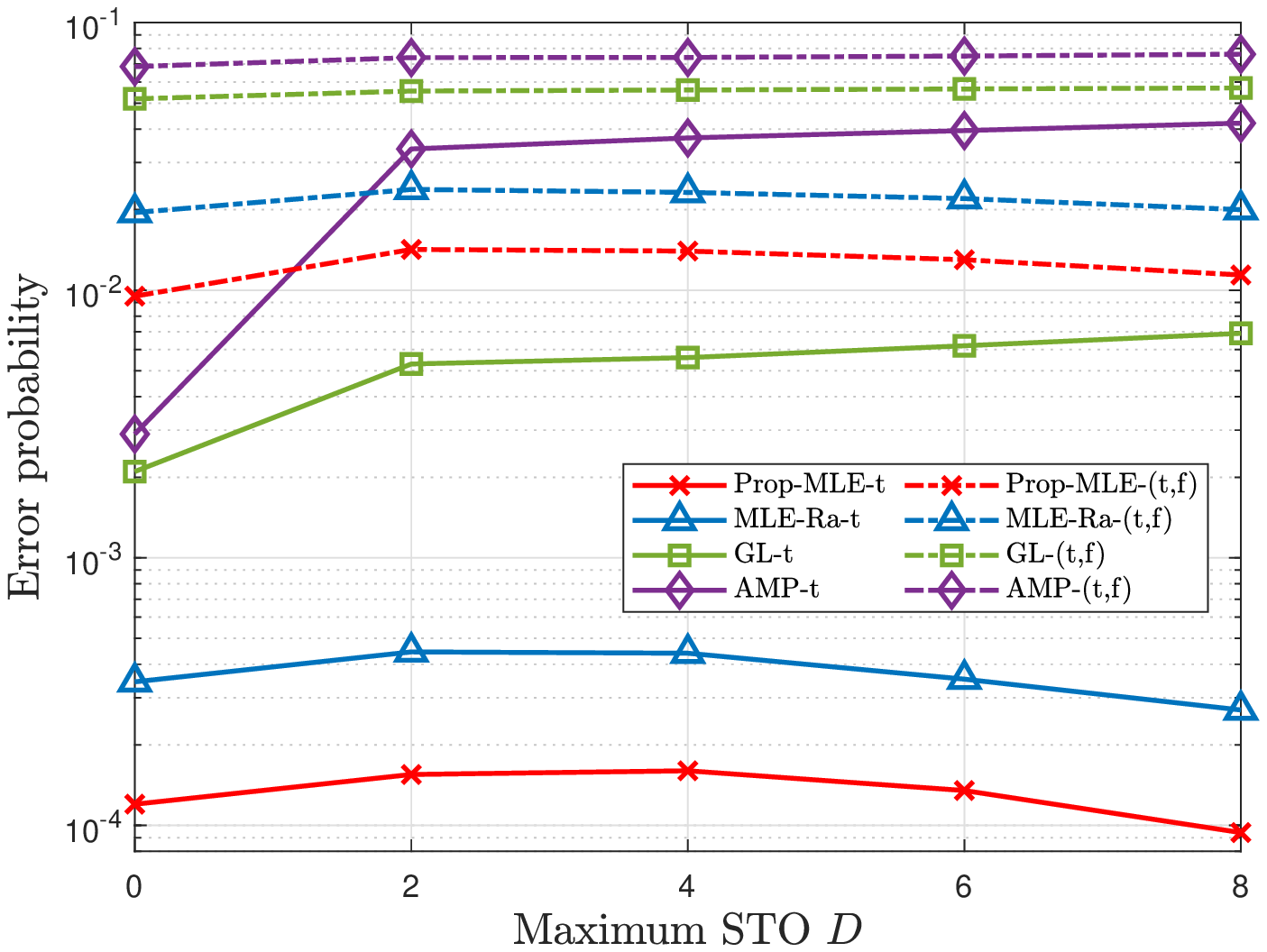}}}   
  		\subfigure[\scriptsize{$\Omega$.}]
  		{\resizebox{\fsize cm}{!}{\includegraphics{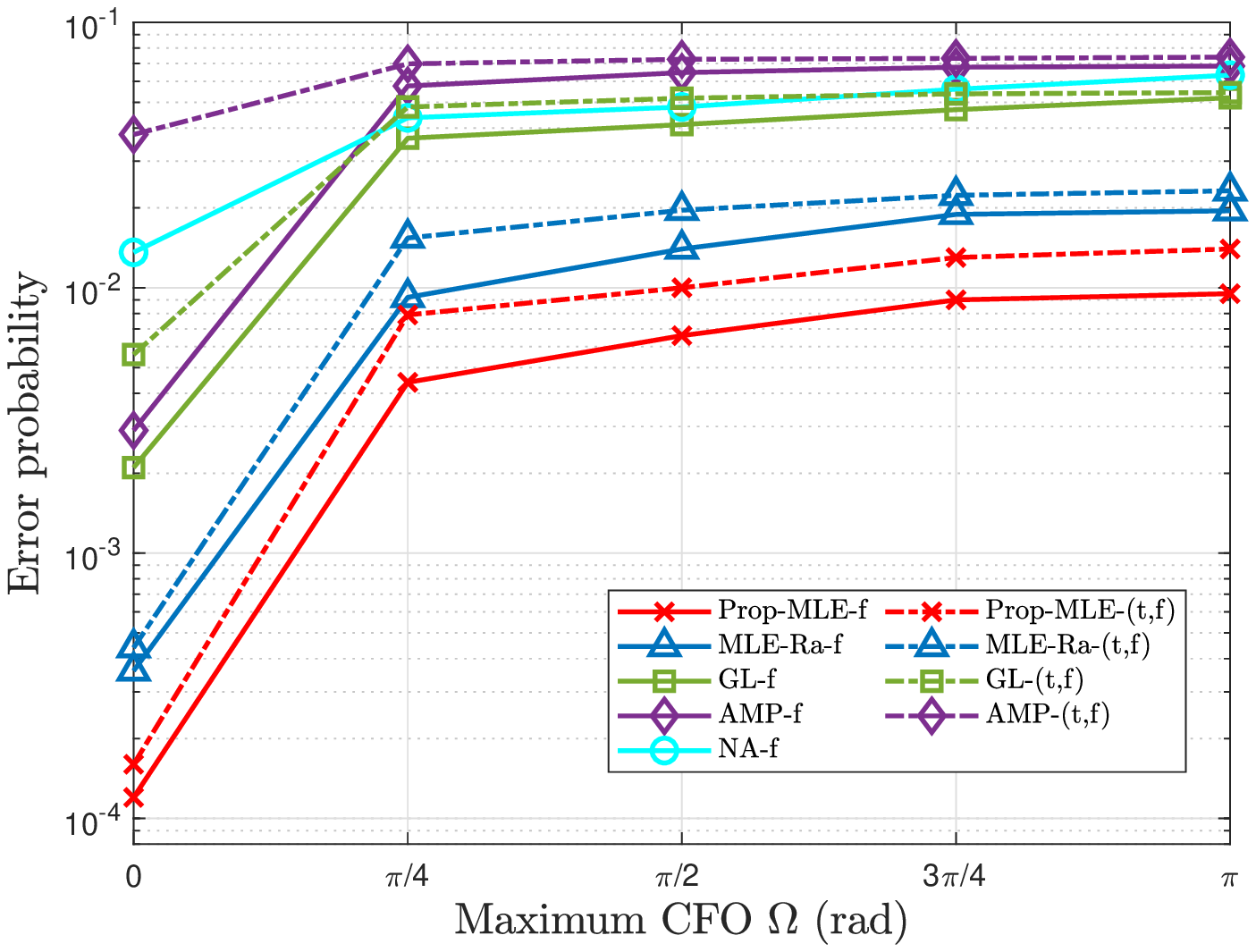}}}
  	\end{center}
  	\vspace{-5mm}
  	\caption{\small{Error probability versus $D$ and $\Omega$.}}
  	\vspace{-5 mm}
  	\label{Figure:3}
  	
  \end{figure}

  \begin{figure}[t]
  	\begin{center}
  			\subfigcapskip = \subcap pt
  		\subfigure[\scriptsize{$D$.}]
  		{\resizebox{\fsize cm}{!}{\includegraphics{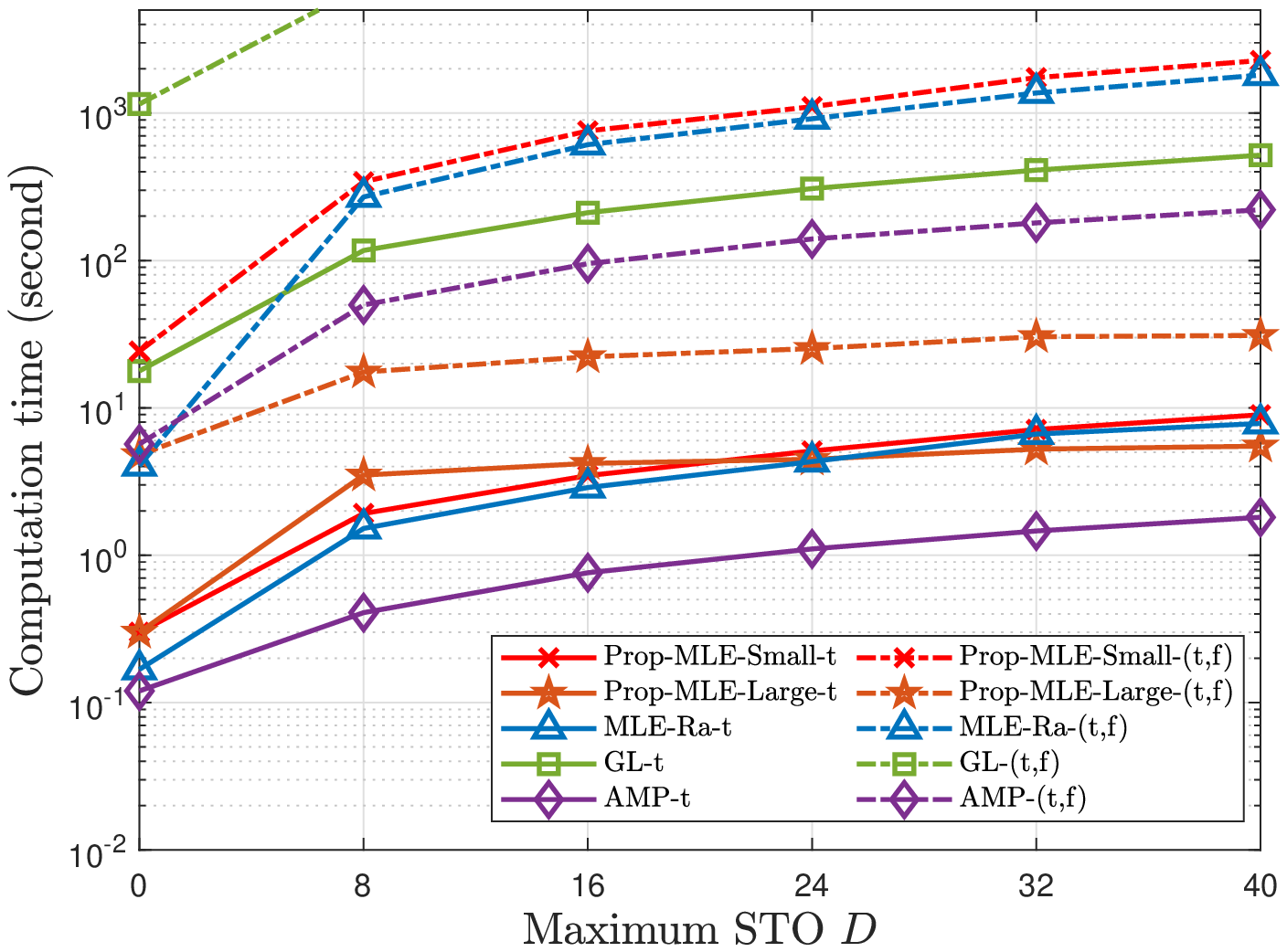}}} 	    
  		\subfigure[\scriptsize{$\Omega$.}]
  		{\resizebox{\fsize cm}{!}{\includegraphics{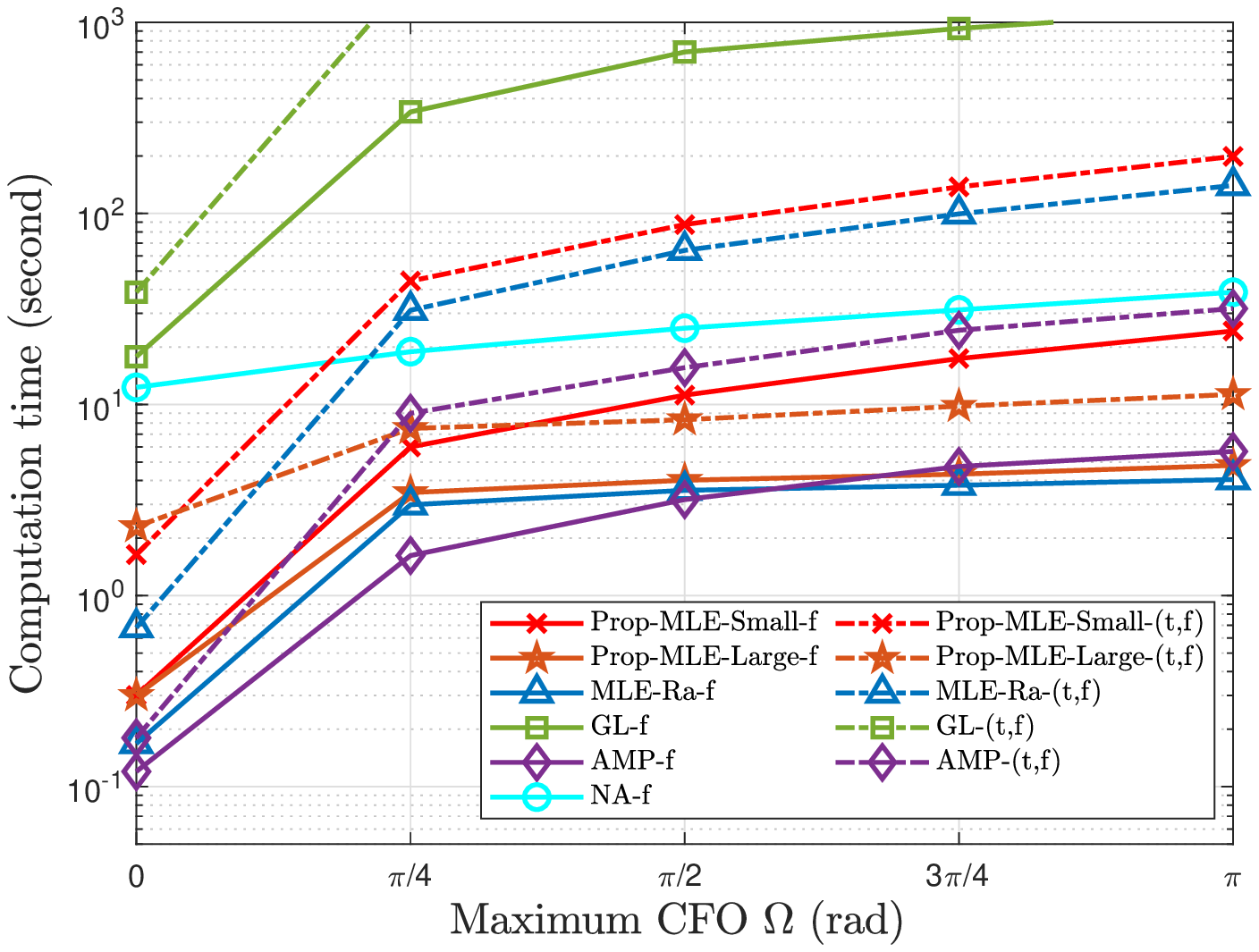}}}

  	\end{center}
  	\vspace{-5mm}
  	 \caption{\small{Computation time versus $D$ and $\Omega$.}}
  	\vspace{\figcapsize mm}
  	\label{Figure:4}
  	
  \end{figure}

\begin{figure}[t]            
	\begin{center}
		
		\subfigcapskip =  \subcap  pt
		\subfigure[\scriptsize{Error probability versus $Q$.}]
	{\resizebox{\fsize cm}{!}{\includegraphics{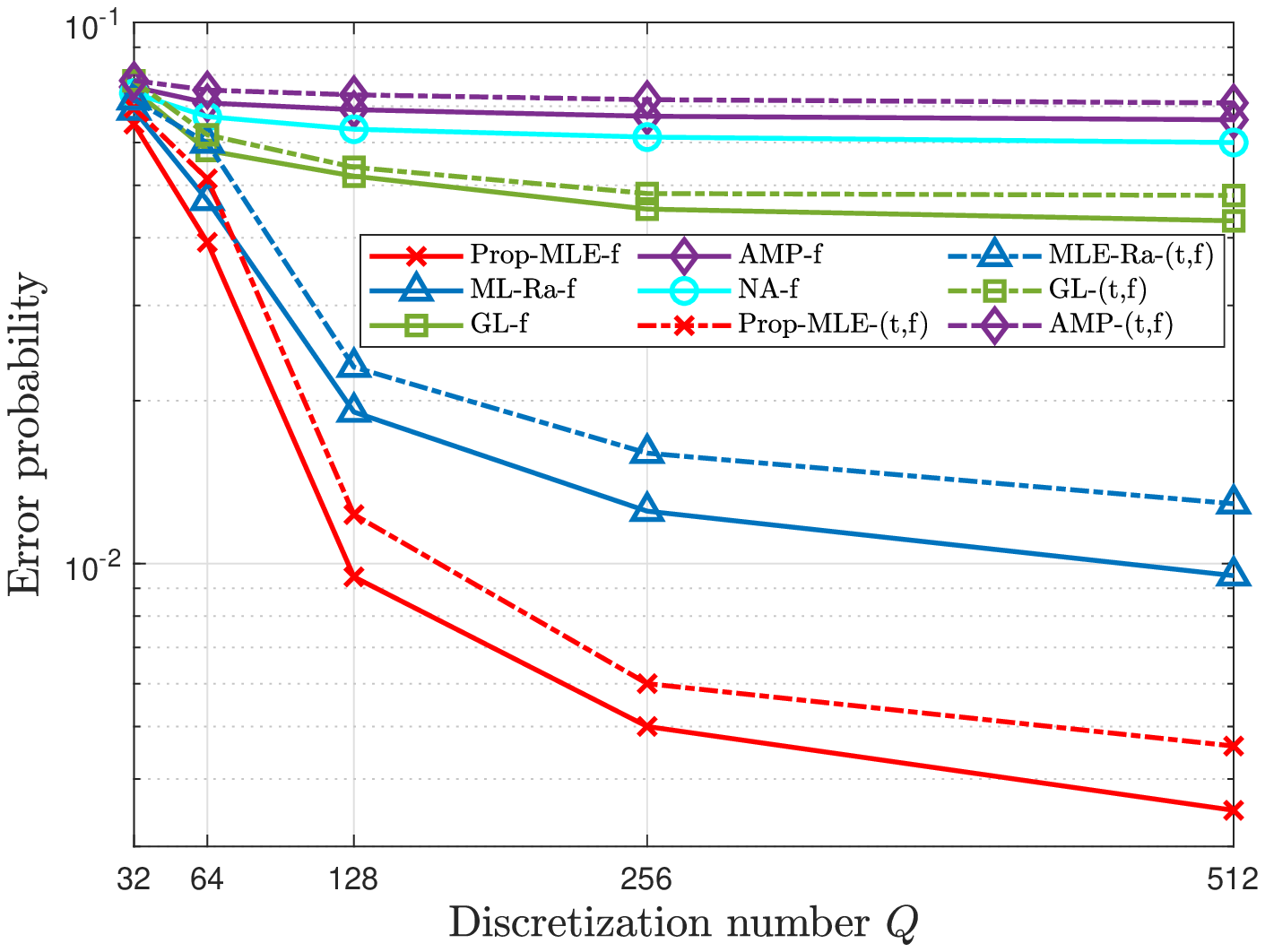}}}  
	\subfigure[\scriptsize{Computation time versus $Q$.}]
{\resizebox{\fsize cm}{!}{\includegraphics{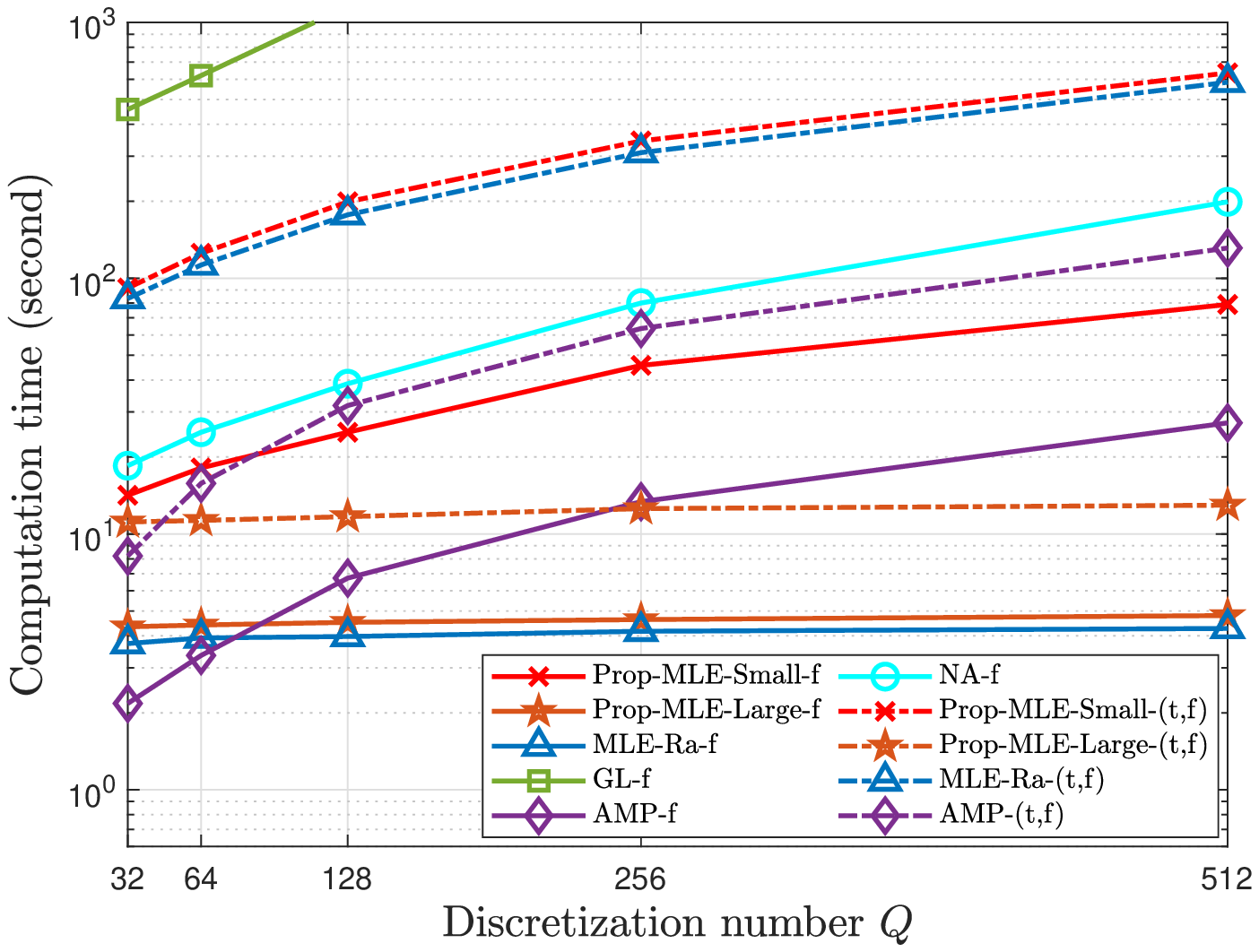}}}
\end{center}
\vspace{-5mm}
\caption{\small{Error probability and computation time versus $Q$.}}
\vspace{ \figcapsize  mm}
\label{R3Q6}
\end{figure}

  Fig.~\ref{Figure:3} plots the error probability versus the maximum STO $D$ and the maximum CFO $\Omega$.\footnote{For $i\in$ \{t, f\}, we run Prop-MLE-Syn, MLE-Ra-Syn, GL-Syn, and AMP-Syn at $D=0$ or $\Omega=0$;
  		for $i=$ (t,f), we run Prop-MLE-f, MLE-Ra-f, GL-f, and AMP-f at $D=0$, and Prop-MLE-t, MLE-Ra-t, GL-t, and AMP-t at $\Omega=0$.\label{foot1}}
The length of measurement vectors $\mathbf Y_{i,:,m}$ is $L_i = L+ \mathbb I(i\neq\text{f}) D  $, for all $i\in$ \{t, f, (t,f)\} and $m\in\mathcal M$, and the number of possible STOs in $\hat {\mathcal X}_{\text{t}}$ and the number of possible CFOs in $\hat {\mathcal X}_{\text{f}}$ are $D+1$ and $ 2  \lfloor \frac{Q\Omega}{2\pi}  \rfloor+\mathbb{I}(\Omega\neq\pi )$, respectively. 
	When $D$ increases, $L_i$ and $\vert \hat {\mathcal X}_{\text{t}}\vert $ increase; when $\Omega$ increases, $L_i$ does not change, and $\vert \hat {\mathcal X}_{\text{f}}\vert $ increases. In general, detection accuracy increases with the number of measurements but decreases with the number of candidates for an unknown parameter to be estimated.
From Fig.~\ref{Figure:3}~(a), we can observe that
for $i\in$ \{t, (t,f)\},
the error probabilities of Prop-MLE-$i$ and MLE-Ra-$i$ (GL-$i$ and AMP-$i$) increase with $D$ when $D\in[0,2]$ ($D\in$ [0,8]), as the impact of the increment in the number of candidates for STOs on detection performance dominants that of the increment in the number of measurements in this regime;
the error probabilities of Prop-MLE-$i$ and MLE-Ra-$i$ decrease with $D$ when $D\in[2,8]$, as the impact of the increment in the number of measurements on detection performance dominants that of the increment in the number of candidates for STOs in this regime.
Besides, from Fig.~\ref{Figure:3}~(b), we can observe that the error probabilities of Prop-MLE-$i$, MLE-Ra-$i$, GL-$i$, AMP-$i$, $i\in$ \{f, (t,f)\}, and NA-f increase with $\Omega$, due to the increment in the number of candidates for CFOs.
  
  We can make the following observations from Fig.~\ref{Figure:1}, Fig.~\ref{Figure:2}, and Fig.~\ref{Figure:3}.
Firstly, in each case,
the proposed method (on average) significantly outperforms all the baseline methods, revealing the proposed method's superiority.
 Specifically, 
 \mcele{Prop-MLE-$i$ reduces the error probability by up to $50.4\%$, $  65.8\%$, $  51.3\%$, and $ 39.6\% $, compared
 to MLE-Ra-$i$ (the MLE-based state-of-the-art) for $i = $ Syn, t, f, and (t,f), respectively, at $\kappa=0.1$.}
The gain of Prop-MLE-$i$ over MLE-Ra-$i$ is because Prop-MLE-$i$ uses the exact channel statistics;
and the gain of Prop-MLE-Syn over MLE-Ri-Syn comes from the fact that Prop-MLE-Syn deals with the original MLE problem, while MLE-Ri-Syn solves an approximate version of the MLE problem.
Secondly, for $i \in $ \{Syn, t, f, (t,f)\}, 
the gain of Prop-MLE-$i$ over MLE-Ra-$i$ increases with $\kappa$, as the error of approximating Rician fading with Rayleigh fading increases with $\kappa$.
\mcele{Thirdly, for $\mu \in$ \{Prop-MLE, AMP, MLE-Ra, LASSO\}, $\mu$-(t,f) underperforms $\mu$-t and $\mu$-f, as the
device activity detection problem in asynchronous case-(t,f) is more challenging than those in asynchronous case-t and asynchronous case-f (due to more unknowns to estimate).}


  Fig.~\ref{Figure:4} plots the computation time versus the maximum STO $D$ and the maximum CFO $\Omega$.\footnote{For $i\in$ \{t, f\}, we run Prop-MLE-Syn, MLE-Ra-Syn, GL-Syn, and AMP-Syn at $D=0$ or $\Omega=0$;
  	for $i=$ (t,f), we run Prop-MLE-Small-f, Prop-MLE-Large-f, MLE-Ra-f, GL-f, and AMP-f at $D=0$, and Prop-MLE-Small-t, Prop-MLE-Large-t, MLE-Ra-t, GL-t, and AMP-t at $\Omega=0$.}
We make the following observations from Fig.~\ref{Figure:4}.
Firstly,
	the computation time of Prop-MLE-Large-$i$ is comparable to that of to AMP-$i$ for all $i\in$ \{Syn, t, f, (t,f)\} and is even (on average) shorter than that of AMP-$i$ for $i\in$ \{f, (t,f)\} and large $D$ and $\Omega$.
This result is appealing as AMP is known to be a highly efficient algorithm. 
Secondly,
 \mcele{Prop-MLE-Large-$i$ has \mcone{a} shorter computation time than Prop-MLE-Small-$i$ for $i=$ t, f, and (t,f) when $D>22$, $\Omega >0.33\pi$, and $D>0$ or $\Omega >0.05 \pi$, respectively, which is \mcele{roughly} in accordance with Lemma~\ref{Lem:FCC}.\footnote{\mcele{In the simulation setup where $L=60$, $M=48$, $Q =128$, $D=4$, and $\Omega =\pi$, 
 	 we have $\underline D_{\text{t}}=29.5$,
 		$\underline \Omega_{\text{f}} = 0.27\pi$, $\underline D_{\text{(t,f)}} = -0.695$, and $\underline \Omega_{\text{(t,f)}} = 0.06 \pi$.}}}
\mcele{Thirdly, compared to MLE-Ra-$i$, Prop-MLE-Large-$i$ can reduce the computation time by up to $ 29.8\%$ and $  92.0\%$ for $i = $ t and (t,f), respectively. \mcele{The \mctht{gains come} from the fact that} the computational complexities of MLE-Ra-$i$ significantly increase with $D$ or/and $\Omega$, whereas the computational complexity of Prop-MLE-Large-$i$ does not change with $D$ or/and $\Omega$, as shown in Table\ref{tab_BL}}.
\mcele{Fourthly, when $D=0$, the computation time of Prop-MLE-Small-t (reduces to Prop-MLE-Syn) is $78.6\%$ longer than that of MLE-Ra-t (reduces to MLE-Ra-Syn), as MLE-Ra-t considers Rayleigh fading as a substitute
	for simplicity.}

Fig.~\ref{R3Q6} shows the error probability and computation time versus $Q$. 
	From Fig.~\ref{R3Q6}, we can see that Prop-MLE-Large-(t,f) (Prop-MLE-Large-f) achieves on average the best detection accuracy with on average the lowest (second lowest) computation time when $Q$ is large, demonstrating the high efficiency of Prop-MLE-Large-$i$, $i\in$ \{f, (t,f)\}. Specifically,
	from Fig.~\ref{R3Q6} (a), we can see that the error probability of each algorithm decreases with $Q$, due to the decrement in the approximation error with $Q$; from Fig.~\ref{R3Q6}~(b), we can see that the computation time of each algorithm increases with $Q$, due to the increment in the number of CFO's candidates.

 In summary, combining the results on detection performance and computational time, we can conclude that the proposed solutions achieve the smallest detection error probabilities with (on average) the shortest computation times in some cases (i.e., asynchronous case-f and asynchronous case-(t,f) at large $D$ and $\Omega$) and relatively short computation times in other cases (i.e., the synchronous case and three asynchronous cases at small $D$ and $\Omega$).

 \vspace{-5mm}

\section{Conclusion}
This paper systematically investigated MLE-based device activity detection under Rician fading for massive grant-free access with perfect and imperfect synchronization.
   The proposed algorithms successfully generalized the existing MLE-based methods for Rayleigh fading and perfect synchronization.
     	Numerical results demonstrated the superiority of the proposed algorithms in detection accuracy and computation time and the importance of explicit consideration of LoS components and synchronization errors in massive grant-free access.
To our knowledge, this is the first work that systematically utilizes FFT and IFFT techniques to accelerate device activity detection algorithms for massive grant-free access.
Besides, this is the first work comprehensively investigating MLE-based device activity detection under Rician fading for massive grant-free access.

\mcele{There are still some key aspects that we leave for future
	investigations. One direction is to study
	MAPE-based device activity detection methods
	under flat Rician fading for massive grant-free with imperfect synchronization by utilizing
	prior information on device activities.	
	Another interesting direction is to investigate MLE and MAPE-based device activity detection under frequency-selective Rayleigh and Rician fading for OFDM-based massive grant-free with imperfect synchronization.}

\vspace{-5mm}

 \section*{Appendix A: Proof for Theorem~\ref{Tem:OptpointBCD_Ri_syn}}

First, consider the coordinate optimization with respect to $a_n$ in (\ref{Prob:cd_Ri_syn}).
By (\ref{LikelySy_Ri_syn}), we have:  
{\small\begin{align}
		& 	\begin{array}{ll}
			\quad f   ( \mathbf a_{-n} , a_n +d_n )	
			\overset{(a)} = \log \left|  \mathbf\Sigma +  \frac{d_ng_n}{1+ \kappa_n} {\mathbf p}_{n}  {\mathbf p}_{n } ^H \right| 
		\end{array} \nonumber \\ 
		&  \begin{array}{ll}
			\quad +     \frac{1}{M}\mathrm{tr} \left(     \Big( \mathbf\Sigma  +   \frac{d_ng_n}{1+ \kappa_n} {\mathbf p}_{n}  {\mathbf p}_{n } ^H  \Big)^{-1} 	\Big(\widetilde {\mathbf Y} -d_n\sqrt{\frac{g_n\kappa_n}{1+ \kappa_n}} {\mathbf p}_{n} \overline{ \mathbf h}_{n}^T \Big) 	\Big(\widetilde {\mathbf Y} -d_n\sqrt{\frac{g_n\kappa_n}{1+ \kappa_n}} {\mathbf p}_{n} \overline{ \mathbf h}_{n}^T \Big)^H    \right)   
		\end{array}
		\nonumber \\
		& \begin{array}{ll}
			& \overset{(b)}= \log \Big( | \mathbf\Sigma | (1+\frac{d_ng_n}{1+ \kappa_n} { \bf p}_n^H
			{\bf\Sigma} ^{-1}     { \bf p}_n )\Big)   \nonumber \\
			&
			+ \frac{1}{M}\mathrm{tr} \Big(   \Big(\mathbf\Sigma^{-1}- \frac{ \frac{d_ng_n}{1+ \kappa_n}\mathbf\Sigma^{-1} {\mathbf p}_{n} {\mathbf p}_{n}^H\mathbf\Sigma^{-1} }
			{1+\frac{d_n g_n}{1+ \kappa_n} { \bf p}_n^H
				{\bf\Sigma} ^{-1}     { \bf p}_n }   \Big) \Big(\widetilde {\mathbf Y} -d_n\sqrt{\frac{g_n\kappa_n}{1+ \kappa_n}} {\mathbf p}_{n} \overline{ \mathbf h}_{n}^T \Big) 	\Big(\widetilde {\mathbf Y} -d_n\sqrt{\frac{g_n\kappa_n}{1+ \kappa_n}} {\mathbf p}_{n} \overline{ \mathbf h}_{n}^T \Big)^H    \Big)    \nonumber \\
			&  \mcfive{
				=  	f   ( \mathbf a  )  
				+\log \Big(   1+\frac{dg_n}{1+ \kappa_n} { \bf p}_n^H
				{\bf\Sigma} ^{-1}     { \bf p}_n  \Big)  
				+ 
				\frac{ d_n^2 g_n\kappa_n}{1+ \kappa_n} { \bf p}_n^H{\bf\Sigma} ^{-1}   { \bf p}_n   
				-
				\frac{d_n }{M}		\sqrt{\frac{g_n\kappa_n}{1+ \kappa_n}}
				\left(\mathrm{tr} \left(	{\bf\Sigma}^{-1}    {\mathbf p}_{n} \overline{ \mathbf h}_{n}^T  \widetilde {\mathbf Y }^H   \right)   +   \mathrm{tr} \left(	{\bf\Sigma}^{-1}   \widetilde {\mathbf Y }  \overline{ \mathbf h}_{n}^*  {\mathbf p}^H_{n}    \right)   \right) 
			} \nonumber \\
			&\quad \mcfive{
				+\frac{				   
					\frac{ d_n^2}{M}		 (\frac{g_n\kappa_n}{1+ \kappa_n})^{\frac{3}{2}}
					\left( \mathrm{tr} \left(     {\bf\Sigma}^{-1} {\mathbf p}_{n} {\mathbf p}_{n}^H{\bf\Sigma}^{-1}
					{\mathbf p}_{n} \overline{ \mathbf h}_{n}^T  \widetilde {\mathbf Y }^H   \right) 
					+  
					\mathrm{tr} \left(     {\bf\Sigma}^{-1}{\mathbf p}_{n} {\mathbf p}_{n}^H{\bf\Sigma}^{-1}
					\widetilde {\mathbf Y }    \overline{ \mathbf h}_{n}^*  {\mathbf p}^H_{n}       \right)   \right) 
				}
				{1+\frac{d_ng_n}{1+ \kappa_n} { \bf p}_n^H
					{\bf\Sigma} ^{-1}     { \bf p}_n }                      
			}\nonumber \\
			& \quad   \mcfive{ - \! \frac{\frac{d_ng_n}{ M(1+ \kappa_n)} { \bf p}_n^H {\bf\Sigma}  ^{-1}   \widetilde {\mathbf Y}    \widetilde {\mathbf Y}^H  	 {\bf\Sigma} ^{-1}   { \bf p}_n    
					+   \frac{1}{M(1+\kappa_n)^2  } d_n^3g_n^2 \kappa_n\mathrm{tr}  \left(     {\bf\Sigma}^{-1} {\mathbf p}_{n} {\mathbf p}_{n}^H{\bf\Sigma}^{-1} 
					{\mathbf p}_{n} \overline{ \mathbf h}_{n}^T \overline{ \mathbf h}_{n}^*  {\mathbf p}_{n}^H         \right) }{1+\frac{d_ng_n}{1+ \kappa_n} { \bf p}_n^H
					{\bf\Sigma} ^{-1}     { \bf p}_n} 
			} \\
			&   \overset{(c)}= 
			f   ( \mathbf a  )  +\log \Big(   1+\frac{ g_n}{1+ \kappa_n} { \bf p}_n^H
			{\bf\Sigma} ^{-1}     { \bf p}_n  d_n\Big)  
			+
			\frac{d_n  
			}
			{1+\frac{ g_n}{1+ \kappa_n} { \bf p}_n^H
				{\bf\Sigma} ^{-1}     { \bf p}_nd_n}  \times    \Big(\frac{ g_n\kappa_n}{1+ \kappa_n} { \bf p}_n^H{\bf\Sigma} ^{-1}     { \bf p}_n d_n \nonumber \\
			&  
			-  \frac{ g_n}{ M(1+ \kappa_n)} { \bf p}_n^H {\bf\Sigma}  ^{-1}   \widetilde {\mathbf Y}   \widetilde {\mathbf Y}^H  	 {\bf\Sigma} ^{-1} { \bf p}_n  
			- \frac{ 2 }{M}		\sqrt{\frac{g_n\kappa_n}{1+ \kappa_n}}
			\mathrm{Re} \left(
			\overline {\bf h}_n^T \widetilde {\mathbf Y}^H        {\bf\Sigma} ^{-1}   { \bf p}_n   \right)   \Big)
		\end{array} 
		\nonumber \\
		&  \!\quad  \begin{array}{ll}  
			\overset{(d)} =  f   ( \mathbf a  ) +  \log( 1+ \alpha_n d_n ) + \frac{  \kappa_n\alpha_n d_n^2 - (\beta_n + \eta_n )d_n     }{1+ \alpha_n d_n},
		\end{array}\label{Appendix_A1}
\end{align}}%
where (a) is due to ${\bf\Sigma} = \mathbf P \mathbf A  \mathbf \Gamma      \mathbf P ^H   + \sigma^2{\bf I}_{L }$
and 
$
\widetilde {\mathbf Y}  =  \mathbf Y -  {\mathbf P} {\bf A} {\bf \Gamma}^{\frac{1}{2}} \overline{\bf H}^T  
$,
(b) is due to the fact that for any positive definite matrix
$ \mathbf\Sigma $, $| \mathbf\Sigma +  \frac{d_ng_n}{1+ \kappa_n} {\mathbf p}_{n}  {\mathbf p}_{n } ^H |= | \mathbf\Sigma| (1+\frac{d_ng_n}{1+ \kappa_n} { \bf p}_n^H
{\bf\Sigma} ^{-1}     { \bf p}_n ) $
and
$
\left(\mathbf\Sigma +  \frac{d_ng_n}{1+ \kappa_n} {\mathbf p}_{n}  {\mathbf p}_{n } ^H \right)^{-1} = \mathbf\Sigma^{-1}- \frac{ \frac{d_ng_n}{1+ \kappa_n}\mathbf\Sigma^{-1} {\mathbf p}_{n} {\mathbf p}_{n}^H\mathbf\Sigma^{-1} }
{1+\frac{d_ng_n}{1+ \kappa_n} { \bf p}_n^H
{\bf\Sigma} ^{-1}     { \bf p}_n }  
$
hold~\cite{TIT_GC_2021},
(c) is due to
\mcfive{\begin{align}
	&	 \frac{1}{M} \mathrm{tr}\Big(	{\bf\Sigma}^{-1}     
	{\mathbf p}_{n} \overline{ \mathbf h}_{n}^T \overline{ \mathbf h}_{n}^*  {\mathbf p}_{n}^H      \Big) =   { \bf p}_n^H
	{\bf\Sigma}^{-1}      { \bf p}_n     ,   \nonumber  \\
	&	   \mathrm{tr} \Big(     {\bf\Sigma}^{-1}   {\mathbf p}_{n} {\mathbf p}_{n}^H{\bf\Sigma}^{-1} 
	{\mathbf p}_{n} \overline{ \mathbf h}_{n}^T  \widetilde {\mathbf Y }^H      \Big)   + \mathrm{tr} \Big(     {\bf\Sigma}^{-1}   {\mathbf p}_{n} {\mathbf p}_{n}^H { \bf\Sigma}^{-1}
	\widetilde {\mathbf Y }     \overline{ \mathbf h}_{n}^*  {\mathbf p}^H_{n}       \Big)   = 2 \left(  { \bf p}_n^H
	{\bf\Sigma}^{-1}    { \bf p}_n \right)    \left(   \mathrm{Re} \Big(
	\overline {\bf h}_n^T \widetilde {\mathbf Y}^H     {\bf\Sigma} ^{-1}    { \bf p}_n   \Big)\right)  ,  \nonumber  \\
	& \frac{1}{M   }\mathrm{tr} \Big(     {\bf\Sigma}^{-1}  {\mathbf p}_{n} {\mathbf p}_{n}^H  {\bf\Sigma}^{-1} 
	{\mathbf p}_{n} \overline{ \mathbf h}_{n}^T \overline{ \mathbf h}_{n}^*  {\mathbf p}_{n}^H         \Big) =    \left(  { \bf p}_n^H
	{\bf\Sigma}^{-1}     { \bf p}_n \right)^2,  \nonumber 
	\end{align}}%
	and
	(d) is due to~(\ref{alpha_syn}), ~(\ref{beta_syn}), and~(\ref{eta_syn}). 
	By (\ref{Appendix_A1}), we have:
	{\small\begin{align}
	&\frac{ \partial f   ( \mathbf a_{-n} , a_n +d_n ) }{\partial d_n}  =\frac{   \kappa_n \alpha_n^2 d_n^2
		+\left(\alpha_n^2 +  2 \kappa_n \alpha_n\right)d_n
		+ \alpha_n -\beta_n-\eta_n }{ (1+\alpha_n d_n)^2} . \nonumber
		\end{align}}%
		\mcfive{Note that the domain of $ f   ( \mathbf a_{-n} , a_n +d_n )$ with respect to $d_n $ is $(-\frac{1 }
{ \alpha_n  }, +\infty)$.
In the following, we obtain the optimal solution of problem in~(\ref{Prob:cd_Ri_syn}) by analyzing the 
monotonicity of $ f   ( \mathbf a_{-n} , a_n +d_n ) $ in the domain $(-\frac{1 }
{ \alpha_n  }, +\infty)\cap [-a_n, 1-a_n]$.
Since solving $   \frac{ \partial f   ( \mathbf a_{-n} , a_n +d_n ) }{\partial d_n} =0 $ is equivalent to solving $ q(d_n) \triangleq \kappa_n \alpha_n^2 d_n^2
+\left(\alpha_n^2 +  2 \kappa_n \alpha_n\right)d_n
+ \alpha_n -\beta_n-\eta_n = 0$, we analyze the
monotonicity of $ f   ( \mathbf a_{-n} , a_n +d_n )  $ by analyzing the roots of $ q(d_n)=0 $ in two cases.
If $\alpha_{n }^2+  4 \kappa_n  (\kappa_n + \beta_n + \eta_n) \leq 0$, then $ q(d_n) =0$ has at most one root (as $ \kappa_n \alpha_n^2 >0$). 
Hence, $ q(d_n) \geq 0$ for all $d_n \in (-\frac{1 }
{ \alpha_n  },  + \infty)$, implying that $ f   ( \mathbf a_{-n} , a_n +d_n )$ increases with $d_n$ when $d_n \in(-\frac{1 }
{ \alpha_n  },  + \infty)$.
Combining with the constraint $d_n \in [-a_n, 1-a_n]$, we have the
optimal solution given in~(\ref{Opt_bc_a_Ri_syn}).
If $\alpha_{n }^2+  4 \kappa_n  (\kappa_n + \beta_n + \eta_n) > 0$, then 
$q (d_n)= 0$ has two roots: one is $\hat d_n$ given by~(\ref{d_n0}), and the other one is $ \frac{-\alpha_n - 2\kappa_n - \sqrt{  \alpha_{n }^2+  4 \kappa_n  (\kappa_n + \beta_n + \eta_n) }}{2 \kappa_n \alpha_n } < -\frac{1 }
{ \alpha_n  }$. 
If $\hat d_n \leq -\frac{1 }
{ \alpha_n  } $, then $ q(d_n) \geq 0$ for all $d_n \in (-\frac{1 }
{ \alpha_n  },  + \infty)$, implying that $ f   ( \mathbf a_{-n} , a_n +d_n )$ increases with $d_n$ when $ d_n\in(-\frac{1 } { \alpha_n  },  + \infty)$;
If $\hat d_n >-\frac{1 }{ \alpha_n  } $, then $ q(d_n)< 0$ for all $d_n \in (-\frac{1 }
{ \alpha_n  },  \hat d_n)$ and $ q(d_n)\geq  0$ for all $d_n \in [ \hat d_n,+ \infty) $, implying that
$f   ( \mathbf a_{-n} , a_n +d_n )$ decreases with $d_n$ when $d_n \in (-\frac{1 }
{ \alpha_n  }, \hat d_n)$, increases with $d_n$ when $d_n \in( \hat d_n, + \infty)$, and achieves its minimum at $d_n =\hat d_n$.
Combining with the constraint $d_n \in [-a_n, 1-a_n]$, we have the
optimal solution given in~(\ref{Opt_bc_a_Ri_syn}).}

  	\vspace{ -4mm}
 
 \section*{Appendix B: Proof for Theorem~\ref{Thm_SP_Ri_syn}}
 
 	
 	Note that
 	 $f(\mathbf a) $ is continuously differentiable with respect to $\mathbf a$, and the constraints for $\mathbf a$ are convex.
 According to the proof of Theorem~\ref{Tem:OptpointBCD_Ri_syn}, each optimal coordinate is uniquely obtained, and $f(\mathbf a_{-n}, a_n)$ is monotonically nonincreasing in the interval from $a_n$ to $a_n+d_n^*$.
 Thus, the assumptions of~\cite[Proposition 3.7.1]{Bertsekas1998NP} are satisfied.
 Therefore, we can show Theorem~\ref{Thm_SP_Ri_syn} by~\cite[Proposition 3.7.1]{Bertsekas1998NP}.
 
  \vspace{ -4mm}
\section*{Appendix C: Proof for Theorem~\ref{Tem:OptpointBCD_app}}
  \vspace{ -2mm}
Note that $\min_{ a
	\in[0,1], x  \in \hat{\mathcal X}_{i} }		 	  f_i  ( \mathbf a_{-n} , a , \mathbf x_{i,-n} , x ) =\min_{ x  \in \hat{\mathcal X}_{i} }\min_{ a
	\in[0,1]} 		 	  f_i  ( \mathbf a_{-n} , a , \mathbf x_{i,-n} , x )  $.
First, we solve $\min_{ a
	\in[0,1]} 		  f_i  ( \mathbf a_{-n} , a , \mathbf x_{i,-n} , x )$ for any given $x\in \hat{\mathcal X}_{i}$.
By~(\ref{LikelySy_Ri_asyn}), we have: 
	{\small \begin{align}
			&	 f_i  ( \mathbf a_{-n} , a , \mathbf x_{i,-n} , x  )      
			\overset{(a)}   =     \log \left|  \mathbf\Sigma_{i,n} +  \frac{a  g_n}{1+ \kappa_n}  { \bf p}_{i,n}(x)  { \bf p}^H_{i,n}(x) \right|  
			+     \frac{1}{M}\mathrm{tr} \Big(     \Big(  \mathbf\Sigma_{i,n} +  \frac{a  g_n}{1+ \kappa_n}  { \bf p}_{i,n}(x)  { \bf p}^H_{i,n}(x)  \Big)^{-1}  
			\nonumber \\ 
			& \times	\Big(\widetilde {\mathbf Y}_{i,n} -a \sqrt{\frac{g_n\kappa_n}{1+ \kappa_n}} \mathbf p_{i,n}(x)\overline{ \mathbf h}_{n}^T \Big) 	\Big(\widetilde {\mathbf Y}_{i,n} - a  \sqrt{\frac{g_n\kappa_n}{1+ \kappa_n}} \mathbf p_{i,n}(x) \overline{ \mathbf h}_{n}^T \Big)^H    \Big)
			\nonumber \\
			&	\overset{(b)} =  f_i  ( \mathbf a_{-n} ,0, \mathbf x_{i,-n} , 0  )     +  \log( 1+ \alpha_{i,n}(x) a  ) + \frac{  \kappa_n\alpha_{i,n}(x) a ^2 - (\beta_{i,n}(x) + \eta_{i,n}(x) )a     }{1+ \alpha_{i,n}(x) a }, \label{fi(a,d)2}	 
	\end{align}}%
	where (a) is due to ${\bf\Sigma}_{i}   = \mathbf\Sigma_{i,n}  +  \frac{a  g_n}{1+ \kappa_n}  { \bf p}_{i,n}(x)  { \bf p}^H_{i,n}(x)$
	and 
	$
	\widetilde {\mathbf Y}_i   =\widetilde{ \mathbf Y}_{i,n} - a \sqrt{\frac{g_n\kappa_n}{1+ \kappa_n}} \mathbf p_{i,n}(x)\overline{ \mathbf h}_{n}^T 
	$, 
	and (b) follows from~(\ref{Appendix_A1}) (by regarding $ { \mathbf p}_{i,n}(x)$, $\mathbf\Sigma_{i,n}$, and $\overline {\mathbf Y}_{i,n} $ in~(\ref{fi(a,d)2}) as ${ \mathbf p}_{n}$, $\mathbf\Sigma$, and $\overline {\mathbf Y} $ in~(\ref{Appendix_A1}), respectively).
	Thus, following the proof for Theorem~\ref{Tem:OptpointBCD_Ri_syn}, we can readily show
	that at any given $x \in  \hat{ \mathcal X}_i$, the optimal solution and optimal value of $	\min_{ a  \in[0,1]  }	   f_i  ( \mathbf a_{-n} ,  a , \mathbf x_{i,-n} , x  )   $ are $ d_{i,n}(x)$ in~(\ref{d_n0_asyn}) and 
	$f_i  ( \mathbf a_{-n} ,  d_{i,n}(x), \mathbf x_{i,-n} , x  )  = f_i  ( \mathbf a_{-n} ,0, \mathbf x_{i,-n} , 0  ) + h_{i,n}(x) $, respectively, where $ h_{i,n}(x)$ is given in~(\ref{h_i_asyn}).
	It remains to solve $\min_{x \in  \hat{ \mathcal X}_i} $ $f_i  ( \mathbf a_{-n} ,  d_{i,n}(x), \mathbf x_{i,-n} , x  )$, which is equivalent to $\min_{x \in  \hat{ \mathcal X}_i} h_{i,n}(x)$.
	Therefore, we have the optimal solution given in~(\ref{Opt_time_t_Ri_asyn}). 
 
   	\vspace{ -3mm}

\section*{Appendix D: Proof for Lemma~\ref{Lem:Compute_t}}
  	\vspace{ -2mm}
For all $\omega \in \mathbb R$, $\mathbf p \in \mathbb C^{K}$, and Hermitian matrix $\mathbf B \in \mathbb C^{K\times K}$, let $J(\omega, \mathbf p, \mathbf B)\triangleq \mathbf p^H 	\mathrm{diag}(( e^{j(k -1) \omega})_{k \in \mathcal K}  )^H   	$ $ \mathbf B  	\mathrm{diag}(( e^{j(k -1) \omega})_{k \in \mathcal K})\mathbf p $. 
Recall that $\boldsymbol\Psi(\mathbf B) = (\boldsymbol\psi_{k} (\mathbf B) )_{k\in\mathcal K}  \in \mathbb C^{K\times K}$,
where
$\boldsymbol\psi_{k} (\mathbf B)\in \mathbb C^{K}$ is given in~(\ref{k_column}).
To prove Lemma~\ref{Lem:Compute_t}, we need the following matrix identities:
\begin{align}
	&		\mathrm{diag}( \mathbf u) \mathbf v =	\mathrm{diag}(\mathbf v )\mathbf u = \mathbf u \odot \mathbf v= \mathbf v \odot \mathbf u , \   \mathbf u , \mathbf v  \in \mathbb C^K,   \label{PEQ_odot}\\
	&			\mathrm{diag}( \mathbf u) \mathbf B \mathrm{diag}( \mathbf v) = \mathbf B \odot (\mathbf u \mathbf v^T), \  \   \mathbf u , \mathbf v  \in \mathbb C^K,  \label{PEQ_dBb}\\
	& (\mathbf w_q^T ( \mathbf u_{t,k}^T \mathbf v_{k}   )_{k \in \mathcal K}  )_{t \in \mathcal K,q\in \mathcal K} =  ( ((\mathbf u_{t,k})_{t\in\mathcal K})^T\mathbf v_{k}  )_{k\in\mathcal K} (\mathbf w_q)_{q\in\mathcal K}, \  \mathbf w_q, \mathbf u_{t,k}, \mathbf v_{k} \in \mathbb C^K, t,q,k\in\mathcal K,\label{PEQ_trans}  \\
	&  (\mathbf B (\mathbf u_k \odot \mathbf v_k) )_{k\in\mathcal K} = \mathbf B \left((\mathbf u_k)_{k\in\mathcal K}\odot (\mathbf v_k)_{k\in\mathcal K} \right), \  \ \mathbf u_k, \mathbf v_k \in \mathbb C^K, k\in\mathcal K. \label{PEQ_odottrans} 
\end{align}
The proofs for~(\ref{PEQ_odot})-(\ref{PEQ_odottrans}) are obvious and hence omitted.
\mcfive{First, we show an equality based on which we will show~(\ref{C1t_Ri_n}) and~(\ref{C2t_Ri_n}):
	\begin{align}
		& J(\omega, \mathbf p, \mathbf B) \overset{(a)}=  
		\left( ( e^{j(k -1) \omega})_{k \in \mathcal K} \right)^H 	\mathrm{diag}(\mathbf p^*  )    	\mathbf B  	\mathrm{diag}(\mathbf p  )    \left( e^{j(k -1) \omega} \right)_{k \in \mathcal K} \nonumber\\
		&	\overset{(b)}=  	\mathrm{tr} \left(  ( \mathbf B   \odot (\mathbf p ^{*}\mathbf p ^{T}) )    \left( ( e^{j(\ell -1) \omega})_{\ell \in \mathcal K} \right)      \left( ( e^{j(k -1) \omega})_{k \in \mathcal K} \right)^H  \right)  
		=  \sum\limits_{\ell=1}^{K} \sum\limits_{k=1}^{K}  e^{j(k-\ell)\omega    }   	( \mathbf B   \odot (\mathbf p ^{*}\mathbf p ^{T}) )_{\ell,k}   \nonumber 
		\\
		& \overset{(c)}= 
		2 \mathrm{Re}\Big(   e^{j0\omega} \sum\limits_{\ell=1 }^{K-0 }
		\Big( \frac{\sqrt{2}}{2}  \mathbf B    \odot \Big(\frac{\sqrt{2}}{2}  \mathbf p ^{*}\mathbf p ^{T} \Big) \Big)_{k, k } + \sum\limits_{k=1}^{K-1}  e^{jk\omega  } 
		\sum\limits_{\ell=1 }^{K-k}  ( \mathbf B     \odot (\mathbf p ^{*}\mathbf p ^{T}) )_{  \ell , \ell+ k} \Big) 
		\nonumber \\
		&
		= 	2 \mathrm{Re}\Big( \left( ( e^{j(\ell -1) \omega})_{\ell \in \mathcal K} \right)^T    
		\Big(  \boldsymbol\psi_{k} ^T(  \mathbf p^{*}   \mathbf p^T    )  \boldsymbol\psi_k(  \mathbf B) \Big)_{k \in \mathcal K}       \Big), \label{Core_EQ}
	\end{align}
	where
	(a) is due to~(\ref{PEQ_odot}), 
	(b) is due to~(\ref{PEQ_dBb}), $ x = \mathrm{tr}(x) $ for $x \in \mathbb C$, and the cyclic property of trace, and
	(c) is due to that $\mathbf B     \odot (\mathbf p ^{*}\mathbf p ^{T})$ is a Hermitian matrix (as $\mathbf B$ and $ \mathbf p ^{*}\mathbf p ^{T} $ are Hermitian matrices).}
Then, by~(\ref{Core_EQ}), we show~(\ref{C1t_Ri_n}) as follows:
 {\begin{align}
 	& \frac{1+ \kappa_n}{ 2g_n }\boldsymbol	\alpha_{\text{t},n }    \overset{(a)} = \frac{1}{2} \left( J(0, \mathbf p_{\text{t},n}(t), \mathbf \Sigma_{\text{t},n}^{-1})\right)_{t \in \mathcal D}  \nonumber \\
 	&\overset{(b)}=  \left(\mathrm{Re} \left(\mathbf 1_{L_{\text{t}}}^T  \left(     \boldsymbol\psi_{ \ell} ^T\Big(  \mathbf p_{\text{t},n}^{*}(t)\mathbf p_{\text{t},n}^{T}(t) \Big) \boldsymbol\psi_{ \ell}  (  {\mathbf\Sigma}_{ \text{t},n}^{-1})    \right)_{ \ell \in \mathcal L_{\text{t}}} \right)  \right)_{t\in \mathcal D}
 	\nonumber \\
 	&  \overset{(c)}=   \mathrm{Re}   \left( \!
 	\left(\!   \left(\!  \Big(  \boldsymbol\psi_{\ell} \Big(  \mathbf p_{\text{t},n}^{*}(t)\mathbf p_{\text{t},n}^{T}(t) \Big) \Big)_{t \in \mathcal D} \right)^T \!\!
 	\boldsymbol\psi_{ \ell}(  {\mathbf\Sigma}_{ \text{t},n}^{-1})     \right)_{\ell \in \mathcal L_{\text{t}}} \!\!\!\!\!
 	\mathbf 1_{L_{\text{t}}} \right)
 	\label{Proofkeyeq1} \\
 	&\overset{(d)}  =   \mathrm{Re}   \Big( 
 	 \Big(  \mathbf F_{L_{\text{t}},\widetilde{\mathcal D}, :}  \mathrm{diag}\Big( \mathbf F_{L_{\text{t}}} 
 	 \boldsymbol\psi_{\ell} \Big(  \mathbf p_{\text{t},n}^{*}(0)\mathbf p_{\text{t},n}^{T}(0) \Big) \Big)       \frac{1}{L_{\text{t}}}\mathbf F_{L_{\text{t}}}^H
 \boldsymbol\psi_{\ell}(  {\mathbf\Sigma}_{ \text{t},n}^{-1})
 	\Big)_{\ell \in \mathcal L_{\text{t}}}
 		\mathbf 1_{L_{\text{t}}} \Big)
 \label{Proofkeyeq2}  \\ 
 	& \overset{(e)} =   \bigg(\mathrm{Re}   \Big(  
 	   \mathbf F_{L_{\text{t}} }   \Big(    
 	\Big(\frac{1}{L_{\text{t}}} \mathbf F_{L_{\text{t}}}^H
 	  \boldsymbol\Psi(  {\mathbf\Sigma}_{ \text{t},n}^{-1}) \Big)   
 	   \odot  \Big( \mathbf F_{L_{\text{t}}} 
 	  \boldsymbol\Psi \big(  \mathbf p_{\text{t},n}^{*}(0)\mathbf p_{\text{t},n}^{T}(0) \big) \Big)\Big)    
 		\mathbf 1_{L_{\text{t}}} \Big) \bigg)_{\widetilde{\mathcal D}}
 	 , \label{Proofkeyeq3} 
 \end{align}}%
where (a) is due to~(\ref{C1_asyn}), (b) is due to~(\ref{Core_EQ}), (c) is due to~(\ref{PEQ_trans}),
	 (d) is due to 
	\begin{align}
		& \left(\left(\boldsymbol\psi_{\ell} \left(  \mathbf p_{\text{t},n}^{*}(t)\mathbf p_{\text{t},n}^{T}(t) \right) \right)_{t \in \mathcal D}  \right)_{k,m} =\left(\boldsymbol\psi_{\ell}    \left(  \mathbf p_{\text{t},n}^{*}(0)\mathbf p_{\text{t},n}^{T}(0) \right) \right)_{1+ ((L_{\text{t}}+k-m )  \bmod  L_{\text{t}}) } , \ k \in {\mathcal L}_{\text{t}},m\in\widetilde{\mathcal D}, \nonumber
		 \end{align}
	   and the eigen-decompositions of circulant matrices,\footnote{If $\mathbf C \in \mathbb C^{K \times K}$ satisfies $ (\mathbf C)_{k,m} = (\mathbf C_{:,1})_{1+(K+k-m )  \bmod K}$ for all $k,m\in\mathcal K$, then $\mathbf C$ is a circulant matrix and its eigen-decomposition is $\mathbf C  = 
	   		\frac{1}{K} \mathbf F_{K}^H \mathrm{diag}\Big( \mathbf F_{K} \mathbf C_{:,1}\Big) \mathbf F_{{K}}$~\cite{CM_Gray_2006}.}
and (e) is due to~(\ref{PEQ_odot}) and~(\ref{PEQ_odottrans}).
Next, by noting that $\boldsymbol\beta_{\text{t},n} =  \frac{ g_n }{1+ \kappa_n} ( J(0, \mathbf p_{\text{t},n}(t), \mathbf \Sigma_{\text{t},n}^{-1}\widetilde{ \mathbf Y}_{\text{t}}  \widetilde{ \mathbf Y}_{\text{t}}^H \mathbf \Sigma_{\text{t},n}^{-1}))_{t \in  \hat {\mathcal X}_{\text{t}}}$, we can show~(\ref{C2t_Ri_n}) following the proof for~(\ref{C1t_Ri_n}).
 Finally, we show~(\ref{C3t_Ri_n}) as follows:
 {\begin{align}
	&  \frac{M }{2} 	\sqrt{\frac{1+ \kappa_n}{g_n\kappa_n } }	\boldsymbol	\eta_{\text{t},n } \overset{(a)} = 	\mathrm{Re} \left(	\Big(
	\Big(  \mathbf p_{\text{t},n}(t)  \Big)_{t\in \mathcal D} \Big)^T
	\Big(	\overline {\bf h}_n^T \overline {\mathbf Y}_{\text{t},n}^H        {\bf\Sigma}_{\text{t} ,n}^{-1}     \Big)^T  \right) 
	 \nonumber \\
	& 
	\overset{(b)}= 	  \mathrm {Re}
	\Big( \mathbf F_{L_{\text{t}}, \widetilde{\mathcal D}, :}   \mathrm{diag}\Big( \mathbf F_{L_{\text{t}}} 
	\mathbf p_{\text{t},n}(0) \Big)  \frac{1}{L_{\text{t}}} \mathbf F_{L_{\text{t}}}^H \Big(	\overline {\bf h}_n^T \overline {\mathbf Y}_{\text{t},n}^H        {\bf\Sigma}_{\text{t} ,n}^{-1}     \Big)^T     \Big)   
	\nonumber \\
	&\overset{(c)}=	 \left( \mathrm {Re} \Big(
	\mathbf 	F_{L_{\text{t}}}
	\left(\frac{1}{L_{\text{t}}}\mathbf
	F_{L_{\text{t}}}^{H}
	(	\overline {\bf h}_n^T  \widetilde {\mathbf Y}_{\text{t},n}^H        {\bf\Sigma}_{\text{t} ,n}^{-1} )^T \right) \odot \Big (\mathbf F_{L_{\text{t}}}  \mathbf p_{\text{t},n}(0 ) \Big)
	\Big)  \right)_{\widetilde{\mathcal D}}, \nonumber
\end{align}}%
where (a) is due to~(\ref{C3_asyn}), 
 (b) is due to 
 $ \left((	\mathbf p_{\text{t},n}(t))_{t\in \mathcal D}  \right)_{k,m} =\left(	\mathbf p_{\text{t},n}(0) \right) _{1+ (L_{\text{t}}+k-m )  \bmod  L_{\text{t}} } $, $k \in {\mathcal L}_{\text{t}}, m\in \widetilde{\mathcal D}$ and the eigen-decompositions of circulant matrices,
 and (c) is due to~(\ref{PEQ_odot}).

 	\vspace{ -2mm}
 \section*{Appendix E: Proof for Lemma~\ref{Lem:Compute_f}}

First, we show~(\ref{C1f_Ri_n}) as follows:
 {\begin{align}
 	&\frac{1+ \kappa_n }{2 g_n}\boldsymbol	\alpha_{\text{f},n }   \overset{(a)}=   \frac{1}{2}\left(J(\omega, \mathbf p_n, \mathbf \Sigma_{\text{f},n}^{-1})  \right)_{\omega \in  \hat {\mathcal X}_{\text{f}}} \nonumber \\
 &	\overset{(b)}=
  	\left( \mathrm{Re}\left( 
  \boldsymbol\tau_{\text{f}}^T \left(\omega^{(q)}  \right)   	\Big(   \boldsymbol\psi_{\ell}^T \left(  \mathbf p_n^{*}   \mathbf p_n^T    \right)  \boldsymbol\psi_{\ell} \left(   {\mathbf\Sigma}_{ \text{f},n}^{-1} \right)\Big)_{\ell \in \mathcal L_{\text{f}}}     \right)  \right)_{{q\in \mathcal Q}}\nonumber \\
 	& \overset{(c)}=    \mathrm{Re}\left( 
 	\left(	\left(\boldsymbol\tau_{\text{f}} (\omega^{(q)}  )\right)_{q\in \mathcal Q} \right)^T
 	  	\Big(  \boldsymbol\psi_{\ell}^T \left(  \mathbf p_n^{*}   \mathbf p_n^T    \right) \boldsymbol\psi_{\ell} \left(   {\mathbf\Sigma}_{ \text{f},n}^{-1} \right) \Big)_{\ell \in \mathcal L_{\text{f}}}     \right)  
 	\nonumber \\
 	& \overset{(d)}=  \mathrm {Re} \left(\mathbf T_{\text{f}, \mathcal Q,: }
 	\Big(     \boldsymbol\Psi \left(  {\mathbf\Sigma}_{ \text{f},n}^{-1}  \right)   \odot     \boldsymbol\Psi \left(  \mathbf p_n^{*}   \mathbf p_n^T    \right)      \Big)^T  \mathbf 1_{L_{\text{f}}} \right)  
 	\nonumber \\
 	&=
 	    \left( \mathrm {Re} \left(  \mathbf
 	T_{\text{f}} 
 	\Big(      \boldsymbol\Psi (  {\mathbf\Sigma}_{ \text{f},n}^{-1})   \odot        \boldsymbol\Psi (  \mathbf p_n^{*}   \mathbf p_n^T    )      \Big)^T  \mathbf 1_{L_{\text{f}}} \right) \right)_{\mathcal Q}
 	,  \nonumber
 \end{align}}%
where 
(a) is due to~(\ref{C1_asyn}), (b) is due to~(\ref{Core_EQ}), 
(c) is due to~(\ref{PEQ_trans}), and (d) is due to 
 $	\left(	\left(\boldsymbol\tau_{\text{f}} (\omega^{(q)}  )\right)_{q\in \mathcal Q} \right)^T = \mathbf T_{\text{f}, \mathcal Q,: } $ and the identity $(\mathbf u^T_{k} \mathbf v_k)_{k \in \mathcal K} = \left((\mathbf v_k)_{k \in \mathcal K} \odot (\mathbf u_k)_{k \in \mathcal K} \right)^T \mathbf 1_K$, for $ \mathbf u_k, \mathbf v_k \in \mathbb C^K, k\in\mathcal K$.
Next, by noting that $\boldsymbol\beta_{\text{f},n} =  \frac{ g_n }{1+ \kappa_n} ( J(\omega , \mathbf p_n, $ $ \mathbf \Sigma_{\text{f},n}^{-1}\widetilde{ \mathbf Y}_{\text{f}}  \widetilde{ \mathbf Y}_{\text{f}}^H \mathbf \Sigma_{\text{f},n}^{-1}))_{\omega \in  \hat {\mathcal X}_{\text{f}}}$, we can show~(\ref{C2f_Ri_n}) following the proof for~(\ref{C1f_Ri_n}).
Finally, we show~(\ref{C3f_Ri_n}) as follows:
{\begin{align}
		&  \frac{M}{2} 	\sqrt{\frac{1+ \kappa_n}{g_n\kappa_n } } \boldsymbol	\eta_{\text{f},n } \overset{(a)}  =  	 \left(\mathrm{Re}  \Big(( \boldsymbol\tau_{\text{f}}  ( \omega  ) )^T
		\Big(	\overline {\bf h}_n^T \overline {\mathbf Y}_{\text{f},n}^H        {\bf\Sigma}_{\text{f},n}^{-1}  \mathrm{diag}(\mathbf p_n )   \Big)^T  \Big) \right)_{\omega \in \mathcal X_{\text{f}}}  \nonumber \\
	&= \mathrm{Re} \left(	\Big(	\Big(\boldsymbol\tau_{\text{f}} (\omega^{(q)}  )\Big)_{q\in \mathcal Q} \Big)^T
		\Big(	\overline {\bf h}_n^T \overline {\mathbf Y}_{\text{f},n}^H        {\bf\Sigma}_{\text{f} ,n}^{-1}  \mathrm{diag}(\mathbf p_n )   \Big)^T  \right)
		\nonumber \\
		& \overset{(b)}= 
		  \left( \mathrm {Re} \left(
		\mathbf  T_{\text{f}}  
		\left((\overline {\bf h}_n^T  \widetilde{\mathbf Y}_{\text{f},n}^H         {\bf\Sigma}_{\text{f} ,n}^{-1})^T    \odot \mathbf p_n  \right)
		\right)  \right)_{\mathcal Q}
, \nonumber
	\end{align}}%
where (a) is due to~(\ref{C3_asyn}), and (b) is due to $	\left(	\left(\boldsymbol\tau_{\text{f}} (\omega^{(q)}  )\right)_{q\in \mathcal Q} \right)^T = \mathbf T_{\text{f}, \mathcal Q,: }  $ and~(\ref{PEQ_odot}).

   	\vspace{ -2mm}
 \section*{Appendix F: Details for Computing Matrix Multiplications with $\mathbf T_i$ by IFFT}
 
For $i\in $ \{f, (t,f)\}, substituting $ \omega^{(q)} = \frac{  q-1 }{Q} 2\pi  $ into~(\ref{tau}), we have $\boldsymbol\tau_{i}(\omega^{(q)}) = (e^{j(\ell-1)(q-1)\frac{  2\pi }{Q} })_{\ell \in  \mathcal L_{i}}  $, and hence $\mathbf	T_{i} $ can be viewed as a submatrix of a $\DeltaQ$-dimensional IDFT matrix, i.e., $\mathbf	T_{i} =  \mathbf F^H_{\DeltaQ, \Tseti ,\mathcal L_{i}} $, where $\DeltaQ\triangleq Q\left\lceil  \frac{L_{i}}{Q}      \right\rceil   \geq L_{i} $, $\Tseti \triangleq \left\{   (q-1)  \left\lceil  \frac{L_{i}}{Q}      \right\rceil   +1 \Big|q \in \{1,2,...,Q\}  \right\} $.
 Hence, the matrix-vector multiplications with $\mathbf	T_{i}$ can be efficiently computed using zero-padded $ \DeltaQ$-dimensional IFFT, i.e.,
  $  \mathbf  T _{i}  \mathbf b_i = \DeltaQ   \left( \mathrm {IFFT} ( [ \mathbf b_i^T  , \mathbf 0^T_{\DeltaQ-L_{i}} ]^T ) \right)_{\Tseti}  \in \mathbb C^{Q} $ for arbitrary $\mathbf b_i \in \mathbb C^{L_{i}}$.

\section*{Appendix G: Proof for Lemma~\ref{Lem:Compute_tf4}}

First, we show~(\ref{C1tf4_Ri_n}) as follows:
 \begin{align}
	&  \frac {1+ \kappa_n}{ 2g_n }\boldsymbol \alpha_{\text{(t,f)},n }   
	  \overset{(a)}
	   =  \frac{1}{2}  \left(J(\omega, \mathbf p_{\text{(t,f)},n}(t,0 ), \mathbf \Sigma_{\text{(t,f)},n}^{-1}) \right)_{t \in \hat {\mathcal X}_{\text{t}}, \omega \in \hat {\mathcal X}_{\text{f}}} 
	    \nonumber \\
	 &\overset{(b)}   =
 	   \bigg( \mathrm{Re}\bigg( 	\boldsymbol \tau_{ \text{(t,f)}}^T( \omega^{(q)})  
	\Big(     \boldsymbol\psi_{ \ell} \Big(  \mathbf p_{\text{(t,f)},n}^{*}(t,0)\mathbf p_{\text{(t,f)},n}^{T}(t,0) \Big)^T  \boldsymbol\psi_{ \ell}  (  {\mathbf\Sigma}_{ \text{(t,f)},n}^{-1})    \Big)_{\ell \in \mathcal L_{\text{(t,f)}}} 
	  \bigg) \bigg)_{t \in \mathcal D,  q\in \mathcal Q} \nonumber \\
& \overset{(c)} =  \mathrm{Re}\bigg(  \!\!
\bigg(  \!\!     \Big(  \boldsymbol\psi_{\ell} \Big(  \mathbf p_{\text{(t,f)},n}^{*}(t,0)\mathbf p_{\text{(t,f)},n}^{T}(t,0) \Big) \Big)_{t \in \mathcal D} ^T \!   \boldsymbol\psi_{ \ell}  (  {\mathbf\Sigma}_{ \text{(t,f)},n}^{-1})    \bigg)_{ \ell \in \mathcal L_{\text{(t,f)}}} 
   \left( \mathbf T_{\text{(t,f)}, \mathcal Q,: }  \right)^T   \bigg) 
  \nonumber \\
 & \overset{(d)}=    \bigg(\mathrm {Re}   \Big(   \Big( \mathbf F_{L_{\text{(t,f)}} }   \Big(    
 \Big(\frac{1}{L_{\text{(t,f)}}} \mathbf F_{L_{\text{(t,f)}}}^H
 \boldsymbol\Psi(  {\mathbf\Sigma}_{ \text{(t,f)},n}^{-1}) \Big)   \odot  \Big( \mathbf F_{L_{\text{(t,f)}}} 	\boldsymbol\Psi  (  \mathbf p^{*}_{\text{(t,f)},n}(0,0 )  \mathbf p^T_{\text{(t,f)},n}(0,0 )  ) \Big)\Big) 
 \Big)_{\widetilde{\mathcal D},:}    \mathbf T_{\text{(t,f)}}^T      \Big) \bigg)_{:,\mathcal Q} ,  \nonumber 
\end{align} %
where (a) is due to~(\ref{C1_asyn}), (b) is due to~(\ref{Core_EQ}), (c)
	is due to~(\ref{PEQ_trans}) and $	 	\left(\boldsymbol\tau_{\text{(t,f)}} (\omega^{(q)}  )\right)_{q\in \mathcal Q}  =	\left( \mathbf T_{\text{(t,f)}, \mathcal Q,: }  \right)^T$, and (d)
follows from~(\ref{Proofkeyeq1})-(\ref{Proofkeyeq3}).
Next, by noting that $\boldsymbol\beta_{\text{(t,f)},n} =  \frac{ g_n }{1+ \kappa_n} \big(J(\omega, \mathbf p_{\text{(t,f)},n}(t,0 ), {\bf\Sigma}_{\text{(t,f)},n }^{-1}    \widetilde {\mathbf Y}_{\text{(t,f)} ,n}    \widetilde {\mathbf Y}_{\text{(t,f)},n }^H   $ $   {\bf\Sigma}_{\text{(t,f)} ,n}^{-1}  \big)_{t \in \hat {\mathcal X}_{\text{t}}, \omega \in \hat {\mathcal X}_{\text{f}}}$, we can show~(\ref{C2tf4_Ri_n}) following  the proof for~(\ref{C1tf4_Ri_n}).
Finally, we show~(\ref{C3tf4_Ri_n}) as follows:
 \begin{align}
 	&  \frac{M}{2} 	\sqrt{\frac{1+ \kappa_n}{g_n\kappa_n } } \boldsymbol	\eta_{\text{(t,f)},n } \overset{(a)} =
  	\mathrm{Re} \left(\!	
 	\Big( \overline {\bf h}_n^T \overline {\mathbf Y}_{\text{(t,f)},n}^H        {\bf\Sigma}_{\text{(t,f)} ,n}^{-1}  \mathrm{diag}(\mathbf p_{\text{(t,f)},n}(t,0) )   \Big)_{t \in \mathcal D}  \!\!	\Big(\boldsymbol\tau_{\text{(t,f)}} (\omega^{(q)}  )\Big)_{q\in \mathcal Q}  \!   \right)  
 \nonumber \\
 & \overset{(b)}= \! \left( \!\!\mathrm {Re} \!\left( \! \left( \!\left( (\overline {\bf h}_n^T  \widetilde{\mathbf Y}_{\text{(t,f)},n}^H         {\bf\Sigma}_{\text{(t,f)} ,n}^{-1})^T \odot \mathbf p_{\text{(t,f)},n}(t,0) \right)_{t \in \mathcal D}\right)^T  \!\! 	\mathbf  T_{\text{(t,f)}}^T   \!  \right) \!\right)_{:,\mathcal Q}, \nonumber
 \end{align} %
where (a) is due to $\mathbf p_{\text{(t,f)},n}(t,\omega ) = \mathrm{diag}(\boldsymbol\tau_{\text{(t,f)}}(\omega))\mathbf p_{\text{(t,f)},n}(t,0 )$ and~(\ref{C1_asyn}), and (b) is due to 
  $ 	\left(\boldsymbol\tau_{\text{(t,f)}} (\omega^{(q)}  )\right)_{q\in \mathcal Q} = 	\left(\mathbf T_{\text{(t,f)}, \mathcal Q,: }  \right)^T $ and~(\ref{PEQ_odot}).

 \section*{Appendix H: Complexity Analysis of Step $8$ of Algorithm~\ref{Alg:Ri_asyn_IM}} 

 \mcone{As $ \mathbf U_i\boldsymbol \Psi (  \mathbf p_{i,n}^{*}(0)\mathbf p_{i,n}^T(0))$, $i \in \text{\{t, f, (t,f)\}}$ and $\mathbf F_{L_{\text{t}}} \mathbf p_{\text{t},n}(0 )    $ are computed 
 	before running Algorithm~\ref{Alg:Ri_asyn_IM},
 	and $	\overline {\bf h}_n^T  \widetilde {\mathbf Y}_{i,n}^H        {\bf\Sigma}_{i,n}^{-1}$ is computed in Step~7, the corresponding computational complexities are not considered in the complexity analysis for Step~$8$ below. 
 	Besides, we need some basic complexity results in the complexity analysis.  
 	For $\mathbf u,\mathbf v,\mathbf w \in \mathbb C^K$, and Hermitian matrix $\mathbf B\in\mathbb C^{K\times K}$, the flop counts for $\boldsymbol\Psi ( \mathbf B)$, $\mathbf u \odot \mathbf v$, $\mathbf u^H \mathbf v$, $\mathbf u^H \mathbf 1_{K}$, $\mathrm{FFT}(\mathbf u)$, and $ \mathrm{IFFT}(\mathbf u)$ are $K^2$, $ 6K$, $8K-2$, $2K-2$, $5K \log_2 K$, and $5K \log_2 K$~\cite{FFT-FLOP}, respectively.
 	Now, we analyze the computational complexity of Step~$8$.
 	First, we compute $\boldsymbol\Psi (  {\mathbf\Sigma}_{ i ,n}^{-1})$
 	and
 	$\boldsymbol\Psi( {\bf\Sigma}_{i,n }^{-1}   \widetilde {\mathbf Y}_{i ,n}    \widetilde {\mathbf Y}_{i,n }^H      {\bf\Sigma}_{i,n}^{-1}   )$ in both $ {L_i^2} $ flops. 
 	Then, we compute $\boldsymbol\alpha_{\text{t},n}$, $\boldsymbol\beta_{\text{t},n}$, and $\boldsymbol\eta_{\text{t},n}$ in $( 5 L_{\text{t}} + 5 )  L_{\text{t}} \log_2  L_{\text{t}} + 8 L_{\text{t}}^2-2 L_{\text{t}}$ flops, $( 5 L_{\text{t}} + 5 )  L_{\text{t}} \log_2  L_{\text{t}} + 8 L_{\text{t}}^2-2 L_{\text{t}}$ flops, and $10L_{\text{t}} \log_2  L_{\text{t}}+6L_{\text{t}} $ flops, respectively;
 	we compute $\boldsymbol\alpha_{\text{f},n}$, $\boldsymbol\beta_{\text{f},n}$, and $\boldsymbol\eta_{\text{f},n}$ in $8  L_{\text{f}}^2 - 2  L_{\text{f}}+  5  Q \log_2 Q $ flops, $8  L_{\text{f}}^2 - 2  L_{\text{f}}+  5  Q \log_2 Q $ flops, and $ 6L_{\text{f}} + 5  Q \log_2 Q $ flops, respectively;
 	we compute $\boldsymbol\alpha_{\text{(t,f)},n}$, $\boldsymbol\beta_{\text{(t,f)},n}$, and $\boldsymbol\eta_{\text{(t,f)},n}$ in $10  L_{\text{(t,f)}}^2 \log_2 L_{\text{(t,f)}} + 6   L_{\text{(t,f)}}^2  +  5  (D+1) Q\log_2 Q $ flops, $10  L_{\text{(t,f)}}^2 \log_2 L_{\text{(t,f)}} + 6   L_{\text{(t,f)}}^2  +  5  (D+1) Q\log_2 Q $ flops, and $ 6(D+1)L_{\text{(t,f)}} + 5(D+1)  Q \log_2 Q $ flops, respectively. 
 	Thus, the total costs of Step $8$ in Algorithm~\ref{Alg:Ri_asyn_IM} for asynchronous case-t, asynchronous case-f, and asynchronous case-(t,f) are
 	$  10 L_{\text{t}}^2  \log_2  L_{\text{t}}+18 L_{\text{t}}^2  + 20 L_{\text{t}}  \log_2 L_{\text{t}} +   2L_{\text{t}}  $ flops,
 	$18 L_{\text{f}}^2  +  2L_{\text{f}}   +15Q \log_2 Q $ flops, and 
 	$ 20L_{\text{(t,f)}}^2   \log_2 L_{\text{(t,f)}} +  14 L_{\text{(t,f)}}^2  + 6(D+1)L_{\text{(t,f)}}+15(D+1) Q \log_2 Q$ flops,
 	respectively.
 	By keeping only dominant terms and eliminating constant multipliers except for $D$ and $\Omega$, we can obtain the computational complexity of Step~8 of Algorithm~\ref{Alg:Ri_asyn_IM}.}
 \vspace{-6mm}


\section*{Appendix I: Proof for Lemma~\ref{Lem:FCC}} 
  	As Algorithm~\ref{Alg:Ri_asyn_ts} and Algorithm~\ref{Alg:Ri_asyn_IM} differentiate with each other only in the computation methods for $\boldsymbol\alpha_{i,n}, \boldsymbol\beta_{i,n}, \boldsymbol\eta_{i,n}$ (cf. Steps~7-10 of Algorithm~\ref{Alg:Ri_asyn_ts} and Steps~7,~8,~12 of Algorithm~\ref{Alg:Ri_asyn_IM}),
  	we only need to compare the flop count of Steps $7$-$10$ of Algorithm~\ref{Alg:Ri_asyn_ts} and the flop count of
  	Steps $7$, $8$, $12$ of Algorithm~\ref{Alg:Ri_asyn_IM}, denoted by $F^{(2)}_{i}$ and $F^{(3)}_{i}$, respectively.
  	As Algorithm~\ref{Alg:Ri_asyn_ts} and Algorithm~\ref{Alg:Ri_asyn_IM} differentiate with each other only in the computation methods for $\boldsymbol\alpha_{i,n}, \boldsymbol\beta_{i,n}, \boldsymbol\eta_{i,n}$ (cf. Steps~7-10 of Algorithm~\ref{Alg:Ri_asyn_ts} and Steps~7,~8,~12 of Algorithm~\ref{Alg:Ri_asyn_IM}),
  	we only need to compare the flop count of Steps $7$-$10$ of Algorithm~\ref{Alg:Ri_asyn_ts} and the flop count of
  	Steps $7$, $8$, $12$ of Algorithm~\ref{Alg:Ri_asyn_IM}, denoted by $F^{(2)}_{i}$ and $F^{(3)}_{i}$, respectively.
  	\mcfive{Besides the basic complexity results introduced in Appendix H, we further need the following result:
  		for $\mu,\rho \in \mathbb R$, $\mathbf u,\mathbf v,\mathbf w \in \mathbb C^K$, and Hermitian matrix $\mathbf B\in\mathbb C^{K\times K}$, the flop count for $ \mathbf B + ((\nu\mathbf u)  \mathbf v^H +\mathbf v  (\nu\mathbf u)^H  )+  (\rho\mathbf w) (\rho\mathbf w)^H $ is $10K^2 + 4K$ (computed utilizing Hermitian symmetry).
  		Now, we characterize $F^{(2)}_{i}$ and $F^{(3)}_{i}$.
  		Similarly to the analysis in Appendix H, we can show that $F^{(2)}_{i} = \vert \hat {\mathcal X}_{i}\vert (8L_i^2 +8L_i M +10{L_i}+6M-5)$, $i \in$ \{t, f, (t,f)\}, 
  		$ F^{(3)}_{\text{t}} = 10 L_{\text{t}}^2  \log_2  L_{\text{t}}+ 90 L_{\text{t}}^2  + 20 L_{\text{t}}  \log_2 L_{\text{t}} + 16  L_{\text{t}}M +  6L_{\text{t}}   $,
  		$ F^{(3)}_{\text{f}} = 90  L_{\text{f}}^2  +   16 L_{\text{f}}M + 6 L_{\text{f}} +15Q \log_2 Q  $,
  		and
  		$ F^{(3)}_{\text{(t,f)}} = 20L_{\text{(t,f)}}^2   \log_2 L_{\text{(t,f)}} + 86 L_{\text{(t,f)}}^2 + 16 L_{\text{(t,f)}}M +  6(D+\frac{5}{3})L_{\text{(t,f)}}+15(D+1) Q \log_2 Q  $,
  		where $L_i$ and $ \vert \hat { \mathcal X}_{i}\vert$ are given in~(\ref{L_i}) and~(\ref{hatX}), respectively.
In what follows, we compare $ F_i^{(2)}$ and $ F_i^{(3)}$ by comparing their lower bounds and upper bounds.
First, we have:
\begin{align}
	&  {\underline F}_i^{(2)} \triangleq\vert \hat {\mathcal X}_{i}\vert (8L_i^2 +8L_i M)   \overset{(a)} <  F^{(2)}_{i}    \overset{(a)}<\vert \hat {\mathcal X}_{i}\vert (12L_i^2 + 10L_i M)
	\triangleq {\overline F}_i^{(2)}  ,\label{A2UBLB} \\ 
	&   {\underline F}_i^{(3)} \triangleq  90 L_{i}^2 + 16 L_{i}M  \overset{(a)}< F^{(3)}_{i}   \nonumber \\
	&\overset{(b)}<    (112+ 20 \log_2 L+   \mathbb{I}\left(i\in\text{\{t, (t,f)\}} \right)20 \frac{  \log_2 (e)}{L}   D ) L_{i}^2  
	+  16\left(M +\mathbb{I}\left(i\in\text{\{f, (t,f)\}} \right) Q \log_2 Q\right) L_{i} \triangleq {\overline F}_i^{(3)}, \label{A3UBLB}
\end{align}
where (a) is due to $L \geq 6$ and $ M\geq 1$, (b) is due to
$L  > 6$ and $ M,Q\geq 1$, and 
$\log_2(L+D) <  \frac{ \log_2 (e)}{L}D+\log_2 L   $ 
and $D+\frac{5}{3} <L+D$.
By~(\ref{A2UBLB}) and~(\ref{A3UBLB}), we can show $ {\overline F}_i^{(2)}< {\underline F}_i^{(3)}$ ($ {\underline F}_i^{(2)}> {\overline F}_i^{(3)}$), implying $  { F}_i^{(2)}<{ F}_i^{(3)}$ ($ {F}_i^{(2)}>{ F}_i^{(3)}$), if the following conditions hold:
(i) Asynchronous case-t: $   D <  \overline D_{\text{t}} = \frac{\sqrt{ (12L+10M-78)^2 + 48(78L+6M ) } +78-12L-10M               }{        24}$ ($  D >  \underline D_{\text{t}} =  \frac{26+5\log_2 L}{2-5 \frac{  \log_2 (e)}{6} }$);
(ii) Asynchronous case-f: $\Omega <  \overline \Omega_{\text{f}}  =      \frac{ (39L +3M)\pi}{ Q(6L+5M) }  $ ($\Omega >\underline \Omega_{\text{f}} =   	 \frac{  ( 120 L+ 20 L \log_2 L + 24M+16Q\log_2 Q  )\pi  }{ 8Q(L+M) }$);
(iii) Asynchronous case-(t,f): $  D< \overline D_{{\text{(t,f)}}} = -\mathbb I (\theta_{\text{(t,f)}}\leq0) + \mathbb I (\theta_{\text{(t,f)}}>0) \big(  \frac{ 90 - S(\Omega)(12L+10M+12) + \sqrt{\theta_{\text{(t,f)}}}  }{        24 S(\Omega) }  \big) $ or $ \Omega < \overline \Omega_{\text{(t,f)}}   =   \frac{\pi}{Q}  \left( \frac{  90(L+D)+16M }{(D+1) (12(L+D) +10M ) }   -1\right)$ ($ D>  \underline D_{\text{(t,f)}} =  \max \left\{\frac{112+20\log_2 L -8 S(\Omega)}{8S(\Omega)-20 \frac{  \log_2 (e)}{L}} ,   \frac{16M + 16 Q\log_2 Q}{8MS(\Omega)} -1 \right\}$ or $\Omega >\underline \Omega_{{\text{(t,f)}}} =\frac{ \pi}{Q} \big(   \frac{  (112+ 20 \log_2 L + 20\frac{  \log_2 (e)}{L} D )(L+D) + 16(M + Q \log_2 Q)  }{8(D+1)(L+D+M)  } +1  \big)$).
Here, $S(\Omega ) \triangleq  2  \lfloor \frac{Q\Omega}{2\pi}  \rfloor	+\mathbb{I}(\Omega\neq\pi ) $ and $\theta_{\text{(t,f)}} \triangleq  (  S(\Omega)(12L+10M+12) -90 )^2 - 48 S(\Omega) (12L  S(\Omega) + 10M  S(\Omega) -90L-16M) $.}

   	\vspace{ -4mm}

\section*{Appendix J: Proof for Lemma~\ref{Lem:FCCL}} 
  	\vspace{ -2mm}

First, by substituting $M =L^s$ and $ Q=L^q$ into the computational complexities in Table~\ref{tab1:Complexity} and keeping the dominant terms, we can obtain the computational complexities in Table~\ref{tab1:Complexity_sup}.
Then, by comparing the computational complexities of Algorithm~\ref{Alg:Ri_asyn_ts} and Algorithm~\ref{Alg:Ri_asyn_IM} in terms of $L$ in each case, we can show the statements in Lemma~\ref{Lem:FCCL}.


\IEEEpeerreviewmaketitle

\ifCLASSOPTIONcaptionsoff
  \newpage
\fi




 \bibliographystyle{IEEEtran}

\end{document}